\documentclass{aa}  

\usepackage{lscape}
\usepackage{graphicx}
\usepackage{dcolumn}

\usepackage[version=4]{mhchem}
\usepackage[varg]{txfonts}

\usepackage{mathtools}

\usepackage{ulem}
\usepackage[separate-uncertainty = true]{siunitx}

\usepackage{txfonts}
\usepackage[colorlinks,bookmarks]{hyperref}
\usepackage{color}
\definecolor{linkblue}{rgb}{0,0,0.8}
\definecolor{linkgreen}{rgb}{0,0.5,0}
\definecolor{mygray}{gray}{0.6}
\hypersetup{pdfpagemode=UseNone, pdfstartview=FitH, linkcolor=linkblue, %
  citecolor=linkblue, urlcolor=linkblue}

\newcommand{\m}{\mathrm}
\newcommand{\angstrom}{\mbox{\normalfont\AA}}

\makeatletter
\renewcommand*\aa@pageof{, page \thepage{} of \pageref*{LastPage}}
\makeatother

\begin{document}

   \title{Equivalent widths of Lyman $\alpha$ emitters in MUSE-Wide and MUSE-Deep\thanks{The catalog described in Appendix B (Table B.1) is only available in electronic form at the CDS via anonymous ftp to \url{cdsarc.u-strasbg.fr} (\url{130.79.128.5})
or via \url{http://cdsweb.u-strasbg.fr/cgi-bin/qcat?J/A+A/}}}
   
   \author{J. Kerutt
          \inst{1}\fnmsep\inst{2}
          \and
          L. Wisotzki \inst{1}
          \and
          A. Verhamme \inst{2}
          \and 
          K. B. Schmidt \inst{1}
          \and 
          F. Leclercq \inst{2}
          \and 
          E. C. Herenz \inst{3}
          \and 
          T. Urrutia \inst{1}
          \and 
          T. Garel \inst{2,5}
          \and 
          T. Hashimoto \inst{6}
          \and 
          M. Maseda \inst{4}
          \and 
          J. Matthee \inst{7}
          \and 
          H. Kusakabe \inst{2}
          \and
          J. Schaye \inst{4}
          \and
          J. Richard \inst{5}
          \and
          B. Guiderdoni \inst{5}
          \and
          V. Mauerhofer \inst{2,5}
          \and 
          T. Nanayakkara \inst{4}
          \and
          E. Vitte \inst{2,3}
          }

   \institute{Leibniz-Institut f\"ur Astrophysik Potsdam (AIP), An der Sternwarte 16, 14482 Potsdam, Germany\\ \email{josephine.kerutt@unige.ch}
         \and
             Observatoire de Genève, Université de Genève, Chemin Pegasi 51, 1290 Versoix, Switzerland
        \and 
            ESO Vitacura, Alonso de Córdova 3107,Vitacura, Casilla 19001, Santiago de Chile, Chile
        \and
            Leiden Observatory, Leiden University, PO Box 9513, NL-2300 RA Leiden, the Netherlands
        \and 
            Univ Lyon, Univ Lyon1, Ens de Lyon, CNRS, Centre de Recherche Astrophysique de Lyon UMR5574, F-69230, Saint-Genis-Laval, France
        \and 
            Tomonaga Center for the History of the Universe (TCHoU), Faculty of Pure and Applied Sciences, University of Tsukuba, Tsukuba, Ibaraki 305-8571, Japan
        \and 
            ETH Z\"urich, Department of Physics, Wolfgang-Pauli-Str. 27, 8093 Z\"urich, Switzerland
            }

   \date{Received 29.07.2021; accepted 15.12.2021}

  \abstract
   {The hydrogen Lyman $\alpha$ line is often the only measurable feature in optical spectra of high-redshift galaxies. Its shape and strength are influenced by radiative transfer processes and the properties of the underlying stellar population. High equivalent widths of several hundred \AA\, are especially hard to explain by models and could point towards unusual stellar populations, for example with low metallicities, young stellar ages, and a top-heavy initial mass function. Other aspects influencing equivalent widths are the morphology of the galaxy and its gas properties.}
   {The aim of this study is to better understand the connection between the Lyman $\alpha$ rest-frame equivalent width ($\m{EW}_0$) and spectral properties as well as ultraviolet (UV) continuum morphology by obtaining reliable $\m{EW}_0$ histograms for a statistical sample of galaxies and by assessing the fraction of objects with large equivalent widths.}
   {We used integral field spectroscopy from the Multi Unit Spectroscopic Explorer (MUSE) combined with broad-band data from the Hubble Space Telescope (HST) to measure $\m{EW}_0$. We analysed the emission lines of $1920$ Lyman $\alpha$ emitters (LAEs) detected in the full MUSE-Wide (one hour exposure time) and MUSE-Deep (ten hour exposure time) surveys and found UV continuum counterparts in archival HST data. We fitted the UV continuum photometric images using the \texttt{Galfit} software to gain morphological information on the rest-UV emission and fitted the spectra obtained from MUSE to determine the double peak fraction, asymmetry, full-width at half maximum, and flux of the Lyman $\alpha$ line.}
   {The two surveys show different histograms of Lyman $\alpha$ EW$_0$. In MUSE-Wide, $20\%$ of objects have $\m{EW}_0 > 240\,\angstrom$, while this fraction is only $11\%$ in MUSE-Deep and $\approx 16\%$ for the full sample. This includes objects without HST continuum counterparts (one-third of our sample), for which we give lower limits for EW$_0$. The object with the highest securely measured EW$_0$ has $\m{EW}_0=589 \pm 193\, \angstrom$ (the highest lower limit being $\m{EW}_0=4464\, \angstrom$). We investigate the connection between EW$_0$ and Lyman $\alpha$ spectral or UV continuum morphological properties.}
  {The survey depth has to be taken into account when studying EW$_0$ distributions. We find that in general, high EW$_0$ objects can have a wide range of spectral and UV morphological properties, which might reflect that the underlying causes for high EW$_0$ values are equally varied.}


   \maketitle
%
%-------------------------------------------------------------------

\section{Introduction} \label{sec:Introduction} 
An important question in galaxy evolution is how stars form in extremely metal poor environments. One way to address this is by observing star-forming galaxies at high redshifts. As already pointed out by \citet{PartridgePeebles}, a tell-tale signature of such galaxies is their hydrogen Lyman $\alpha$ emission.

Young, massive, hot stars in star-forming regions produce hydrogen ionising radiation and when the hydrogen recombines, the most likely outcome of the recombination process is a Lyman $\alpha$ photon. Roughly $\sim 2/3$ of the recombination events lead to the emission of such a Lyman $\alpha$ photon, assuming case B and a temperature of $10^4\,$K (e.g. \citealp{Dijkstra2014}). The higher the number of such massive, hot stars, the more ionising photons are produced, but stars that are bright in the Lyman continuum (LyC) are also the shortest lived, which means the starburst age, star formation rate, and stellar initial mass function (IMF) regulate the strength of the Lyman $\alpha$ radiation. A low metallicity in the gas forming the stars is typically expected to influence both the IMF, which is assumed to be universal except for metallicities of $Z<10^{-3} Z_{\odot}$, and the number of ionising photons produced in such stars (\citealp{Schaerer2002, Raiter2010, Pallottini2015, Sobral2015}). 

Other effects to be considered for the production efficiency of ionising photons are the rotation of the stars (see e.g. \citealp{Haemmerle2017}) and the evolution of binaries, which can strip stars of their envelopes resulting in hot, compact Helium stars which have a higher production of ionising photons (e.g. \citealp{Stanway2016, Gotberg2017}), as well as the possible existence of super-massive stars with masses of $\m{M}\sim 10^4\,\m{M}_{\odot}$ (\citealp{Denissenkov2014}).
Thus a young starburst with a high star-formation rate, top-heavy IMF, and low metallicity can result in the production of a high number of Lyman $\alpha$ photons. 
The ratio between the luminosity of the Lyman $\alpha$ line and the continuum luminosity density at $\lambda_{\m{Lyman} \alpha} = 1215.67\,\angstrom$ is defined as the equivalent width (EW, \citealp{Schaerer2003}), which is usually corrected to the rest-frame equivalent width $\m{EW}_0~=~\m{EW}~/~(~1~+~z~)$. This traces in principle the production rate of ionising photons per stellar mass. It is thus sensitive to the physical properties of a galaxy discussed above. Studying the distribution of EW$_0$ in Lyman $\alpha$ emitting galaxies, or Lyman Alpha Emitters (LAEs), therefore enables us to put statistical constraints on those properties. 

It has often been quoted that there is an upper limit to the EW$_0$ of $240\, \angstrom$ (\citealp{Charlot1993, Malhotra2002, Schaerer2002, Laursen2013}), although this value\footnote{The number of $240\,\angstrom$ first appears in \citet{Malhotra2002}, probably based on a plot in \citet{Charlot1993}, who give $\m{EW}_0~\gtrsim~200\,~\angstrom$ as a limit. The limit of $240\,\angstrom$ is used in this paper as a way to compare our results to the literature.} is based on specific assumptions on the underlying stellar population (such as a constant SFR, an age of $\sim10^7$ years, an IMF with a slope of 0.5, and an upper cut-off of the IMF of $80\,\m{M}_{\odot}$, \citealp{Charlot1993}) and does not take into account radiation transfer effects through complex interstellar medium (ISM) kinematics and morphology, which can boost the EW$_0$ in addition. While a smaller rest-frame equivalent width can be explained by normal stellar populations or a lower escape fraction of Lyman $\alpha$ photons, a value exceeding this number implies unusual conditions in the star-forming regions of the galaxy on top of a high transmission through the ISM, circum galactic medium (CGM) and intergalactic medium (IGM) in the line of sight. Such unusual conditions can be sub-solar metallicities, high star-formation rates, young stellar ages, a top-heavy initial mass function, and a departure from case B (\citealp{Schaerer2003, Raiter2010, Maseda2020}). While a young starburst even at solar metallicity can explain EW$_0 \sim500\,\angstrom$ (\citealp{Raiter2010}), a low metallicity of $Z/Z_{\odot}\lesssim 1/20$, even for a normal IMF and older ages, can increase EW$_0$ by a factor of $\approx 70\%$.
Thus selecting objects with large EW$_0$ is the most efficient way to find objects that are likely interesting to study when searching for example for young massive stars, low metallicity, or even population III stars. 

Apart from photoionisation from stars and recombination, there are other channels that can produce Lyman $\alpha$ photons, even from outside the galaxy, which therefore do not correlate with stellar properties. 
In particular, gas falling in the potential well of galaxies, or shocks in the interstellar gas, can be cooled through de-excitation of collisionally excited hydrogen atoms, which produces Lyman $\alpha$ photons (a process that is not yet well understood, see reviews \citealp{Dijkstra2017, Faucher2017}). From numerical simulations, collisions are predicted to provide less than\,$10\%$ of the intrinsic Lyman $\alpha$ emission produced in the galaxy, but could still represent $40\%$ of the escaping radiation (e.g. \citealp{Dayal2010,Mitchell2021}).
Observationally it is still difficult to determine the contribution of gravitational cooling of infalling gas to the production of Lyman $\alpha$ photons (e.g. \citealp{Leclercq2017}). Studies of Lyman $\alpha$ halos have pointed in a direction where this scenario is not favoured over the production in the star-forming regions of the galaxy or in satellite galaxies (\citealp{Momose2016,Xue2017}).

A third production channel is fluorescence, which is a photoionisation and recombination event where the ionising photon exciting the hydrogen atom was not produced inside the galaxy itself, but comes from the external ultraviolet (UV) background (e.g. \citealp{Cantalupo2005, Kollmeier2010, Dijkstra2017}). This fluorescent emission always happens, but is predicted to be much fainter than the other production channels, except in the vicinity of a strong ionising source (\citealp{Cantalupo2005}). However, in so-called dark galaxies that are invisible in the UV but bright in Lyman $\alpha$, it has been proposed that the main production mechanism can indeed be fluorescence, although stacking analysis shows that there is still ongoing star formation (e.g. \citealp{Maseda2018}).
In both scenarios (collisional excitation and fluorescence), the strength of the Lyman $\alpha$ emission is not expected to relate to the strength of the UV continuum, and extreme EW$_0> 240\angstrom$  have been proposed as a way to identify dark galaxies, shining through Lyman $\alpha$ fluorescence triggered by a neighbouring quasar (\citealp{Cantalupo2007,Cantalupo2012, Marino2018}). 

On their way to the observer the Lyman $\alpha$ photons pass through the ISM, CGM and IGM. Since the photons are resonantly scattered by neutral hydrogen both in real and in frequency space, the strength and shape (e.g. \citealp{Verhamme2006, Henry2015}) of the Lyman $\alpha$ line will be affected by the neutral hydrogen column density (e.g. \citealp{Shibuya2014b, Yang2017}) as well as the dust content, the morphological properties, and kinematics of the gas (e.g. \citealp{Atek2008, Trainor2016}). 

There are only a few effects that can potentially increase the observed EW$_0$, but most radiative transfer effects result in a decrease. 
For the former, there have been several studies concerned with the influence of radiative transfer effects on boosting the Lyman $\alpha$ equivalent width, for example through the preferential escape of Lyman $\alpha$ emission over the UV continuum (\citealp{Neufeld1991}) or Lyman $\alpha$ beaming (\citealp{Behrens2014a, Behrens2014b, Verhamme2015}). Numerical simulations have shown however, that the gain in EW$_0$ is marginal (e.g. \citealp{Laursen2013, Gronke2014}). It thus seems that large EW$_0$ values are more likely explained by the properties of the underlying stellar populations or through external effects like collisions and fluorescence (which would not influence the UV radiation), than through radiative transfer effects.  

However, there are several reasons why we can expect the observed Lyman $\alpha$ EW$_0$ to be lowered. Since neutral hydrogen will cause the Lyman $\alpha$ photons to scatter, their path length is increased and thus also the probability to hit a dust grain and be destroyed or to get scattered out of the line of sight (see \citealp{Dijkstra2014} for a review), an effect that increases with increasing neutral hydrogen column density. The morphological and kinematic properties of the CGM can also play an important role in the radiative transfer of Lyman $\alpha$ photons and it has been shown that most Lyman $\alpha$ emitters have halos of extended Lyman $\alpha$ emission of ten times the size of the UV continuum (e.g. \citealp{Hayashino2004, Steidel2011, Momose2014, Momose2016, Wisotzki2016, Leclercq2017, Wisotzki2018, Leclercq2020}). Given a high enough surface brightness sensitivity of the data, it is possible that such halos are even ubiquitous (see Kusakabe et al. in prep.).

The morphological structure of the galaxy itself and the angle at which we see it can also have an effect on the Lyman $\alpha$ line (e.g. \citealp{Venemans2005, Gronwall2011, Bond2012, Jiang2013, Kobayashi2016, Paulino-Afonso2018}). A clumpy morphology could be caused by a merger, multiple star-forming regions or satellite galaxies, each of which could boost the Lyman $\alpha$ emission. Paths driven by supernovae can allow Lyman $\alpha$ emission to escape in a preferential direction (\citealp{Behrens2014a}), while the inclination of the galaxy can also lead to different EW$_0$ values (\citealp{Verhamme2012,Behrens2014b}), with face-on galaxies having potentially higher EW$_0$ (\citealp{Laursen2009, Shibuya2014a}).

The radiative transfer processes in the emitting galaxy can be imprinted on the shape of the Lyman $\alpha$ line, as a high neutral hydrogen column density in the ISM can broaden the Lyman $\alpha$ line or result in a blue bump (e.g. \citealp{Verhamme2015}). The neutral hydrogen in the IGM can then absorb any photons that are redshifted to the position of the Lyman $\alpha$ line wavelength, meaning any galactic emission to the blue of the Lyman $\alpha$ rest-frame wavelength, including the blue bump or the left side of the broadened line (e.g. \citealp{Laursen2011,Garel2021}), potentially changing its shape even after the photons left the galaxy. 

Therefore, to get the Lyman $\alpha$ EW$_0$ directly after the galaxy, some studies assume a symmetric, usually single-peaked Lyman $\alpha$ line (after leaving the ISM and potentially CGM), the blue part of which would be attenuated by the IGM, which would lead to an apparent EW$_0$ of half the intrinsic value (e.g. \citealp{Kashikawa2012, Zheng2014}), where we define the intrinsic EW$_0$ as the rest-frame EW we would observe without IGM attenuation. The caveat with this idea is that first, the Lyman $\alpha$ line can be double peaked after passing through neutral hydrogen in the ISM or CGM (\citealp{Hu2016, Matthee2018, Songaila2018, Meyer2020, Hayes2020}) and contrary to the assumption that the IGM absorbs all emission to the blue of the intrinsic Lyman $\alpha$ wavelength, blue bumps at high redshifts have indeed been observed (e.g. \citealp{Hayes2020}). Second, it has been shown that the typical asymmetric line profile can be explained on the basis of the gas kinematics alone (\citealp{Verhamme2006}) and the Lyman $\alpha$ line is often shifted with respect to the systemic redshift (e.g. \citealp{Kulas2012, Hashimoto2015, Verhamme2018, Muzahid2020}), even before the emission reaches the IGM (e.g. in the low-redshift LAE analogues in the Lyman Alpha Reference Sample, see \citealp{Ostlin2014, Rivera-Thorsen2015}, but also in Green Peas, see \citealp{Henry2015}). Although it is important to discuss the effects of the IGM on the Lyman $\alpha$ radiation when deriving the EW$_0$, it is not possible to correct for them for individual objects due to the redshift dependence and the high line of sight stochasticity of the opacity of the IGM (e.g. \citealp{Inoue2014, Byrohl2020, Bassett2021}). 

In previous studies, different fractions of objects with large EW$_0$ have been determined among high-redshift LAEs. Typically, LAEs are found using the narrow-band technique (e.g.\ \citealp{Hu1998, Malhotra2004, Ouchi2008}), which however needs spectroscopic follow-up observations to rule out the possibility of low redshift interlopers (e.g.\ \citealp{Erb2014}). There is no consensus yet on the fraction of high $\m{EW}_0$ values among high redshift LAEs, which is what we want to analyse in this paper.

Although purely narrow-band selected samples can produce a large number of LAE candidates, they need spectroscopic follow-up observations. In addition, they often lack the observational depth of deep Hubble Space Telescope (HST) data, since they get their large number of LAEs through their large surface area as they are restricted in the redshift range by the wavelength range given by the narrow-band. This makes it difficult to get good constraints on the UV continuum and thus on the EW$_0$ of Lyman $\alpha$. What is more, \citet{Maseda2020} argue that surveys based on a pre-selection by UV continuum detections of LAEs in imaging are biased against finding high EW$_0$ objects.

Moreover, taking the full extent of Lyman $\alpha$ haloes into account when determining the EW$_0$ of Lyman $\alpha$ lines is important and often difficult for narrow-band surveys or even slit-spectroscopy. When measured to large enough radii and to lower surface brightness limits, many galaxies that otherwise would not have been classified as such, are actually LAEs (\citealp{Steidel2011}). The contribution of the Lyman $\alpha$ halo is typically around $\sim65\%$ (\citealp{Leclercq2017}), which means it often even dominates the Lyman $\alpha$ emission.
For these reasons it is ideal to use integral field spectroscopy, to include the full Lyman $\alpha$ halo flux and correctly identify galaxies as LAEs.

Therefore we use data from the integral field spectrograph Multi Unit Spectroscopic Explorer (MUSE, \citealp{Bacon2010}) at the Very Large Telescope (VLT) in Chile, which is ideal for the detection of LAEs (e.g.\ \citealp{Bacon2015, Wisotzki2016}), combined with deep HST data to constrain the EW$_0$ as best as possible. MUSE has been used in previous studies of Lyman $\alpha$ EW$_0$ distributions, such as by \citet{Hashimoto2017}, who find a rather low fraction of high EW$_0>200\,\angstrom$ LAEs (12 out of 417) using MUSE data of the Hubble Ultra Deep Field (HUDF, \citealp{Beckwith2006, Bacon2015}). This study mostly probes the low luminosity part of the LAE luminosity function, which is why we use data from both the MUSE-Wide (\citealp{Herenz2017, Urrutia2019}) and MUSE-Deep (\citealp{Inami2017}) surveys, containing a large sample of LAEs (around $2000$) over a wide redshift range ($3\lesssim z\lesssim6$) and large survey area (100 pointings) of different observation depths. 

We build on the work of \citet{Hashimoto2017}, who construct a Lyman $\alpha$ EW$_0$ distribution with the LAEs from the MUSE-Deep survey only and we add the MUSE-Wide detections to get better statistical constraints and probe a larger field-of-view and a larger range of Lyman $\alpha$ and UV luminosities.

This paper is structured as follows: In Sect.~\ref{sec:Data} we describe the data from the MUSE-Wide and MUSE-Deep surveys, as well as the ancillary deep HST broad-band data. In Sect.~\ref{Sample Selection} we construct a sample of LAEs and describe the detection and classification of emission lines in MUSE data. We determine the UV continuum counterparts in the HST data and measure the UV continuum flux density as described in Sect.~\ref{The UV continuum}. In Sect.~\ref{The Lyman alpha line} we explain the line flux measurements as well as the fitting of the Lyman $\alpha$ lines.
We combine the line fluxes measured from the MUSE data with the continuum flux densities measured from HST to obtain EW$_0$ in Sect.~\ref{Equivalent Widths}, where we show the EW$_0$ distribution and discuss the differences between the MUSE-Wide and the MUSE-Deep surveys. In Sect.~\ref{Connecting Equivalent Widths to other Properties} we show connections between EW$_0$ and other properties of the LAEs and in Sect.~\ref{Discussion} we discuss the properties of the LAEs with the highest measured EW$_0$ and give a summary and conclusion in Sect.~\ref{Summary and Conclusion}. Throughout this paper we use AB magnitudes, physical distances and assume flat $\Lambda\m{CDM}$ cosmology with $\m{H}_0 = 70\,\m{km}\,\m{s}^{-1}\,\m{Mpc}^{-1}$, $\Omega_{\m{m}} = 0.3$ and $\Omega_{\Lambda}=0.7$.

%%%%%%%%%%%%%%%%%%%%%%%%%%%%%%%%%%%%%%%%%%%%%%%%%%%%%%%%%%%%%%%%%%%%%%%%%%%%%%%%%%%%%%%%%%%

\section{Data} \label{sec:Data} 

Measuring EW$_0$ can be dissected into two parts: determining the Lyman $\alpha$ line flux and the rest-frame UV continuum flux density. For both measurements we used different kinds of data, which made it possible to combine the power of integral field spectroscopy (for the line flux and other emission line properties) from MUSE (\citealp{Bacon2010}) with the depth of HST broad-band photometry (for the UV continuum). Therefore, special care had to be taken to properly combine the information gained from both data sets. 

We used the spectroscopic data from MUSE to detect the LAEs and measure the line properties such as line flux, FWHM, asymmetry, and spectral peak separation (in case of a double peak in the spectrum, which is common for Lyman $\alpha$ lines). However, even for the MUSE-Deep data, the exposure time is not long enough to reliably measure the rest-frame UV continuum for all LAEs directly from the MUSE data, which is needed for the determination of the EW$_0$. Therefore, we used deep, broad-band data from HST at wavelengths longer than the redshifted Lyman $\alpha$ line so as not to contaminate the HST bands with the Lyman $\alpha$ emission itself. 

\subsection{Spectroscopic information from MUSE}

MUSE is an integral field spectrograph and as such has two spatial and one spectral dimensions. It has a field of view (FoV) of one arcmin$^2$ and a spatial sampling of $0.2$ arcseconds. The spectral range covers $4750\, \angstrom$ to $9350\, \angstrom$ which allows for the detection of Lyman $\alpha$ at $1215.67\, \angstrom$ in the redshift range of $2.9<z<6.7$. Since this makes MUSE ideal to detect high-redshift LAEs, there are several surveys within the MUSE consortium which are aimed at studying their properties. In this paper we use data from the MUSE-Wide\footnote{For MUSE-Wide, the data and data products for the first 44 fields, such as cut-outs, mini-cubes and extracted spectra as well as emission line catalogues are publicly available and can be found at \url{https://musewide.aip.de/project/}.} (\citealp{Herenz2017, Urrutia2019}) and MUSE-Deep\footnote{A catalogue of objects in the MUSE-Deep fields (presented by \citealp{Inami2017}) is available at \url{http://cdsarc.u-strasbg.fr/viz-bin/qcat?J/A+A/608/A2}.} (\citealp{Bacon2017, Inami2017}) surveys. The MUSE-Wide survey, located in the CANDELS-Deep (\citealp{Giavalisco2004, Koekemoer2011}) as well as COSMOS (\citealp{Scoville2007}) fields, is specifically aimed at discovering a large number of bright LAEs around $L^*$ of the LAE luminosity function (LF, see \citealp{Herenz2019}), while the MUSE-Deep survey samples more sub-$L^*$ LAEs. 

The two surveys, both taken at the ESO-VLT during guaranteed time observations (GTOs) of the MUSE consortium, complement each other, as they have different depths and sizes. The MUSE-Wide survey provides a shallower approach, with one hour exposure time for each field, but covers a large survey area with a total of 100 fields. Part of the data used in this paper is already publicly available, see the data release of the first 44 fields (\citealp{Urrutia2019} which also contains a footprint of the MUSE-Wide survey). 
The pointings overlap by $\approx4\arcsec$, which reduces the total area covered to slightly less than\,$100\, \m{arcmin}^2$, covering a comoving volume of roughly $\sim 10^7\, \m{Mpc}^3$. 
The main part of MUSE-Wide is in the GOODS-S/CDFS and CANDELS-COSMOS areas with 60 fields. Eight additional fields are in the two HUDF parallel fields and 23 fields are in the COSMOS area. The nine HUDF fields of the MUSE-Deep survey area complete the 100 fields of MUSE-Wide (see Fig.~1 by \citealp{Urrutia2019} for a footprint of the MUSE-Wide survey).
The data reduction process for both MUSE-Wide and MUSE-Deep is described by \citet{Urrutia2019}, using the data reduction pipeline by \citet{Weilbacher2020}.

The detection limit for the Lyman $\alpha$ luminosity $\m{L}_{\m{Ly}\alpha}$ varies with wavelength due to the sensitivity curve of MUSE, the luminosity distance and the sky spectrum, as can be seen for example in Figs.~6 and 7 by \citet{Herenz2019}, which show the selection function for MUSE-Wide LAEs in the first 24 fields. At a redshift of $z \sim 3$ the $15\%$ completeness limit is roughly $\m{log_{10}(L}_{\m{Ly}\alpha}\, \m{[erg/s]}) \approx 42.2$ and goes up to $\m{log_{10}(L}_{\m{Ly}\alpha}\,\m{[erg/s]}) \approx 42.5$ at a redshift of $z\approx6.5$ between the skylines. \citet{Herenz2019} find a characteristic Lyman $\alpha$ luminosity for their luminosity function of $\m{log}L^*\m{[erg/s]}=42.2^{+0.22}_{-0.16}$.

The MUSE-Deep survey focuses on the HUDF, with the MUSE-Mosaic consisting of nine fields of ten hours exposure time each with a final contiguous area of $9.92\, \m{arcmin}^2$ (\citealp{Bacon2017}). The third and deepest tier in this construction is the ultra deep field (UDF) 10 (or MUSE-UDF10), located in the deepest part of the HUDF and reaching a total exposure time of $31$~hours (see \citealp{Bacon2017} for more information on the data reduction and \citealp{Bacon2021} for the latest, deepest MUSE surveys). 

We included the ten HUDF fields in this study (the nine mosaic fields and the deeper UDF10 field) to better understand the influence of the survey depth on the distribution of measured Lyman $\alpha$ EW$_0$ and to increase the sample size. 
Since the MUSE-Deep fields have a longer exposure time and thus go deeper in Lyman $\alpha$ luminosity and since the Lyman $\alpha$ luminosity function is steep, the number of objects in the 91 MUSE-Wide fields and in the ten MUSE-Deep fields including the UDF10 is similar, with 1051 objects from MUSE-Wide and 869 from MUSE-Deep.
As the MUSE-Deep area has a different orientation and slightly overlaps with several other MUSE-Wide fields, objects that fall into both the MUSE-Wide and the MUSE-Deep parts of the survey were found by positional matching, and in case of duplicates only the MUSE-Deep information was used here.

From Fig.~5 of \citet{Drake2017b} the $15\%$ completeness limit for the detection of Lyman $\alpha$ emission in MUSE-Deep is roughly  $\m{log_{10}(L}_{\m{Ly}\alpha}\,\m{[erg/s]}) \approx 41$ at a redshift of $z\approx3$ and goes up to $\m{log_{10}(L}_{\m{Ly}\alpha}\,\m{[erg/s]}) \approx 42$ at $z\approx6.5$ (for the nine MUSE-Deep mosaic fields). The seeing conditions for MUSE-Wide were around $1\farcs0$ for most of the MUSE-Wide fields (the full width at half maximum, FWHM, of the point spread function) and $\approx 0\farcs6$ at $7750\,\angstrom$ for MUSE-Deep.

Combined, these three depths represent different Lyman $\alpha$ luminosities which allows for example to study the Lyman $\alpha$ luminosity function of LAEs, probing the bright end (with MUSE-Wide, see \citealp{Herenz2019}) as well as the faint end (with MUSE-Deep, see \citealp{Drake2017b}) but also the EW$_0$ distribution for UV-faint LAEs (\citealp{Hashimoto2017}), which is a precursor to the present study. However, \citet{Hashimoto2017} only consider LAEs in their sample that are detected with at least $2\,\sigma$ significance in at least two HST bands, while we also considered HST undetected objects.

\subsection{Photometry from HST} \label{Photometry}

We used the ACS F775W, ACS F814W, WFC3 F125W, and WFC3 F160W bands (\citealp{Giavalisco2004, Koekemoer2011, Grogin2011}) to measure the UV continuum flux density, the ACS F814W band to determine the UV continuum counterparts (and the ACS F775W band for the HUDF parallel fields where the ACS F814W was not available), and ACS F435W and ACS F606W to check for low-redshift interlopers\footnote{The HST data can be found at \url{http://arcoiris.ucolick.org/candels/data_access/GOODS-S.html} and \url{https://archive.stsci.edu/prepds/goods/}. We do not include the data from \citet{Rafelski2015} to assure a homogeneous depth of the HST data.}. This process is described further in Sect.~\ref{Counterparts}.

\begin{table*}
\begin{center}
\caption{Properties of HST filter bands used in this paper, for the CANDELS GOODS-S and COSMOS area.}
\begin{tabular}{ l l l l l }
\hline\hline
  HST filter & sensitivity & spatial resolution & spatial sampling & exposure time  \\ 
  & & (arcsec) & (per pixel) & (seconds)\\ \hline
ACS F435W & 27.8 $^{\m{a,c}}$ & $0\farcs125$ $^{\m{a,b}}$ & $0\farcs03$ & 7200 $^{\m{a}}$ \\
ACS F606W & 27.8 $^{\m{a,c}}$ & $0\farcs125$ $^{\m{a,b}}$ & $0\farcs03$ & 3040 $^{\m{a}}$ \\
ACS F775W & 27.1 $^{\m{a,c}}$ & $0\farcs125$ $^{\m{a,b}}$ & $0\farcs03$ & 3040 $^{\m{a}}$ \\
ACS F814W & 28.32 $^{\m{d,e}}$ & $\sim0\farcs1$ $^{\m{d,b}}$ &  $0\farcs03$ & 5761 $^{\m{d}}$  \\
WFC 3 F125W & 27.04 $^{\m{d,e}}$ & $\sim0\farcs17 - 0\farcs19$ $^{\m{d,b}}$ & $0\farcs06$ & 1963 $^{\m{d}}$\\
WFC 3 F160W & 27.15 $^{\m{d,e}}$ & $\sim0\farcs17 - 0\farcs19$ $^{\m{d,b}}$ & $0\farcs06$ & 3480 $^{\m{d}}$\\ \hline
\end{tabular}\label{tab:bands_overview}
\end{center}
\tablefoot{\\ $^{\m{a}}$ Information taken from \citealp{Giavalisco2004}. $^{\m{b}}$ The FWHM of the point spread function (PSF). $^{\m{c}}$ $10\,\sigma$ point source sensitivity in AB magnitudes in an aperture with a diameter of $0\farcs2$. $^{\m{d}}$ Information taken from \citealp{Koekemoer2011}. $^{\m{e}}$ $5\,\sigma$ magnitude for a point source using an aperture of three detector pixels for ACS F814W and four detector pixels for WFC 3 F125W and F160W for the CANDELS UDS field. \\
} 
\end{table*}

%%%%%%%%%%%%%%%%%%%%%%%%%%%%%%%%%%%%%%%%%%%%%%%%%%%%%%%%%%%%%%%%%%%%%%%%%%%%%%%%%%%%%%%%%%%

\section{Sample selection} \label{Sample Selection}

In this section we describe the process of assembling a sample of LAEs from the MUSE-Wide and MUSE-Deep survey areas, including the detection of emission lines and their classification and report on a comparison with the catalogue by \citealp{Inami2017}.

\subsection{Detection and classification}

Integral field spectroscopy is ideal for detecting LAEs and we used the full two-dimensional spatial and combined spectral information of MUSE to detect emission line objects with \texttt{LSDCat}\footnote{\texttt{LSDCat} is available via the Astrophysics Source Code Library: \url{http://www.ascl.net/1612.002}} (\citealp{HerenzWisotzki2017}), a tool using a matched-filtering approach to maximise the signal-to-noise (S/N) of the detections. For the first 24 fields we estimated the variance from 100 empty sky positions in the MUSE data directly (since the propagated noise from the pipeline underestimates the uncertainties due to the resampling when constructing the cube), but for the rest of the fields we used an updated version of the effective variance measurements (see \citealp{Urrutia2019}), taking into account the spectral resampling, which makes up a factor of $1.25$ in the noise level, correcting the initial threshold for the first 24 fields of $\m{S/N}=8$ to $\m{S/N}=6.4$. For the following fields the threshold was lowered to five, which kept the rate of false positives at $\approx 5\%$ when compared to the MUSE-Deep catalogue by \citet{Inami2017} (see explanation by \citealp{Urrutia2019}). This means that since the selection of LAEs in this paper was based on the detection of the Lyman $\alpha$ line in the MUSE spectra, we did not have a cut in EW$_0$ to define an LAE, but only in the S/N of the emission line. The full 100 MUSE-Wide fields with an updated data reduction and a consistent cut in S/N for emission line detection will be published by Urrutia et al. in prep.

After their detection, the emission lines were then grouped together by \texttt{LSDCat} into individual objects where the line with the highest S/N is referred to as the lead line. For consistency and to create a homogeneous sample, we used detections and line flux measurements from \texttt{LSDCat} for both the MUSE-Wide and MUSE-Deep fields (but using the full observational depth of the MUSE-Deep data). Therefore our sample of objects for MUSE-Deep is slightly different from that presented in the catalogue of \citet{Inami2017} and used for the EW$_0$ distribution study by \citet{Hashimoto2017}. An additional benefit of using \texttt{LSDCat} for measuring the Lyman $\alpha$ line fluxes is the inclusion of the full emission, even for objects with extended Lyman $\alpha$ halos (see the discussion on the influence of the halo size in Appendix \ref{Halo Sizes}).

After detecting all emission line object candidates with \texttt{LSDCat}, the next step was to classify the objects, which was done with the help of the custom graphical user interface (GUI) \texttt{QtClassify}\footnote{\texttt{QtClassify} is available via the Astrophysics Source Code Library: \url{http://ascl.net/code/v/1628}}, (\citealp{Kerutt2017}, see also the appendix of \citealp{Herenz2017}). \texttt{QtClassify} enables the user to load a full MUSE datacube, as well as a catalogue created by \texttt{LSDCat} and ancillary HST data if needed. For a detailed discussion of the procedure see \citet{Herenz2017} and \citet{Urrutia2019}. For the nine MUSE-Deep fields and the UDF10 we used the same procedure as for the rest of the MUSE-Wide fields to create a homogeneous sample with the same detection and classification methods.

Even though MUSE provides spectra for each object, allowing to look for additional emission lines that could confirm or rule out the classification of an LAE, there is still the possibility of false classifications (see \citealp{Urrutia2019} for a discussion), which can have an influence on the measured characteristic EW$_0$ values $w_0$ of the histograms of equivalent widths (see Sect.~\ref{Distribution of Equivalent Widths}). To assess the reliability of the classification, we introduced a classification confidence between zero (the lowest confidence) and three (the highest confidence). This confidence was based on several aspects of the object, that each individual classifier used to base their decision on. If multiple lines matching a common redshift were detected in the spectrum, the confidence was three. If there was only one detected line but others were visible by eye (at a S/N below the detection threshold) using \texttt{QtClassify}, the confidence was set to two or three. If only one emission line was visible, the shape of the line was considered. Especially Lyman $\alpha$ lines often have a characteristic asymmetric and/or double peaked shape which is a clear indicator and resulted in a confidence of two to three. In the case of double peaks, we made sure that the peak separation and shape did not match with the \rm{O\textsc{ii}} doublet. If there was only one line, the S/N was low, and/or the line shape was not characteristic, a confidence of one was given. A confidence of zero was only given in extreme cases where no estimate of the redshift could be made. The process of classification was performed by two people who consolidated their results with a third person.

\subsection{Comparison to the MUSE-Deep catalogue}

For the MUSE-Deep part of our sample, a catalogue of objects using a different detection and classification method has been published by \citet{Inami2017} with a subsequent analysis of Lyman $\alpha$ EW$_0$ by \citet{Hashimoto2017}. In contrast to the MUSE-Wide approach this catalogue is not based solely on an emission line selection. Instead, several different methods were used to construct the catalogue, including an automated emission line detection software similar to \texttt{LSDCat} based also on a matched filtering approach (ORIGIN, \citealp{Mary2020}) and MUSELET (\citealp{Piqueras2017}) based on {\textsc SExtractor} (\citealp{SExtractor}), using narrow-band images created by collapsing five wavelength layers and subtracting the continuum (\citealp{Inami2017}). To classify the detected emission lines, the software MARZ (\citealp{Hinton2016}) was modified and the classification was done semi-automatically, aided by a team of human classifiers. 

We matched the LAEs found with the MUSE-Wide approach to the LAEs in the \citet{Inami2017} catalogue for comparison, with a maximum allowed separation of $0\farcs5$ and chose the closest counterpart within this area (given the redshifts in both catalogues matched as well). We chose this distance based on \citet{Herenz2017}, where it was found that the $3\,\sigma$ positional uncertainty between the MUSE-Wide position and HST catalogues was below $0\farcs5$. Of the 807 LAEs in the \citet{Inami2017} catalogue\footnote{\citet{Inami2017} give a number of 692 LAEs with high confidence and an additional 115 LAEs with lower confidence}, 124 could not be matched to detections with the MUSE-Wide method, while 214 objects were found in addition to the LAEs by \citet{Inami2017}. The latter had on average a lower confidence level (with a mean of 1.7 compared to 2.1 for all \texttt{LSDCat} detected objects in MUSE-Deep), which means mostly weak lines were missed before. 
The 124 objects that were found in \citet{Inami2017} but could not be matched to our catalogue can be explained by the different catalogue creation methods. While \citet{Inami2017} not only searched for emission lines directly in the MUSE data, but also used the HST-based UVUDF catalogue (\citealp{Rafelski2015}) as priors, the MUSE-Wide method using \texttt{LSDCat} is a blind emission line search only\footnote{A new data reduction version, DR2, of the MUSE-Deep survey will be presented in Bacon et al. in prep. Similarly, the final MUSE-Wide data release with all 100 fields will be described in Urrutia et al. in prep. There, we will investigate the differences between the two detection methods and the discrepancy between the catalogues in more detail.}. Objects that are visible in HST but have emission lines below our detection threshold would therefore be missed. Conversely, \texttt{LSDCat} detected additional emission line objects that do not have an HST counterpart, which accounts for the 214 objects found with the MUSE-Wide method that are not in \citet{Inami2017}. It should also be noted that we are only comparing LAEs here, which means that part of the discrepancy can be attributed to a difference in the classification, meaning LAEs that are missing in either catalogue might be present in the other but classified as something else.

\subsection{Final sample of LAEs}

We excluded three AGN from our sample, two of which (IDs 104014050 and 115003085) were already mentioned in \citet{Urrutia2019}, one AGN (ID 214002011) is in the COSMOS field. For the MUSE-Deep part we excluded two objects (IDs 1841655 and 1381485), the first based on \citet{Hashimoto2017}, the second is a CIV emitter (both are also discussed in \citealp{Bacon2021}).

The total sample of LAEs used in this paper consists of 1920 LAEs. We excluded 35 objects that might be superpositions (two objects at different redshifts that overlap spatially), based either on a spectral energy distribution (SED) that does not match expectations for high redshift LAEs, which means no drop in the HST band to the blue of the Lyman $\alpha$ line (see Sect.~\ref{Counterparts}) or on the presence of other emission lines in the MUSE data. In the latter case, we could not assign each emission line in the spectrum to one single redshift, which means that within the MUSE resolution, we were not able to disentangle the superposed objects. Since the reason for this is usually a low redshift interloper and to avoid any possible contamination, we excluded such objects, even if they were separated in the HST data.
We thus got $\sim11$ LAEs per field in the MUSE-Wide survey (and therefore also roughly per square arcminute) and $\sim96$ LAEs per field for MUSE-Deep. To keep the methods consistent, we not only used the same detection method for MUSE-Wide and MUSE-Deep, but we also selected the UV continuum counterparts and measured the UV continuum flux density for MUSE-Deep in the same way as for the MUSE-Wide objects.

%%%%%%%%%%%%%%%%%%%%%%%%%%%%%%%%%%%%%%%%%%%%%%%%%%%%%%%%%%%%%%%%%%%%%%%%%%%%%%%%%%%%%%%%%%%

\section{The UV continuum} \label{The UV continuum} 

As mentioned above, for measuring the EW$_0$ of the LAEs found in the MUSE data, the rest-frame UV continuum flux density was obtained from the deeper broad-band HST data instead of directly from the extracted MUSE spectra, since the latter were not deep enough for all objects. An additional advantage of the HST data is the higher resolution, allowing for a more detailed analysis of the UV continuum morphological properties. For this purpose we determined the UV continuum counterparts and fitted them with \texttt{Galfit} (\citealp{Galfit, Galfit2}) where possible, a process which we describe in this section.

\subsection{Identification of UV continuum counterparts} \label{Counterparts}

We determined the UV continuum counterparts of the LAEs in our sample in the HST data using the filter band ACS F814W for each object where possible (both for MUSE-Wide and MUSE-Deep, except for the eight HUDF parallel fields where we use the HST filter ACS F775W instead). We used the reddest available ACS filter band to have a high spatial resolution (twice better than the WFC 3 bands, see Table \ref{tab:bands_overview}) and to make sure as many LAEs as possible would be detectable. Since the HST filter bands ACS F775W and ACS F814W overlap significantly (while ACS F814W is slightly deeper), using ACS F775W for the parallel fields was an adequate solution where ACS F814W was not available. We did not use any existing catalogues for the determination of the UV continuum counterparts as we wanted to be as unbiased as possible.
Any signal within a radius of $0\farcs5$ (measured from the maximum S/N in the Lyman $\alpha$ detection in MUSE) was taken into consideration as a counterpart (as this distance was found to be the $3\,\sigma$ positional difference by \citealp{Herenz2017} when comparing the MUSE-Wide catalogue from \texttt{LSDCat} and the catalogue from \citealp{Skelton2014}). This criterion was used as a starting point, as we expected that not all Lyman $\alpha$ emission lines we find in the MUSE data have a UV continuum counterpart that is bright enough to be visible in the HST data. In case of more than one counterpart candidate within the $0\farcs5$, additional HST filters were examined by at least two people to visually determine if the SED matches with what we would expect from LAEs, namely that there is little to no flux to the blue of the Lyman $\alpha$ line (as seen in the HST bands ACS F435W and ACS F606W), the spectrum declines towards the red, and there is an increase of flux in the band containing the Lyman $\alpha$ line. 

\begin{figure*}%[H]
\centering
\includegraphics[width=0.21\textwidth,trim={3cm 0 3.5cm 0},clip]{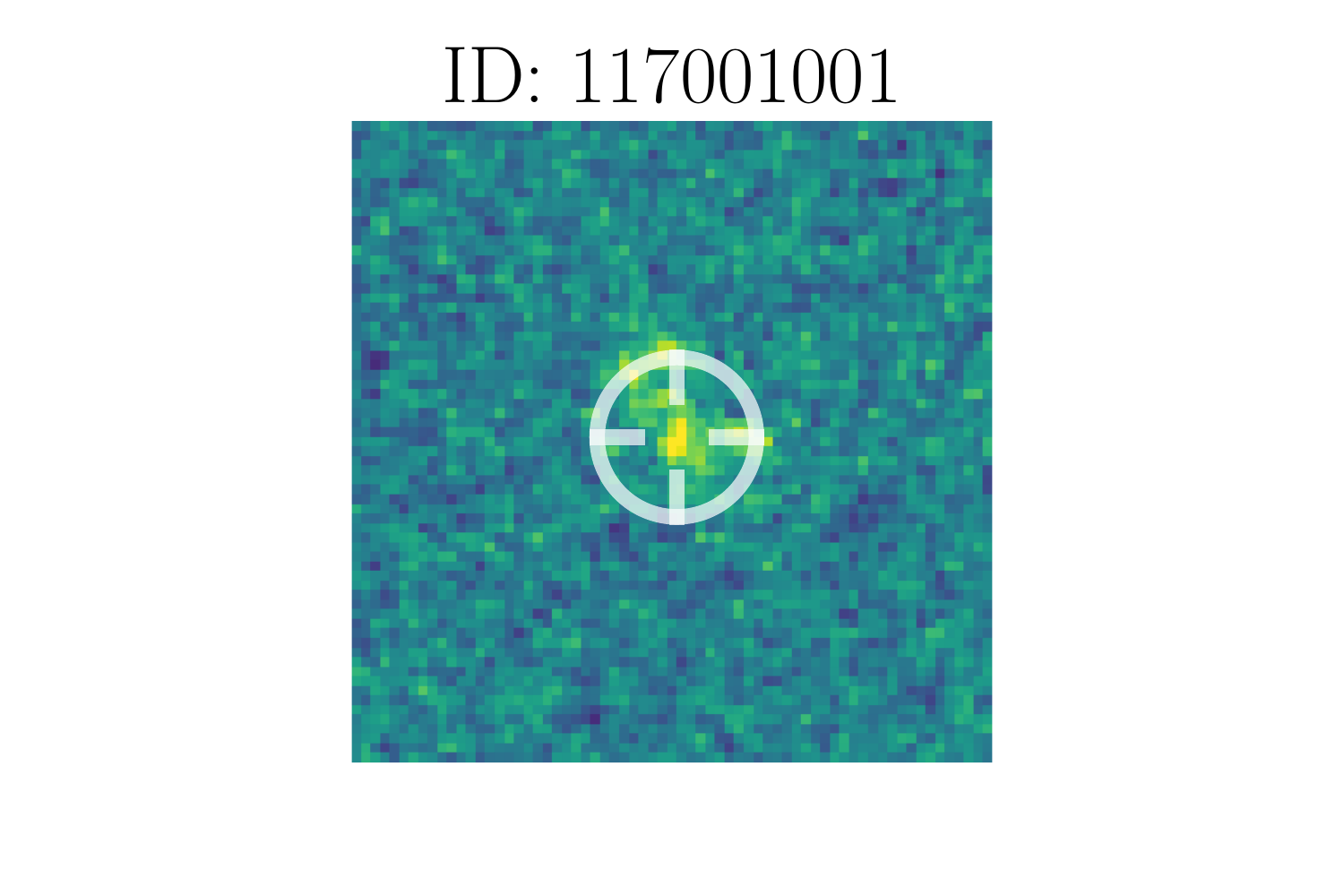}
\includegraphics[width=0.21\textwidth,trim={3cm 0 3.5cm 0},clip]{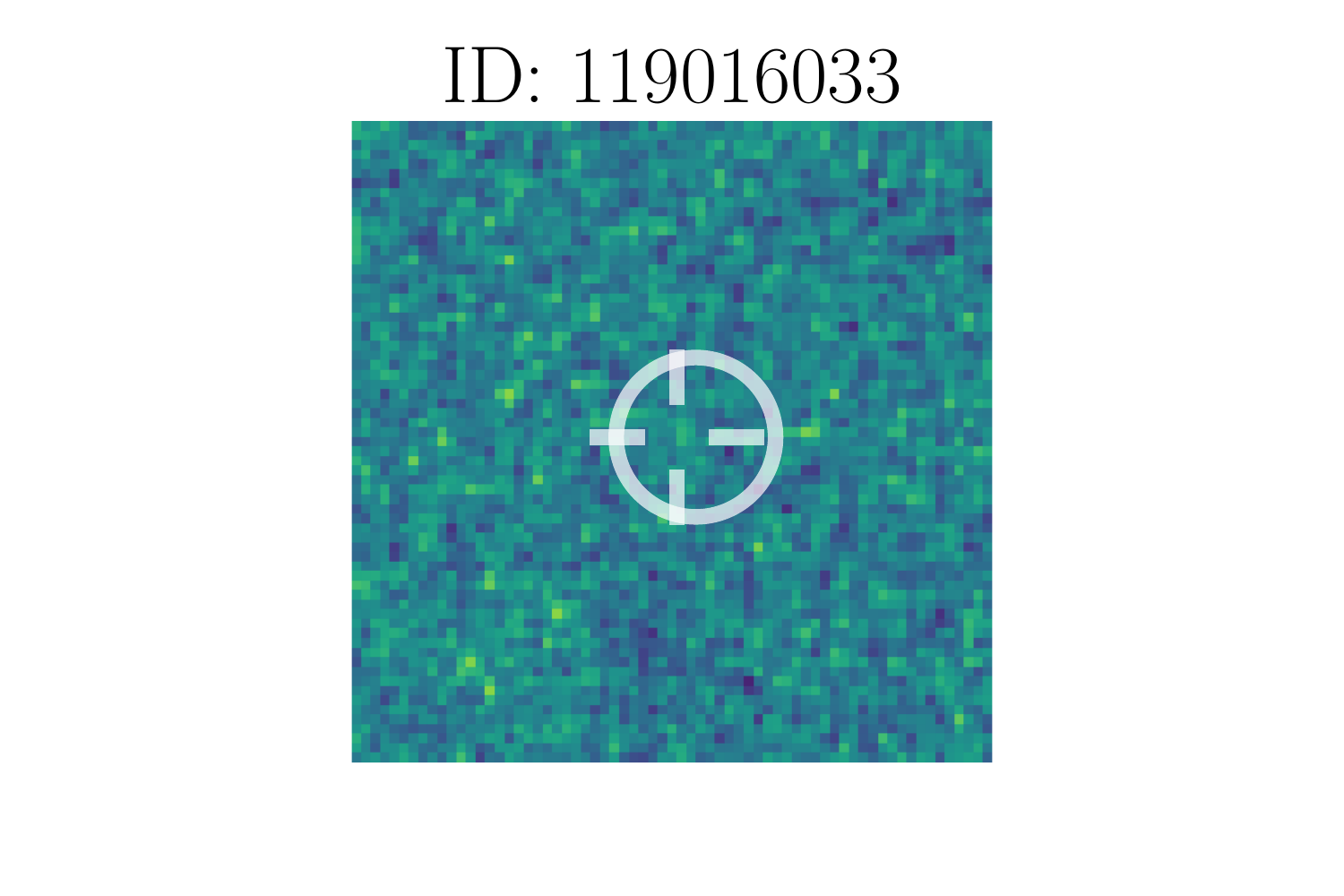}
\includegraphics[width=0.29\textwidth,trim={3cm 0 0 0},clip]{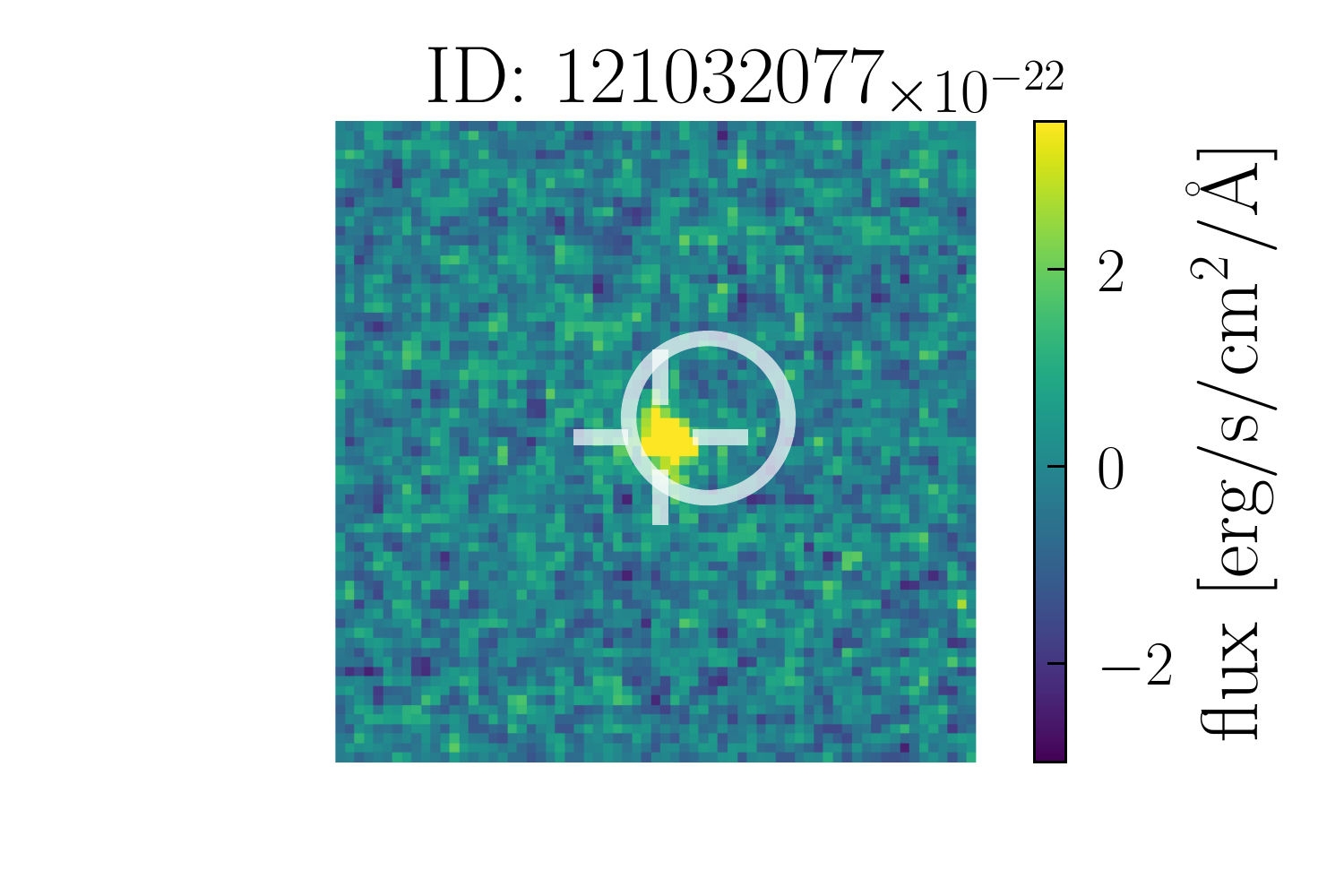}
\caption{Examples of LAEs with different kinds of UV continuum counterparts, shown in $2\arcsec\times2\arcsec$ cutouts of the filter band ACS F814W. The white crosses are centred on the UV continuum position, the white circles are centred on the Lyman $\alpha$ position from \texttt{LSDCat} and have a diameter of $0\farcs5$. Left image: Object with multiple components or clumps in the counterpart. Middle image: Object without a significant detection in the filter band ACS F814W (the centre of the cross is set on a noise peak/bright pixel). Right image: Counterpart with only one component, but slightly offset with respect to the Lyman $\alpha$ position.} 
\label{fig:example_cutouts}
\end{figure*}

Due to the IGM absorption, the emission will likely drop in HST filter bands to the blue side of the Lyman $\alpha$ line, although this depends on the assumed SED models. 
In cases where the possible counterparts were close and seemed to have the same positions and brightness in all HST bands considered, there is a possibility that it is really only one counterpart consisting of multiple components or clumps, which belong to the same object (see Fig.~\ref{fig:example_cutouts} for examples of different configurations). This could be caused by the clumpy nature of high redshift galaxies, due to distinct star formation areas or satellite galaxies. In this case, all components (within $0\farcs5$) were considered to belong to the same Lyman $\alpha$ emission (as e.g. in the left panel of Fig.~\ref{fig:example_cutouts}). When matching the Lyman $\alpha$ positions to the UV continuum counterparts, there is often a slight shift between the two positions (see also Claeyssens et al. in prep for a study using lensed LAEs), as the Lyman $\alpha$ emission might escape preferentially through outflows or holes in the ISM of the galaxy (e.g. \citealp{Shibuya2014a}). 

\begin{figure}%[H]
\centering
\includegraphics[width=0.45\textwidth]{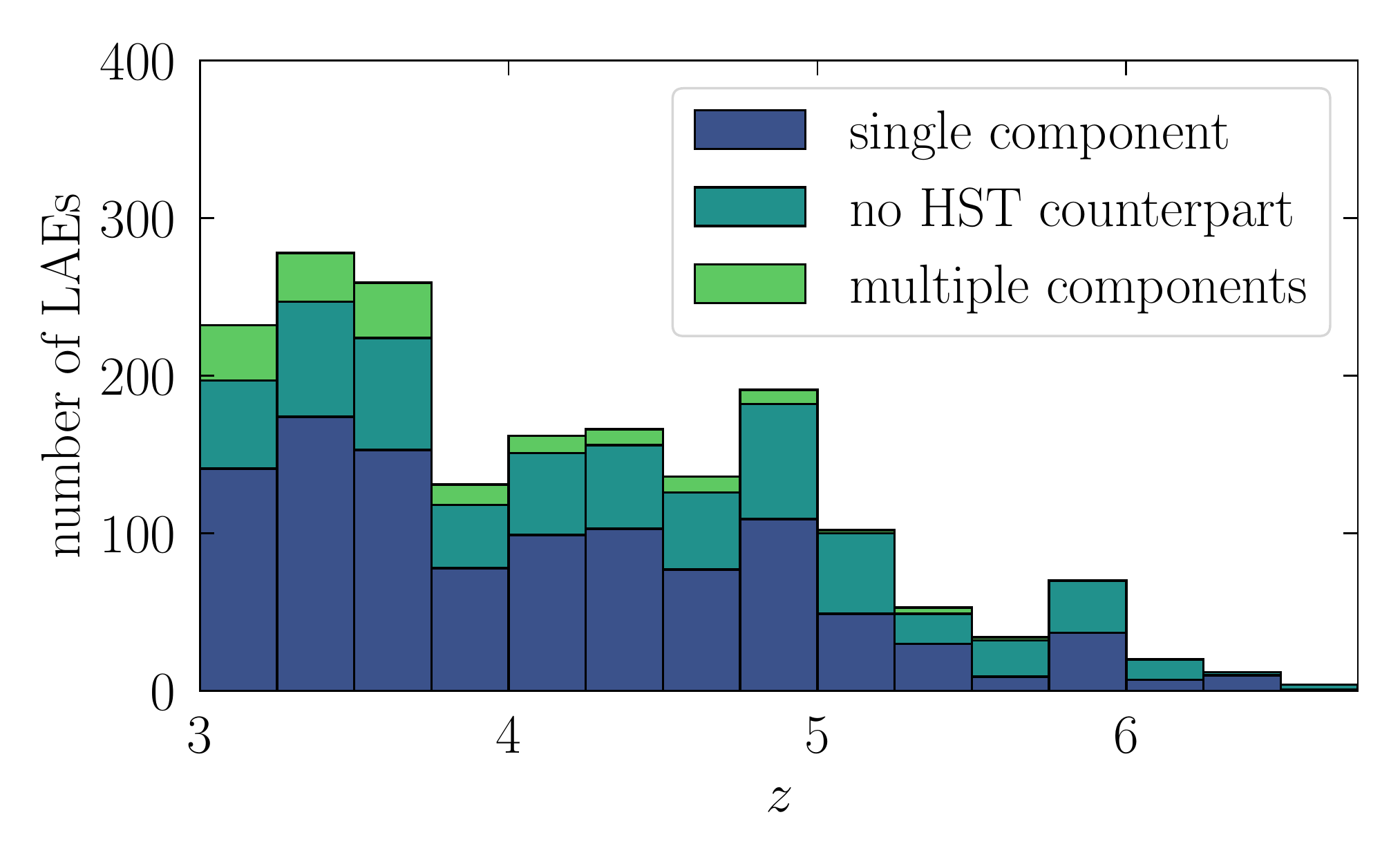}
\caption{Redshift histogram of the full sample of LAEs, split into groups according to their UV continuum counterparts in HST, in bins of $\Delta z = 0.25$. The dark blue histogram shows objects with UV counterparts which are not resolved into multiple components. The turquoise histogram shows objects without a visible UV counterpart in the available HST imaging and the light green histogram shows objects with UV counterparts consisting of more than one component. In these cases, the HST counterpart could not be fitted with a single S\'ersic profile, as it was clumpy or asymmetric.} 
\label{fig:z_dist}
\end{figure}

We show the redshift distribution of our sample in Fig.~\ref{fig:z_dist} (we use redshifts corrected using the equations in \citet{Verhamme2018} based on the information of the FWHM and peak separation of the line, see Sect.~\ref{Lyman alpha Line Properties}). 
The constant decrease in the number of LAEs with high redshifts is influenced by the increasing luminosity distance and the resulting detection of more luminous objects, which are less numerous (see \citealp{Herenz2019} for a discussion on the selection function and Lyman $\alpha$ luminosity function). At redshifts $z>5$ the number of objects with and without a UV continuum counterpart is almost the same, while at lower redshifts most objects can be seen in the HST data. This is caused by the decline in the intrinsic UV continuum brightness at higher redshifts as well as cosmological surface brightness dimming, making it harder to detect the counterparts, commonly referred to as the Malmquist bias (\citealp{Malmquist1922,Malmquist1925}), which also contributes to the decline in LAE numbers at higher redshifts. \\

\subsection{Fitting the UV continuum using \texttt{Galfit}} \label{Fitting the UV continuum}

After assigning UV continuum counterparts to all LAEs where possible, we fitted them with \texttt{Galfit}, a fitting algorithm created for HST images that is fast and efficient at fitting multiple components simultaneously (\citealp{Galfit, Galfit2}). From these fits we gained information not only on the magnitudes, but also on the sizes, number of components, and axis ratio of the UV continuum counterparts where they are resolved in HST. Another advantage of using \texttt{Galfit} over simple fixed apertures is the possibility to model (and thus subtract) neighbouring objects as well, to obtain more reliable magnitude measurements. As the LAEs we want to fit are at high redshifts and thus small even in the HST data, we used \texttt{Galfit} to fit them with a simple S\'ersic (\citealp{Sersic1968}) profile (per component):

\begin{equation}
    \m{log} \left( \frac{I(R)}{I_{e}} \right) = -b_n \left[ \left( \frac{R}{R_e} \right)^{\frac{1}{n}} -1 \right] .
\end{equation}

With $I(R)$ the surface brightness, $I_e$ the surface brightness at $R_e$ (the effective radius containing half of the luminosity), $b_n \approx 2n-0.331$ (\citealp{Galfit}) making sure that half of the total flux will be within one effective radius, and $n$ the S\'ersic (power-law) index. Depending on the S\'ersic index $n$, the profile can become a Gaussian for $n=0.5$, an exponential profile at $n=1$, and a de Vaucouleurs profile (\citealp{deVaucouleurs1948}) with $n=4$.

\begin{figure*}%[H]
\centering
\includegraphics[width=0.8\textwidth]{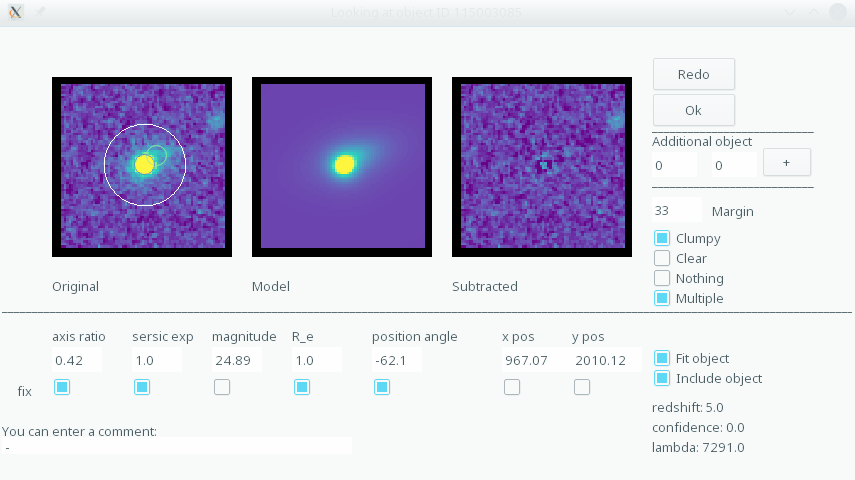}
\caption{Screenshot of the \texttt{Galfit} wrapper used for fitting and measuring the UV continuum in different bands. The left image shows a cutout of the data in the HST filter band ACS F814W, the middle image shows the (in this case) two models for the object and the right image shows the residuals. Clicking in the right panel will allow the user to add a \texttt{Galfit} model for an additional component, which creates a green circle to mark the place. Below the panels are morphological measurements as well as magnitude and position, each of which can be fixed if needed. In the example, the object was fitted with two S\'ersic models, each of which can be selected by clicking on the small green circles. The check boxes on the right allow the user to asses whether the object is clumpy and if there are multiple components in the counterpart or no counterpart at all, as well whether to include the model in the object.}
\label{fig:Screenshot_Galfit_Wrapper_viridis}
\end{figure*}

To facilitate the fitting, we used a wrapper with a GUI that automatically generates input files for \texttt{Galfit}, first based on initial priors\footnote{We set the priors for all objects to the following values: mag $=27$, $\m{R}_{\m{e}}=3$ pixels, $n=1$, axis ratio $=0.5$ and position angle $=0$.}, then on previous runs, and stores the output data in a coherent way (see Fig.~\ref{fig:Screenshot_Galfit_Wrapper_viridis} for a screenshot of the wrapper). We iterated the fitting using the results form the previous run as new initial priors until the parameters were stable, in cases where the fitting was difficult. The procedure was to fit the morphological properties of the objects first in the ACS F814W filter (or the ACS F775W for the HUDF parallel fields), which is the deepest in the MUSE-Wide area, and then use them as fixed values for the other filter bands (ACS F606W, ACS F775W, WFC3 F125W and WFC3 F160W) with the magnitude as a free parameter. That way fainter flux at larger wavelengths could be captured within the same area as in the deepest HST filter. Another advantage of fixing parameter values was that even if the object was faint and its morphological parameters could not be fit well with \texttt{Galfit}, we could capture the magnitude by reducing the number of free parameters, facilitating the fitting. For faint objects, the axis ratio and S\'ersic index $n$ could often not be constrained in the fit. In this case we fixed $n=1$ and b/a=1, that is we fitted only a circular exponential profile, so that at least the effect radius and the objects' magnitude are recovered. In some cases the objects were still too small to be resolved in HST and we used a fixed aperture to capture the magnitudes. Of the 1284 objects that have an HST counterpart, 538 have a reliable measurement of the effective radius.

In cases where the object was very faint or maybe even not visible, we also used a fixed radius of the size of the point spread function (PSF) for the Galfit model. If the flux in this aperture was not above $1\,\sigma$ in any of the HST bands, we considered the object undetected in HST.

For the HST PSF we used a Moffat function (\citealp{Moffat1969}) measured using \texttt{Galfit}, from stars in the corresponding HST filter bands. With the help of the wrapper, nearby objects that might influence the fit of the LAE can be fit as well, so they do not artificially increase the measured magnitude of the LAE. In the same way, multiple components can be fit and included in the total flux of the UV continuum counterpart. In these cases, we added the continuum flux density of all components for the EW$_0$ estimation, assuming that the Lyman $\alpha$ emission is produced from the entirety of the components. To exclude the possibility of multiple components being superpositions of different objects, we checked for additional emission lines in the MUSE spectra and for different SEDs in the separate components of the counterparts in the HST data (see Sect.~\ref{Counterparts} above).
For the morphological properties of the LAEs, that is the effective radius and axis ratio, we used the parameter uncertainty estimates from the \texttt{Galfit} models, while we used random apertures for the errors on the magnitudes (see Sect.~\ref{Methods: Limits and Magnitude Errors} below).

Using this method of carefully checking and measuring the UV continuum counterpart properties of each individual object we can go beyond existing catalogues (e.g. \citealp{Guo2013,Skelton2014,Rafelski2015}) and include even very faint sources that had been missed before, but that we know to be real through the Lyman $\alpha$ line we found with MUSE. If we compare objects for which we did find an HST counterpart to the catalogues by \citet{Guo2013} and \citet{Skelton2014}, we find that $42\%$ of our LAEs have a match in the \citet{Guo2013} catalogue and $59\%$ have a match in the \citet{Skelton2014} catalogue. The discrepancy can be explained by the fact that these catalogues are based on near-IR detections, while our LAEs are most likely UV-dominated.

%%%%%%%%%%%%%%%%%%%%%%%%%%%%%%%%%%%%%%%%%%%%%%%%%%%%%%%%%%%%%%%%%%%%%%%%%%%%%

\subsection{Continuum flux densities} \label{Continuum Flux Density} 
For the EW$_0$, we need the continuum flux density at the position of the Lyman $\alpha$ line (for objects where a counterpart is detected in the HST data). In principle the flux density at $\lambda_{\m{Ly}\alpha}$ could be obtained directly from the HST band containing Ly$\alpha$. However, this requires a correction for the contribution of the Lyman $\alpha$ flux to the band, which introduces uncertainties that are hard to quantify. This is because of the halo component that is often undetected at the depth of HST. Moreover, the necessary correction for IGM absorption to the blue side of Lyman $\alpha$ introduces an additional complication, since we do not know how much of the measured flux in the HST filter band can be attributed to the UV continuum redwards of the line. 
Therefore we decided not to attempt a correct for the Lyman $\alpha$ line flux or the IGM absorption and instead we used the flux density from the HST filter band to the red of the line that has zero throughput at the Lyman $\alpha$ wavelength. 

This means for objects below a redshift of $\sim 4.7$ we used the HST bands ACS F775W, ACS F814W, WFC3 F125W and WFC3 F160W. For objects above this redshift we used the HST bands WFC3 F125W and WFC3 F160W to measure the magnitudes which are then used to estimate the flux at the Lyman $\alpha$ position.

To extrapolate the measured flux at the effective wavelength of the HST filter to the Lyman $\alpha$ line position, we need to know the UV continuum slope ($\beta$) of the spectrum, which correlates the flux density $f$ at a certain wavelength $\lambda$ to the wavelength via $ f_{\lambda} \propto \lambda^{\beta}$. If we know the flux density $f$ at two or more different wavelengths $\lambda_1$ and $\lambda_2$, we can derive the UV continuum slope, assuming the continuum is a power law:

\begin{equation}
\beta = \frac{\log_{10}\left(\frac{f_{\lambda_{1}}}{f_{\lambda_{2}}}\right)} {\log_{10}\left(\frac{\lambda_{1}}{\lambda_{2}}\right)}
\label{beta1} .
\end{equation}

If we have detections in two or more HST filter bands, we can fit a linear relation to the logarithm of the measured flux density values. For this we used HST bands to the red side of the emission line (as mentioned above) and fitted a simple linear relation to the continuum flux density measurements. For objects that have a detection in only one HST band or even no counterpart at all, it is not possible to measure the $\beta$ parameter. For this reason, and since the measured $\beta$ values scatter significantly and have large errors especially for faint objects, we used our median value of $\beta = -1.97$ of the entire sample as a fixed value for each individual object. 

The caveat with this approach is that LAEs with high EW$_0$ possibly have different properties from LAEs with lower EW$_0$, which means their $\beta$ values could be systematically different from each other. The same could apply for LAEs that are undetected in the UV continuum for which it is not possible to measure the continuum slope at all from individual objects. If such objects had a bluer continuum slope due to less dust or younger ages, that would mean that their real EW$_0$ would be lower than what we assume here (see \citealp{Maseda2020}). Therefore the lower limits we are quoting here for EW$_0$ of HST undetected objects are only lower limits assuming a fixed $\beta = -1.97$. Ideally, for objects that are individually undetected in the UV continuum, one could stack the HST photometry to obtain $\beta$ values from the stacks (see Maseda et al. in prep.). 

In the study on EW$_0$ distributions using data from MUSE-Deep alone, \citet{Hashimoto2017} measured the UV continuum slope from three (or two) adjacent HST bands where available. They show that using a fixed value for $\beta$ has little impact on the derived characteristic EW$_0$ values $w_0$, but caution that at lower (higher) redshifts, the EW$_0$ values can be overestimated (underestimated) when using a fixed $\beta$ value of $\beta = -2$ due to the redshift evolution of the UV continuum slope. 

Most literature studies find $\beta$ values around $\beta \leqslant -2$ (\citealp{Bouwens2009, Castellano2012}) for Lyman Break Galaxies (LBGs), but \citet{Karman2017} and \citet{Santos2020} find even steeper slopes for faint LAEs at similar redshifts. A larger UV continuum slope indicates a redder spectrum, corresponding to dust absorption and metallicity in the galaxy. The median of the UV continuum slope derived from multiple HST bands $\beta = -1.97$ used here thus matches well with results found in the literature. To understand the influence of different $\beta$ parameters on the measured characteristic EW$_0$ values of the EW$_0$ histograms, we compare the median value to the upper $\beta=-1.57$ and lower $\beta=-2.29$ quartiles of the distribution of measured $\beta$ values (see Sect.~\ref{Distribution of Equivalent Widths}). 

%%%%%%%%%%%%%%%%%%%%%%%%%%%%%%%%%%%%%%%%%%%%%%%%%%%%%%%%%%%%%%%%%%%%%%%%

\subsection{Limits and magnitude errors} \label{Methods: Limits and Magnitude Errors}

Any object not detected in a particular HST filter above a $1\,\sigma$ detection significance was assigned the limiting flux density (assuming a point source) in that HST band as the continuum flux density (see Table~\ref{tab:lim_mags} for median values of the limiting magnitudes in each band) and the HST filter was excluded from the estimations of EW$_0$ and the UV continuum slope. Objects with no significant detection in any HST filter band have lower limit EW$_0$ which we determined using the limiting flux in the deepest band (ACS F775W for the parallel fields and ACS F814W for all other bands) in combination with the median UV slope $\beta = -1.97$. 
In this paper, in plots showing the full sample including objects with lower limits for EW$_0$ (which is always the case except if explicitly mentioned in the figure caption), those limiting values are shown as if they were detections at their respective limits.

Since we used different fields that have different HST depths and since the depth can also vary over the field, we measured the limiting flux density for each object using 100 random apertures of $0\farcs5$ close (within a radius of $2\arcsec$) to the Lyman $\alpha$ position of the LAE in question. To avoid neighbouring objects, we excluded apertures with a measured flux density above the standard deviation of the 100 apertures as well as apertures that are outside the field of view of the specific MUSE pointing. 
In order to be consistent with the error measurements of the magnitudes, we used the same method for the $1\,\sigma$ errors on the flux densities of objects that do have measureable counterparts in the HST photometry.

\begin{table*}
\begin{center}
\caption{Median values of limiting magnitudes and flux densities in different HST filter bands. }
\begin{tabular}{ l l l l l }
\hline\hline
  HST filter & mag$_{\m{AB,lim}}$ (MW) & mag$_{\m{AB,lim}}$ (MD) & flux density (MW) & flux density (MD)  \\ \hline
ACS F435W & 29.03 & 29.03 & 1.41 & 1.41\\
ACS F606W & 29.07 & 29.22 & 0.71 & 0.62 \\
ACS F775W & 28.63 & 28.64 & 0.65 & 0.64 \\
ACS F814W & 28.92 & 29.16 & 0.46 & 0.37 \\
WFC 3 F125W & 28.82 & 29.06 & 0.21 & 0.17 \\
WFC 3 F160W & 28.65 & 28.85 & 0.16 & 0.13 \\ \hline
\end{tabular}\label{tab:lim_mags}
\end{center}
\tablefoot{The flux densities are given in $10^{-20}\, \m{erg}\, \m{s}^{-1}\, \m{cm}^{-2}\, \angstrom^{-1}$ for a $2\,\sigma$ detection, separated into MUSE-Wide (MW) and MUSE-Deep (MD). To convert the magnitudes to flux densities we use the effective wavelengths computed from transmission curves of the individual filters, which can be found at \url{https://www.stsci.edu/hst/instrumentation/acs/data-analysis/system-throughputs}.} 
\end{table*}

\subsection{Results: objects with no HST counterpart} \label{Objects with no HST counterpart}

\begin{figure}
\centering
\includegraphics[width=0.49\textwidth]{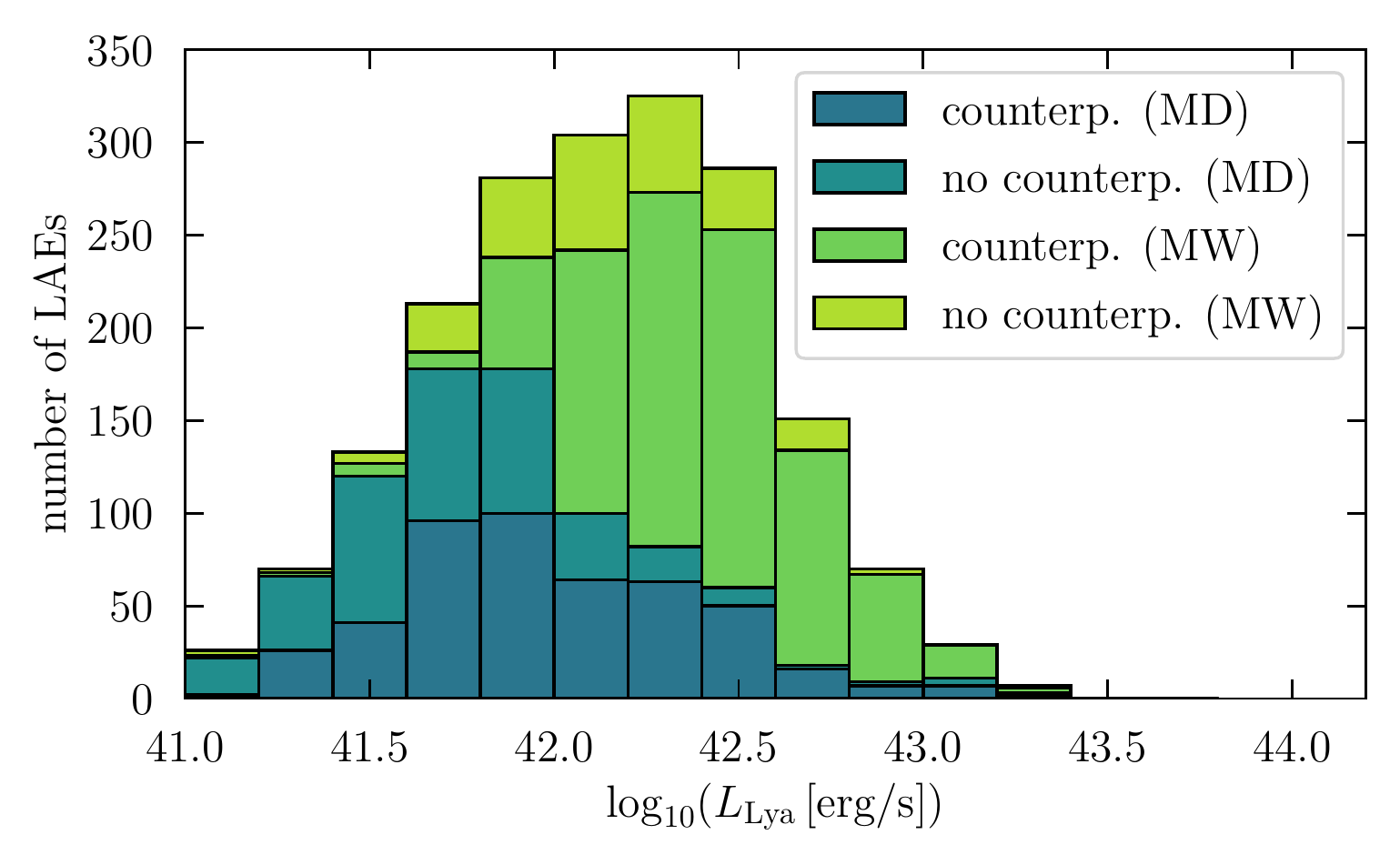}
\caption{Histogram showing the logarithmic Lyman $\alpha$ luminosity of the LAEs in the MUSE-Wide (light green, labelled MW) and MUSE-Deep surveys (dark green, labelled MD). The samples are divided into objects with UV continuum counterparts and without counterparts. The number of objects per sample are stacked, which means every histogram bar gives the total number of objects in that luminosity bin.} 
\label{fig:logLLya_dist_MW_MD}
\end{figure}

In total, of our sample of 1920 LAEs, 636 (around $33\%$ of the full sample) have either no visible counterpart in the UV continuum or only a faint one that was not distinguishable from noise (see \citealp{Bacon2017, Marino2018, Maseda2018, Maseda2020} for studies of dark galaxies not visible in the rest-frame UV continuum). The 869 MUSE-Deep LAEs have fractionally more UV continuum non-detections ($44\%$) than the 1051 MUSE-Wide LAEs ($24\%$). This is mainly caused by the increase in emission line depth in the MUSE data of MUSE-Deep, making it easier to detect objects in Lyman $\alpha$ that are too faint to see in the HST photometry (which does not have a big difference in depth in the different fields, see Table~\ref{tab:lim_mags}). This can be seen in Fig.~\ref{fig:logLLya_dist_MW_MD}, where the Lyman $\alpha$ luminosity histograms of MUSE-Wide and MUSE-Deep objects with and without counterparts are shown. The MUSE-Deep objects occupy the fainter range of Lyman $\alpha$ luminosity values and the objects without UV continuum counterparts have a slight shift to lower line luminosities as well (see Sect.~\ref{The Influence of Survey Depth on Equivalent Widths}).

\subsection{Discussion: morphological properties of the UV continuum}

Most objects (1118 or $58\%$) have a counterpart with only one component visible in the HST data, but 166 ($\approx 8.6\%$) have a counterpart consisting of multiple components (see Fig.~\ref{fig:z_dist} and left panel of Fig.~\ref{fig:example_cutouts}). LAEs consisting of multiple components could be intrinsically clumpy or irregular, which could have several causes: It could be that 1) a merger is underway, which could potentially boost the star formation and thus the Lyman $\alpha$ EW$_0$, 2) multiple star-forming regions give an irregular shape and outshine the rest of the galaxy or 3) multiple objects or satellite galaxies at close range that all contribute to the Lyman $\alpha$ emission (e.g. \citealp{Venemans2005, Gronwall2011, Bond2012, Jiang2013, Kobayashi2016, Paulino-Afonso2018}). Galaxies at high redshifts do not necessarily follow the same morphological patterns we see in the nearby universe and \citet{Gronwall2011} show that while LAEs are heterogeneous in their rest-frame UV morphological properties, they have more in common with high-z LBGs than with local galaxies and tend to be on average rather compact. However, the smaller sizes of LAEs could be the result of a possible redshift evolution, which is found by \citet{Shibuya2019,Ferguson2004}, but for example not by \citet{Paulino-Afonso2018}. 
Even though most of our LAEs are not resolved in HST, which could be caused by their compact nature, it is interesting to see a significant fraction of LAEs at redshifts $3<z<6$ being clumpy or possibly merging. 

%%%%%%%%%%%%%%%%%%%%%%%%%%%%%%%%%%%%%%%%%%%%%%%%%%%%%%%%%%%%%%%%%%%%%%%%%%%%%%%%%%%%%%%%%%%

\section{The Lyman \texorpdfstring{$\alpha$}{Lg} line} \label{The Lyman alpha line}

\begin{figure*}
\centering
\includegraphics[width=\textwidth]{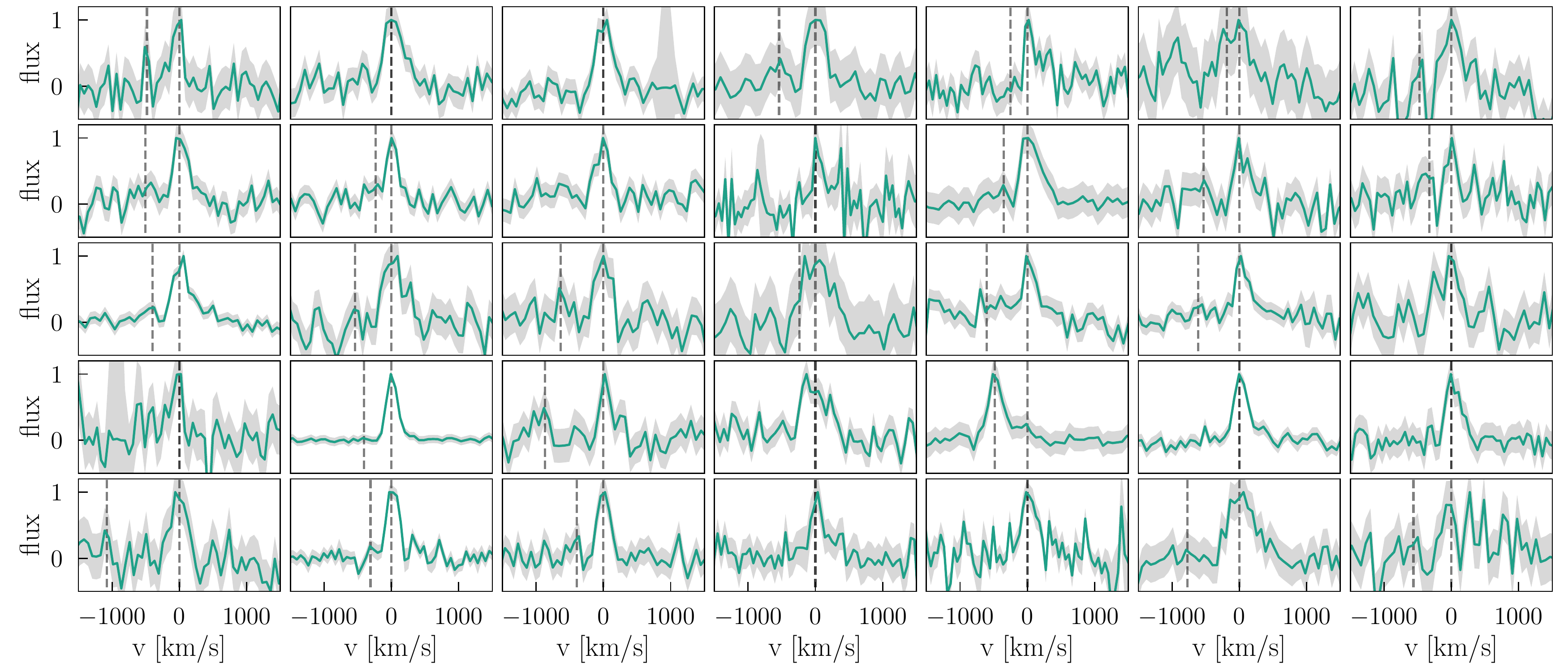}
\caption{Overview of the Lyman $\alpha$ emission lines of the 35 LAEs with the highest measured EW$_0$, sorted by EW$_0$ in descending order (left to right, top to bottom). We note that the y-axis shows the flux normalised to the maximum of each peak for better comparison. The vertical dashed lines indicate the positions of the Lyman $\alpha$ lines and the blue bump if there is one. The grey shaded area shows the standard deviation from the MUSE spectra. The shown objects are taken both from the MUSE-Wide and -Deep surveys and have been continuum subtracted.}
\label{fig:overview}
\end{figure*}

Another ingredient to measuring the EW$_0$ is the line flux of the Lyman $\alpha$ line (Sect.~\ref{Line Fluxes}). Its spectral properties can give us information on possible radiative transfer processes and the neutral hydrogen column density and we find a large variety of different Lyman $\alpha$ line shapes, as can be seen in Fig.~\ref{fig:overview}, which we fitted with an asymmetric Gaussian (Sect.~\ref{Lyman alpha Line Properties}).

%%%%%%%%%%%%%%%%%%%%%%%%%%%%%%%%%%%%%%%%%%%%%%%%%%%%%%%%%%%%%%%%%%%%%%%%%%%%

\subsection{Line fluxes} \label{Line Fluxes} 

The line flux was measured directly from the MUSE data cubes using \texttt{LSDCat} (\citealp{HerenzWisotzki2017}) in spatial apertures of three Kron radii (\citealp{KronRadius}, see \citealp{Herenz2017}), which gives the first-order moment of the light distribution (\citealp{Bond2009}). The spectral range for the line flux measurement of \texttt{LSDCat} includes all voxels (volume pixels) above the analysis threshold (see \citealp{Herenz2017}, especially Fig. 8).
This way we used the full three dimensional information from the MUSE cube to include the entire Lyman $\alpha$ line flux. In appendix~\ref{Halo Sizes} we show a comparison between our measurements for line fluxes and measurements from \citet{Leclercq2017} including a modelling of the extended Lyman $\alpha$ halo. Since both values match well, we conclude that our line flux measurements capture well the flux in the extended Lyman $\alpha$ halo (except for extreme cases).

This extended emission is often not taken into account when measuring Lyman $\alpha$ EW$_0$ if the line flux is measured from slit spectroscopy or narrow-band images where the object is assumed to have the same size in Lyman $\alpha$ as in the complementary broad-band images. Since $40 - 90\%$ of the line flux can be found in the Lyman $\alpha$ halo when comparing the Lyman $\alpha$ emission to the extent of the UV continuum (\citealp{Wisotzki2016}), a large portion of the line flux could be omitted.
In cases where the Lyman $\alpha$ line is double peaked (see Sect.~\ref{Lyman alpha Line Properties} below), \texttt{LSDCat} sometimes does not include the full flux of the blue bump, which is why we corrected the line flux measurements for the fraction missed in this way. For this correction we fitted a linear combination of two asymmetric Gaussian functions to one-dimensional spectra (see Sect. \ref{Lyman alpha Line Properties} below), took into account the wavelength window that LSDCat used for the line flux, measured the ratio of the part of the line that was missed to the part that was included, and corrected the line flux accordingly.

Another potential correction of the Lyman $\alpha$ line flux is the absorption in the IGM (e.g. \citealp{Stark2011, Laursen2011, Caruana2014, Kusakabe2020, Hayes2020}). While many studies on Lyman $\alpha$ EW$_0$ assume a generic flux correction, the line shape can be influenced by radiative transfer processes and gas kinematics even before absorption in the IGM. In addition to this, the stochasticity and redshift-dependence of the IGM absorption (see e.g. \citealp{Thomas2017,Thomas2020} who show the large dispersion in the IGM transmission along different lines of sight) make it impossible to correct at an object by object basis, also given the fact that LAEs could reside in ionised bubbles (e.g. \citealp{Roberts-Borsani2016, Castellano2016, Stark2017, Castellano2018, Mason2018, Tilvi2020,Jung2020,Endsley2021}) and we do not know the systemic redshifts. Therefore we did not correct for IGM absorption, so as not to overestimate the Lyman $\alpha$ line flux and thus the $\m{EW}_0$. In extreme cases, assuming the Lyman $\alpha$ line is symmetric around the systemic redshift (which is unlikely though due to radiative transfer processes) and the IGM absorption in the sightline is $100\%$, we could thus underestimate the real Lyman $\alpha$ flux by half. 

%%%%%%%%%%%%%%%%%%%%%%%%%%%%%%%%%%%%%%%%%%%%%%%%%%%%%%%%%%%%%%%%%%%%%%%%%%%%%

\subsection{Lyman \texorpdfstring{$\alpha$}{Lg} line shape properties} \label{Lyman alpha Line Properties}

In this section we quantify the line shape properties, which are the asymmetry, FWHM (corrected for the line spread function, LSF, of MUSE), double peak fraction, and peak separation from one-dimensional spectra extracted from the MUSE datacubes. These spectra were obtained by weighted summation in each spectral layer. As weighting function we used the wavelength dependent Moffat (\citealp{Moffat1969}) profile that describes the PSF of our observations (see \citealp{Urrutia2019}). This was done to get the highest possible S/N in the emission line for our spectral fits, by weighing down the more noisy outer parts of the emission. This way we did not retain the information of the total line flux in these spectra, which is why we used the line flux measurements directly from \texttt{LSDCat}. We also did not model any specific spatially extended shape of the Lyman $\alpha$ emission for these one dimensional spectra that we used for fitting the Lyman $\alpha$ line. This assumes that there are no spatial variations in the line shape properties and the line has the same properties in the halo as in the central part of the LAE. This simplification is sufficient for the purpose of this paper, but it should be noted that a spatial variation of Lyman $\alpha$ line properties of our high-redshift LAEs is possible (see e.g. \citealp{Erb2018} for an example at $z=2.3$, \citealp{Claeyssens2019}  and \citealp{Leclercq2020} for Lyman $\alpha$ halos in the MUSE Lensing Clusters survey, \citealp{Richard2021}, and the MUSE-Deep survey).

To determine which Lyman $\alpha$ lines have a blue bump, we used a visual inspection tool that also fits the spectra with an asymmetric equation (explained below, equation~\ref{asymmetry}). We opted for a manual detection of blue bumps for this work to find even unusual multiple peaks, but an automatic determination of double peaks is a possibility as well (see Vitte et al. in prep.).
As mentioned above, we used the fits to the two parts of the Lyman $\alpha$ line to correct the line fluxes to include the blue bumps, since the line flux is taken from \texttt{LSDCat} in a certain wavelength window, which not always includes the full blue bump, especially for high peak separations. 

In order to gauge the reliability of the visual inspection regarding the presence of blue bumps, we used a Monte-Carlo type approach by creating 1000 randomised spectra with varying S/N values (of the red, main peak compared to the noise). The basis for this were ten spectra with intrinsically high S/N ratios (all with a S/N$>20$ for the main, red peak), five of them with clear blue bumps and five without blue bumps, taken from the first 24 fields of the MUSE-Wide survey. Their peak separations range from $308\pm39\,\m{km/s}$ to $477\pm89\,\m{km/s}$ and their blue bump to total line flux ratios range from $9\%$ to $29\%$. These spectra were then artificially degraded to lower S/N values between 0.5 and 10 of the main line and again analysed in the same way as the real spectra, which means determining visually whether an object has a double peak or not. However, since we know from the original spectra which objects should have double peaks, we can now analyse up to which S/N value we are able to accurately recover the double peaks.

At the S/N values used as the detection limit in MUSE-Wide (6.4 and 5 respectively for the first 24 fields and the rest of the fields), we already reach an accuracy of over $80\%$, which means in $80\%$ of cases we could correctly determine whether the spectrum has a blue bump or not (where the accuracy is the sum of determined true positives and true negatives divided by the sum of actual positives and negatives). This also depends on the peak ratio, since a low blue bump to main peak ratio in combination with a low S/N of the main peak will result in a blue bump that is harder to detect than for the same S/N with a higher peak ratio. Another aspect is the peak separation of the chosen objects. We chose objects that had obvious double peaks, which lead to the rather narrow range of peak separations. For closer peaks, the accuracy of the classification is expected to be lower.

We fitted the Lyman $\alpha$ line using the asymmetric Gaussian function described by \citet{Shibuya2014b}:

\begin{equation}\label{asymmetry}
 f(\lambda) = A \exp{\left(- \frac{(\lambda-\lambda_{0}^{\m{asym}})^2}{2 \sigma_{\m{asym}}^2} \right)} + f_0 .
\end{equation}

Here, $f_0$ is the continuum level, $A$ is the amplitude, and $\lambda_{0}^{\m{asym}}$ is the peak wavelength of the line. For the latter, we took the \texttt{LSDCat} measurements as a first guess. The asymmetric dispersion is $\sigma_{\m{asym}}$, consisting of $\sigma_{\m{asym}} = \m{a}_{\m{asym}}(\lambda-\lambda_0^{\m{asym}})+d$. Here, $d$ is the typical width of the emission line and $a_{\m{asym}}$ is the asymmetry parameter. A positive asymmetry value suggests a line with a red wing, which is the case for most of the red (main) Lyman $\alpha$ lines, a negative value means the line has a blue wing. As described above, a Lyman $\alpha$ line can also be double peaked. If that is the case and the blue bump was fit individually, the asymmetry only refers to the red peak. If there is a close but unresolved blue bump, the measured asymmetry will be smaller than if the double peak could have been resolved and fit individually or there could even be a blue wing.
After fitting the line, we derived the FWHM value from the asymmetric Gaussian fit, which is given by (see also \citealp{Claeyssens2019}):

\begin{equation}
    \m{FWHM} = \frac{2\sqrt{2\,\m{ln}2}\,d}{1-2\,\m{ln}2\,\m{a}_{\m{asym}}^2} .
\end{equation}

We corrected the FWHM of the Lyman $\alpha$ line for the spectral line spread function (LSF) of MUSE, which can be approximated by a Gaussian. The FWHM of the LSF is wavelength dependent and we used the value for the MUSE-Deep Mosaic fields (with ten hours exposure time) given by \citet{Bacon2017}, which follows $F_{\m{mosaic}}(\lambda\,[\angstrom]) = 5.835 \, 10^{-8} \lambda^2 -9.080 \, 10^{-4} \lambda + 5.983$. This correction assumes that the Lyman $\alpha$ line can be approximated by a Gaussian as well, which is not always the case. Therefore the LSF correction itself is also just an approximation (for a detailed discussion of this problem see \citealp{Childs2018}). It should be kept in mind as well, that the measure of the asymmetry becomes unreliable for narrow lines which are dominated by the LSF.

\citet{Verhamme2015,Verhamme17} have predicted and shown that the line shape properties are connected with the Lyman $\alpha$ escape fraction and EW$_0$ of the LAEs. There is a correlation between the peak separation and the shift of the red peak with respect to the systemic redshift as well as between the FWHM of the line and the shift of the red peak (\citealp{Verhamme2018}). Thus, we used both the peak separation and the FWHM to more accurately estimate the systemic redshift of the LAEs (according to equations one and two in \citealp{Verhamme2018}). We use these corrected redshifts throughout the paper.

%%%%%%%%%%%%%%%%%%%%%%%%%%%%%%%%%%%%%%%%%%%%%%%%%%%%%%%%%%%%%%%%%%%%%%%%%%%%%%%%%%%%%%%%%%%

\section{Equivalent widths} \label{Equivalent Widths}

To investigate the strength of the Lyman $\alpha$ lines and later compare it to other properties, we measure EW$_0$. With the measurements of the Lyman $\alpha$ line flux $F^{\m{line}}_{\m{Ly}\alpha}$ and UV continuum flux density $f^{\m{cont}}_{\m{Ly}\alpha}$ at the wavelength of the Lyman $\alpha$ line described above, we can now determine the Lyman $\alpha$ EW$_0$ as the fraction between the two:
\begin{equation} \label{eq:EW}
 \m{EW} = \int\limits^{\lambda_{1}}_{\lambda_{0}} \frac{ f^{\m{line}}_{\m{Ly}\alpha} - f^{\m{cont}}_{\m{Ly}\alpha} }{f^{\m{cont}}_{\m{Ly}\alpha}} \, \m{d}\lambda \approx \frac{F^{\m{line}}_{\m{Ly}\alpha}}{f^{\m{cont}}_{\m{Ly}\alpha}} . 
\end{equation} 

Here, $\lambda_0$ and $\lambda_1$ define the range of integration, which is the width of the Lyman $\alpha$ line, while the flux density in the line is $f^{\m{line}}_{\m{Ly}\alpha}$. The approximation used here on the right side of the equation is valid given $f^{\m{cont}}_{\m{Ly}\alpha}<<f^{\m{line}}_{\m{Ly}\alpha}$.

The rest-frame EW is then given as $\m{EW}_0~=~\m{EW}~/~(1+z)$. As explained above, we did not correct the measured EW$_0$ for potential IGM absorption. For all plots showing EW$_0$ measurements (except the histograms) we include both the error on the Lyman $\alpha$ line flux as well as on the UV continuum measurements in the error bars.

\subsection{Results: histograms of equivalent widths} \label{Distribution of Equivalent Widths}

\begin{figure}
\centering
\includegraphics[width=0.49\textwidth]{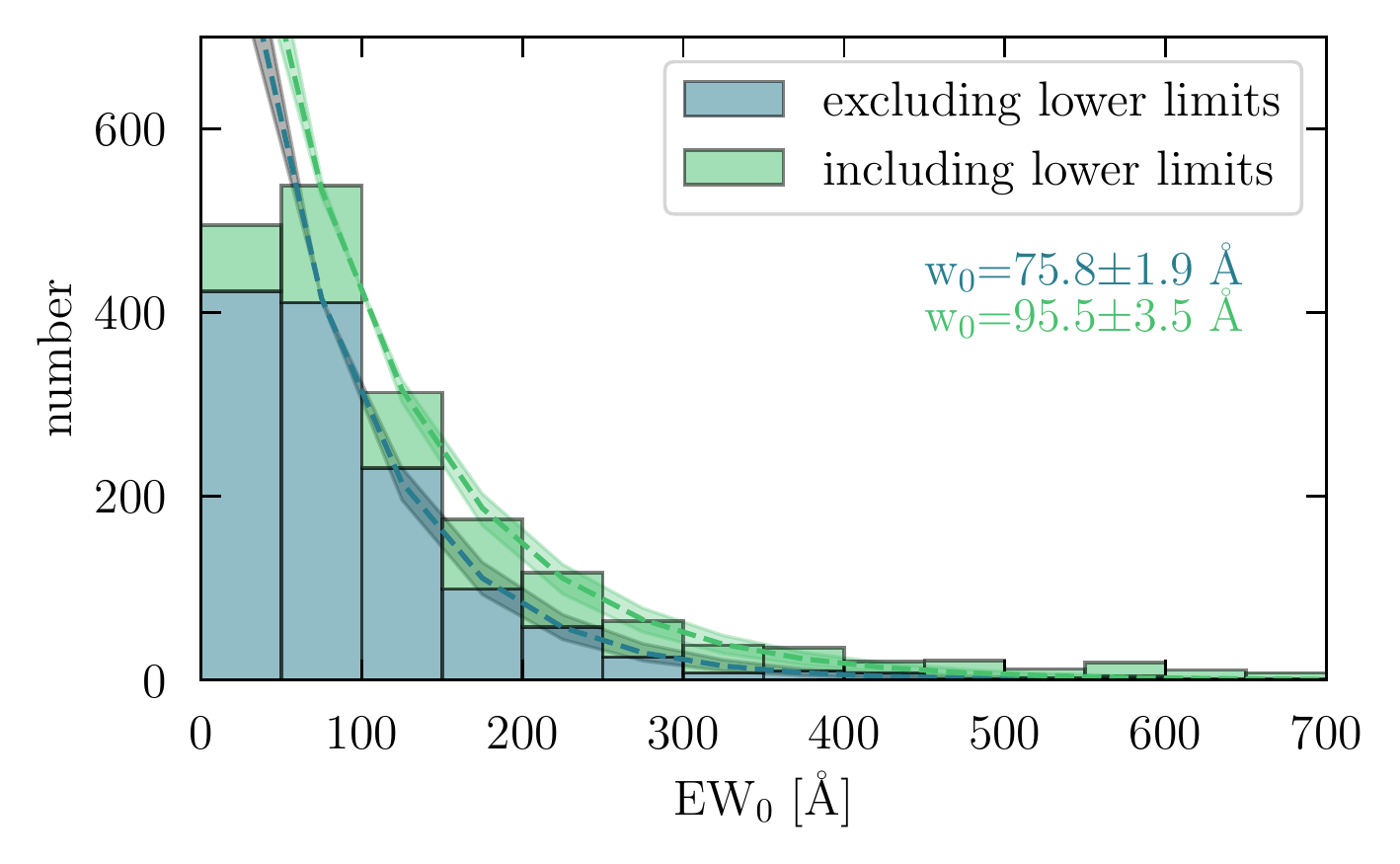}
\caption{Histograms of the Lyman $\alpha$ EW$_0$ of the full sample, including both MUSE-Wide and MUSE-Deep objects. The blue histogram shows the objects with a secure measurement (excluding lower limits), the green bars on top show the objects with a limiting $\m{EW}_0$. The exponential fits $N=N_0\, \m{exp}(-\m{EW}_0/\m{w}_0)$ are shown as the dashed lines, blue for the objects with secure measurements and light green for the full sample including lower limits. The bin width in EW$_0$ is $50\, \angstrom$. For the measurement of the characteristic EW$_0$ w$_0$ (given below the legend and in Table~\ref{tab:scale_lengths}) the smallest EW$_0$ bin was excluded as it is likely incomplete (as Lyman $\alpha$ lines with small or zero line fluxes are hard to detect and preferentially result in small EW$_0$). The shaded areas around the dashed lines indicate the exponential fits for the EW$_0$ distributions using UV continuum slopes of $\beta = -1.57$ and $-2.29$, the first and last quartiles of the distribution of measured $\beta$ values.} 
\label{fig:EW_dist}
\end{figure}

Histograms of EW$_0$ are shown in Fig.~\ref{fig:EW_dist} with exponential fits to determine the scaling factor. Objects where only a lower limit for the EW$_0$ could be measured are shown in the histogram as the green bars. The best-fit values (using a least-squares fit) for the exponential fit $N=N_0\, \m{exp}(-\m{EW}_0/\m{w}_0)$ are $\m{w}_0=75.8\pm1.9\,\angstrom$ and $\m{w}_0=95.5\pm3.5\,\angstrom$ for the characteristic EW$_0$, and $N_0=804.2\pm20.3\,\angstrom$ and $N_0=901.4\pm5.6\,\angstrom$ (full sample and for the sample excluding lower limits, respectively). Results in the literature on $\m{w}_0$ vary but generally fit well with our measurements. Recently, \citet{Jung2018} found $\m{w}_0\sim 60-100\,\angstrom$ for a sample of LAEs with redshifts $0.3<z<6$. In a previous study on LAEs in the MUSE-Deep survey, \citet{Hashimoto2017} divided their sample in three redshift bins, finding $\m{w}_0=113\pm14\,\angstrom$, $\m{w}_0=68\pm13\,\angstrom$ and $\m{w}_0=134\pm66\,\angstrom$ for redshifts $z\sim3.6$, $z\sim4.9$ and $z\sim6.0$, respectively.

Table~\ref{tab:dist_stats} gives an overview of the number of objects in the full sample as well as in the MUSE-Wide and MUSE-Deep samples that have  $\m{EW}_0 > 100\, \angstrom$ and $\m{EW}_0 > 240\, \angstrom$. The strength of the MUSE-Wide survey in detecting extreme LAEs is evident when looking at the statistics: including the lower limits for $\m{EW}_0$ (for objects that are undetected in the HST data), around $20\%$ of LAEs in MUSE-Wide have $\m{EW}_0$ higher than predicted by stellar population models (which set the limit at $240\, \angstrom$, assuming solar metallicity, a constant SFR, and an upper cut-off of the IMF of $80\,\m{M}_{\odot}$, \citealp{Charlot1993}), while only $\sim 11 \%$ of objects in MUSE-Deep have EW$_0>\,240\,\angstrom$. 

To estimate the influence of the choice of $\beta$ on the EW$_0$ histograms and the characteristic EW$_0$, we used the first and last quartile values of our measured UV continuum slopes in addition to the median value of $\beta=-1.97$ and show the result as the shaded range in Fig.~\ref{fig:EW_dist} and in Table~\ref{tab:scale_lengths}. This highlights that, within a reasonable range, the choice of UV continuum slope does not affect the measured characteristic EW$_0$ much.

\begin{table*} 
\begin{center} 
\caption{Overview of the numbers and fractions of high $\m{EW}_0$ LAEs in the different samples.} 
\begin{tabular}{ l | l l | l l | l l } 
\hline\hline 
& \multicolumn{2}{c|}{full sample} & \multicolumn{2}{c|}{MUSE-Wide} & \multicolumn{2}{c}{MUSE-Deep} \\ \hline 
& total & without lower limits & total & without lower limits & total & without lower limits \\ \hline 
\# of measurements & 1920 & 1284 & 1051 & 800 & 869 & 484 \\ 
 $\m{EW}_0>100\, \angstrom$ & 887 (46$\%$) & 450 (35$\%$) & 555 (53$\%$) & 346 (43$\%$) & 332 (38$\%$) & 104 (22$\%$) \\ 
$\m{EW}_0>240\, \angstrom$ &  306 (16$\%$) & 71 (6$\%$) & 211 (20$\%$) & 56 (7$\%$) & 95 (11$\%$) & 15 (3$\%$) \\  \hline 
\end{tabular}\label{tab:dist_stats} 
\end{center} 
\tablefoot{The numbers are given for the total sample of objects (split into MUSE-Wide and MUSE-Deep in the last two columns) including both objects with and without UV continuum counterparts and for the sample of objects with UV continuum counterparts (in this case excluding objects with lower limits on $\m{EW}_0$).} 
\end{table*} 

By using the largest, homogeneous dataset of MUSE-identified LAEs, combined with the deepest HST broad-band data, we are able to make accurate measurements of EW$_0$ for $\sim2000$ galaxies, making this sample over an order of magnitude larger than previous studies. Our measurements establish the existence of high EW$_0$ objects with EW$_0>240\,\angstrom$ and even several $10^2\,\angstrom$, and their occurrence rate is not low, so the physical conditions allowing for the production of such a huge number of Lyman $\alpha$ photons per UV magnitude seem rather common in $3<z<6$ star-forming galaxies. Other studies often correct for the Lyman $\alpha$ absorption of the IGM, but as explained in Sect.~\ref{Line Fluxes}, we did not correct for the IGM. This means that some of our measured EW$_0$ could intrinsically be even larger (if we assume that the intrinsic EW$_0$ is what comes out of the galaxy without IGM attenuation).

\citet{Gawiser2006} caution that non-detections on broad-band data used for the continuum flux density can lead to extremely large $\m{EW}_0$ for spurious detections in narrow-band images. This mostly applies to narrow-band selected LAE samples and since the MUSE-Wide and MUSE-Deep samples were constructed using the spectroscopic information of MUSE where we can confirm the presence of an emission line (and thus classify the objects correctly), this danger is lower in our study. 

In Table~\ref{tab:scale_lengths} we show an overview of the measured $\m{w}_0$ values for different confidence levels of the classifications (see Sect.~\ref{Sample Selection}) and in Table~\ref{tab:conf_frac} we show the fractions of high EW$_0$ (EW$_0>100\,\angstrom$ and EW$_0>240\,\angstrom$) for different confidence levels and $\beta$ parameters.
If we exclude objects with a confidence below 2, the characteristic EW$_0$ values of the EW$_0$ distribution do not change much and the fraction of high EW$_0$ (EW$_0>100\,\angstrom$ and EW$_0>240\,\angstrom$) stays almost the same. Using the highest confidence objects reduces our sample to 455 LAEs. The biggest effect is the exclusion of objects with faint (or low S/N) emission lines with potentially smaller EW$_0$. Although the fraction of HST undetected objects among the highest confidence sample is only $\approx 12\%$, compared to around a third for the entire sample, using the high confidence objects results in a similar characteristic EW$_0$ value $\m{w}_0$ and fraction of high EW$_0$ objects (both for EW$_0>100\,\angstrom$ and EW$_0>240\,\angstrom$). \\
The reason why the fraction of HST undetected objects is lower for LAEs with a high confidence is that their Lyman $\alpha$ lines are usually stronger, with a higher flux allowing for a higher classification confidence. A higher Lyman $\alpha$ flux usually comes along with a brighter UV continuum, making the object more likely to be detectable in the HST data as well (see Sect.~\ref{Differences between MUSE-Wide and MUSE-Deep} and Fig.\ref{fig:logLLya_dist_MW_MD}). 

This study demonstrates the existence of LAEs with high EW$_0$ (above EW$_0>240\,\angstrom$) and that they are quite numerous. Notably, one of the highest securely measured $\m{EW}_0$ (with an error that indicates a $3\,\sigma$ confidence) we find is at $\m{EW}_0 = 588.9 \pm 193.4\ \angstrom$, which is a clear indication of an unusual underlying stellar population (see Sect.~\ref{The highest EW LAE}).

\begin{table}
\begin{center}
\caption{w$_0$ for different sub-samples}
\begin{tabular}{ l | l l }
\hline\hline
   sub-sample &  w$_0$ with lower limits &  w$_0$ without lower limits  \\ \hline
   total & $95.5\pm3.5\,\angstrom$ & $75.8\pm1.9\,\angstrom$ \\
   conf. $>1$ & $92.1\pm3.0\,\angstrom$ & $76.3\pm2.1\,\angstrom$ \\ 
   conf. $>2$ & $86.6\pm2.9\,\angstrom$ & $78.0\pm1.1\,\angstrom$ \\ 
   $\beta = -1.57$ & $104.9\pm4.5\,\angstrom$ & $85.0\pm1.6\,\angstrom$ \\
   $\beta = -2.29$ & $85.3\pm3.0\,\angstrom$ & $67.1\pm1.3\,\angstrom$ \\ \hline
\end{tabular}\label{tab:scale_lengths}
\end{center}
\end{table}

\begin{table*}
\begin{center}
\caption{Fraction of high EW$_0$ for different confidence sub-samples and $\beta$ values}
\begin{tabular}{ l | l l l l}
\hline\hline
    &  conf. $>1$ & conf. $>2$ & $\beta = -2.29$ &  $\beta = -1.57$  \\ \hline
    \# of measurements & 1407 & 455 & 1920 & 1920 \\
   $\m{EW}_0>100\, \angstrom$ & 651 ($46\%$) & 212 ($47\%$) & 794 ($41\%$) & 984 ($51\%$) \\
   $\m{EW}_0>240\, \angstrom$ & 191 ($14\%$) &  55 ($12\%$) & 256 ($13\%$) & 370 ($19\%$) \\ \hline
\end{tabular}\label{tab:conf_frac}
\end{center}
\end{table*}

%%%%%%%%%%%%%%%%%%%%%%%%%%%%%%%%%%%%%%%%%%%%%%%%%%%%%%%%%%%%%%%%%%%%%%%%%%%%%%%%

\subsection{The influence of survey depth on equivalent widths} \label{The Influence of Survey Depth on Equivalent Widths}

We now want to understand the influence of cuts in Lyman $\alpha$ luminosity on the measured EW$_0$ and histograms. Thus, to estimate the effect of different survey depths, we used a statistical experiment, creating a sample of $10\,000$ objects with UV magnitudes drawn randomly from a luminosity function based on the Schechter parameterisation $ \phi^* \, \m{ln}(10)\, 0.4 \times  10^{-0.4(M-M^*)(\alpha+1)} e^{-10^{-0.4(M-M^*)}}$ with parameters according to \citet{Bouwens2015} for objects at redshift $z\approx3.8$ with a characteristic magnitude $M^* = -20.88$, a normalisation $\phi^* = 1.97 \times 10^{-3}\,\m{Mpc}^{-3}$ and a slope of $\alpha = -1.64$, in the absolute magnitude range of $-23 < M_{\m{UV}}< -17 $. For the distribution of EW$_0$ of these objects we assumed $\m{w}_0=80\,\angstrom$, close to what we measured for our full sample of real LAEs. We drew EW$_0$ values randomly for the $10\,000$ objects and obtained the Lyman $\alpha$ line flux by converting the absolute magnitudes from the luminosity function of \citet{Bouwens2015} to continuum flux densities and multiplying that with the EW$_0$. 

\begin{figure}
\centering
\includegraphics[width=0.49\textwidth]{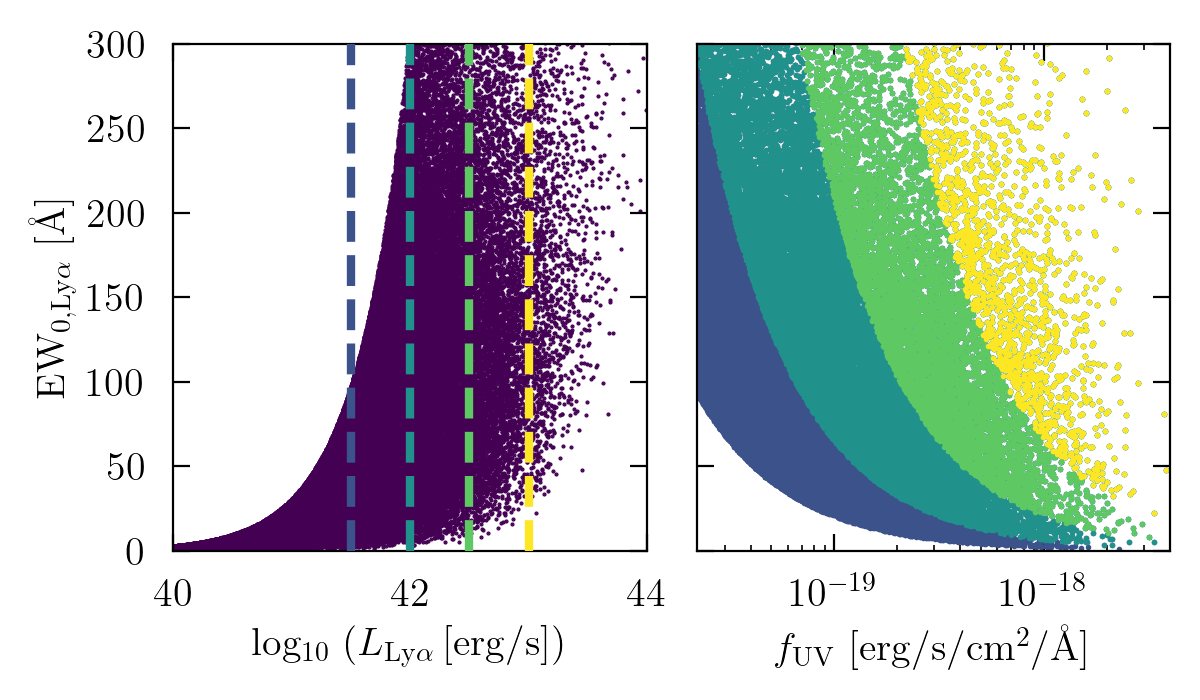}
\caption{Statistical experiment to understand the cuts of Lyman $\alpha$ luminosity on EW$_0$. Left panel: Lyman $\alpha$ EW$_0$ over Lyman $\alpha$ line luminosity for $10\,000$ simulated objects (purple dots). The dashed lines indicate different luminosity cuts in the Lyman $\alpha$ line at $\m{log}_{10}(L_{\m{Ly}\alpha}[\m{erg/s}])=[41.5,42,42.5,43]$. 
Right panel: Lyman $\alpha$ EW$_0$ over the UV continuum luminosity. The different colours correspond to the different line luminosity cuts in the left panel.} 
\label{fig:stat_LUVC_LLya_Ando_combo} 
\end{figure}

\begin{figure}
\centering
\includegraphics[width=0.49\textwidth]{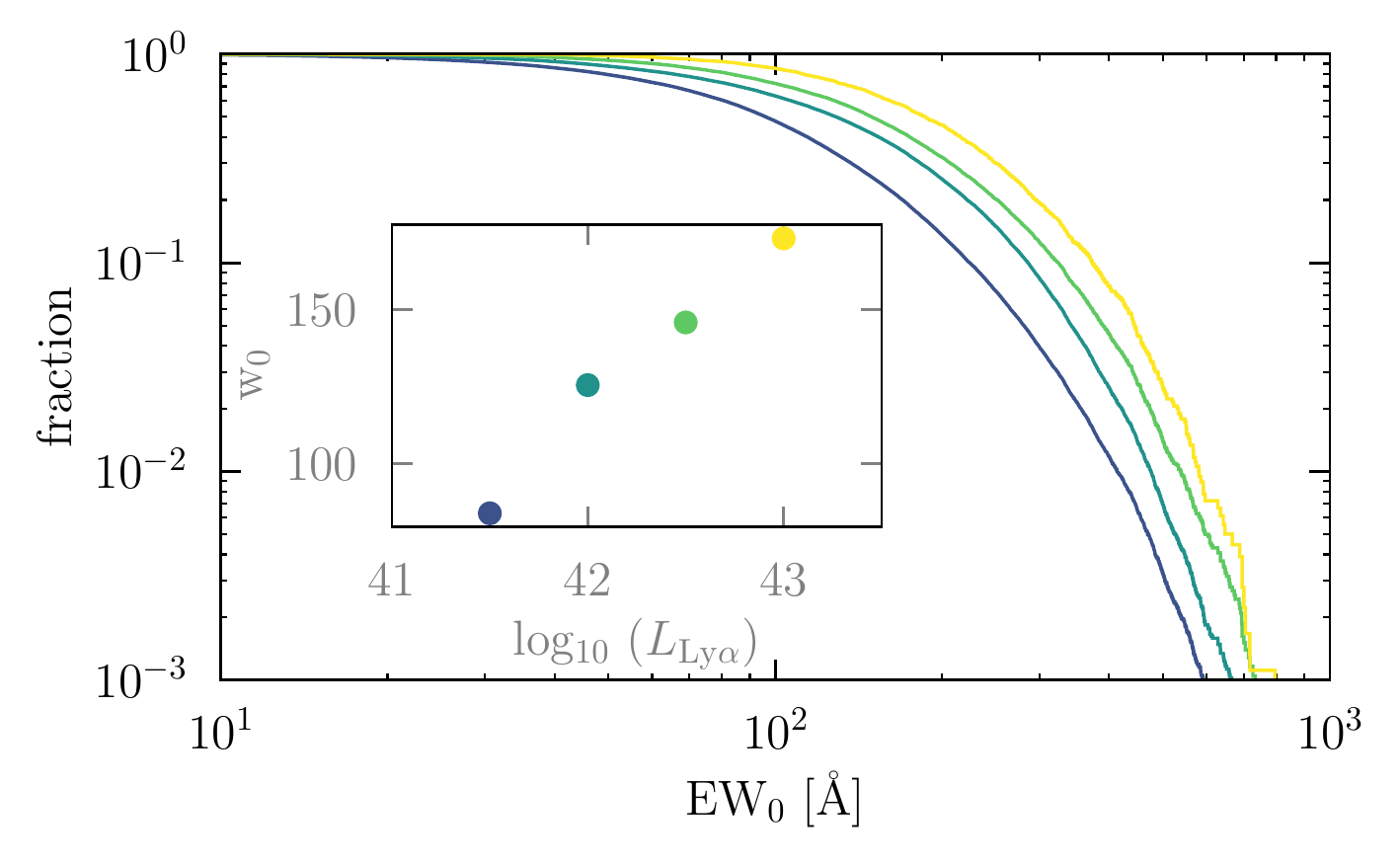}
\caption{Cumulative distributions of EW$_0$ corresponding to the different line flux cuts in Fig.~\ref{fig:stat_LUVC_LLya_Ando_combo}. The inset shows the measured $\m{w}_0$ values of exponential fits to histograms of EW$_0$ resulting from the luminosity cuts.} 
\label{fig:stat_frac_ews_cum} 
\end{figure}

The results of this experiment can be seen in Figs.~\ref{fig:stat_LUVC_LLya_Ando_combo} and \ref{fig:stat_frac_ews_cum}. Lyman $\alpha$ luminosities versus EW$_0$ are shown in the left panel in Fig.~\ref{fig:stat_LUVC_LLya_Ando_combo}. The hard cut on the left in the left panel is caused by the UV continuum magnitude range that we assumed. If we went to fainter magnitudes, it would be possible to populate the area more with higher EW$_0$ objects for the lower range of Lyman $\alpha$ line fluxes.

Next, we simulated different survey depths by introducing luminosity cuts and re-derived the EW$_0$ with these new Lyman $\alpha$ luminosities and the corresponding continuum flux densities. The result can be seen in Fig.~\ref{fig:stat_frac_ews_cum}, where we show the different cumulative distributions for the luminosity cuts in Fig.~\ref{fig:stat_LUVC_LLya_Ando_combo}. Inset in Fig.~\ref{fig:stat_frac_ews_cum} are the measured characteristic EW$_0$ values $\m{w}_0$, which increase with increasing luminosity cut. The same result remains (although with more scatter), if we reduce the number of objects to $1000$ and $100$, to see if a smaller survey size would change the distribution.

Thus, for a population where the EW$_0$ distribution is independent of UV luminosity, a survey with shallower depth in Lyman $\alpha$ will preferentially contain a larger fraction of higher EW$_0$ LAEs than a sample of larger depth. Our result is in agreement with \citet{Nilsson2009}, who showed that UV bright selected samples exhibit a lower fraction of high EW$_0$ LAEs than UV faint selected samples.

%%%%%%%%%%%%%%%%%%%%%%%%%%%%%%%%%%%%%%%%%%%%%%%%%%%%%%%%%%%%%%%%%%%%%%%%%

\subsection{Results: differences between MUSE-Wide and MUSE-Deep} \label{Differences between MUSE-Wide and MUSE-Deep}

\begin{figure}
\centering
\includegraphics[width=0.42\textwidth]{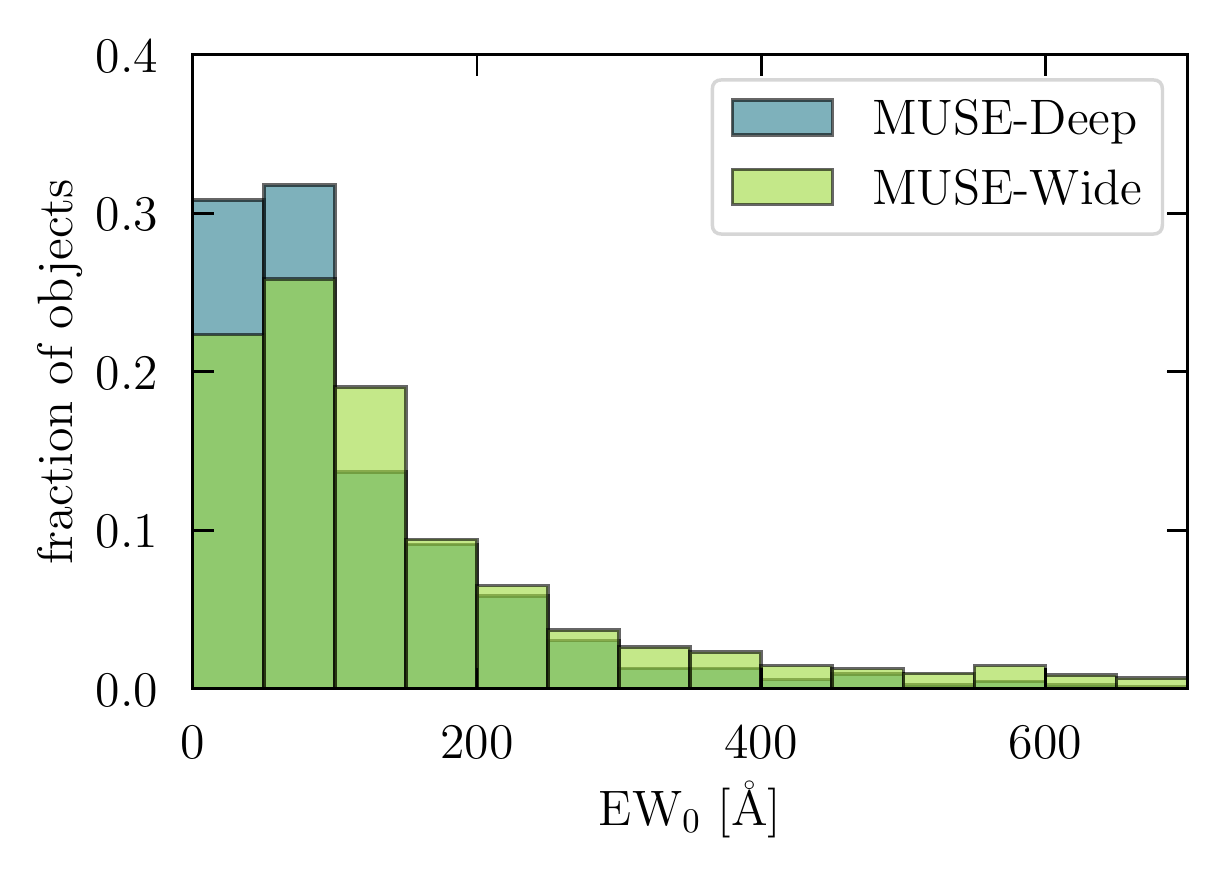}
\caption{Histograms of EW$_0$ for MUSE-Wide (green) and MUSE-Deep (blue) in bins of $50\,\angstrom$ normalised to the total number of objects in each survey. The histograms show the distributions including objects with no detectable UV continuum counterpart in HST.}
\label{fig:EW_dist_shifted_MW_MD_one}
\end{figure}

The influence of survey depth on the EW$_0$ statistics (as described above) can be demonstrated by comparing the measured EW$_0$ distributions from MUSE-Deep with the MUSE-Wide sample, shown in the histograms in Fig.~\ref{fig:EW_dist_shifted_MW_MD_one}. As can be seen, MUSE-Wide indeed contains a larger fraction of high and extreme EW$_0$ objects ($20\%$ with EW$_0>240\,\angstrom$ compared to MUSE-Deep having $11\%$ with EW$_0>240\,\angstrom$, see Table~\ref{tab:dist_stats}), whereas MUSE-Deep picks up a larger fraction of low-EW$_0$ objects ($62\%$ with EW$_0<100\,\angstrom$ compared to MUSE-Wide having $47\%$ with EW$_0<100\,\angstrom$).

The reason for this discrepancy are the characteristics of the datasets from the two surveys. MUSE-Wide covers a larger survey area, which leads to the discovery of more extreme and thus rare objects with very high EW$_0$. MUSE-Deep has longer exposure times and thus a lower Lyman $\alpha$ detection flux limit, making it possible to detect fainter Lyman $\alpha$ luminosities. The question is whether the LAEs that are fainter in Lyman $\alpha$ are also fainter in the UV continuum, which would lead to a similar EW$_0$ distribution between the brighter and the fainter Lyman $\alpha$ objects, keeping in mind that the LAEs were detected based on their Lyman $\alpha$ emission line (although the HST depth is similar in both surveys). 
It can be seen from the distribution of EW$_0$ that objects with lower EW$_0$ are more numerous, especially in the survey going deeper in Lyman $\alpha$ emission. 

\begin{figure}
\centering
\includegraphics[width=0.49\textwidth]{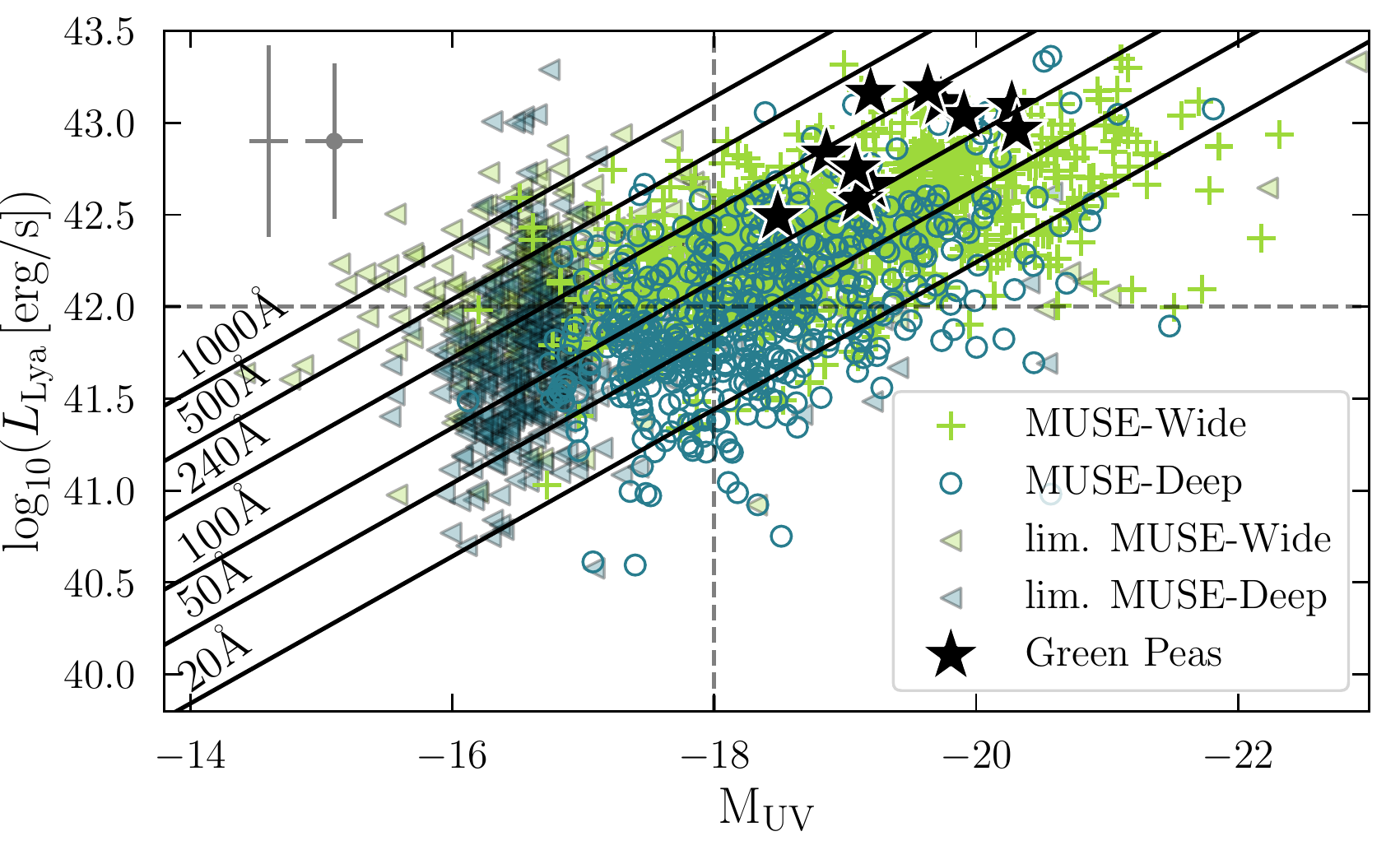}
\caption{Logarithmic Lyman $\alpha$ luminosity $\m{log}_{10}(L_{\m{Ly}\alpha}[\m{erg/s}])$ plotted against the UV continuum magnitude $\m{M}_{\m{UV}}$, divided into MUSE-Wide (light green crosses and triangles) and MUSE-Deep (dark green open dots and triangles). The triangles show upper $1\,\sigma$ limits of $\m{M}_{\m{UV}}$ for HST undetected objects. The diagonal black lines are lines of equal equivalent widths (at the median redshift of $z\approx 4$ and using a $\beta$ value of $-1.97$) to guide the eye.
The grey dashed lines indicate $\m{M}_{\m{UV}} = -18$ and $\m{log}_{10}(L_{\m{Ly}\alpha}[\m{erg/s}]) = 42$. The grey crosses in the top left corner show the median sizes of the errors (for objects with HST counterparts). The black stars show Green Pea galaxies from \citet{Izotov2016a,Izotov2016b, Izotov2018a,Izotov2018b} that were all found to be leaking Lyman continuum.}
\label{fig:MW_MD_logLLya_M_UV}
\end{figure}

To investigate the differences between the two surveys, the UV continuum magnitudes (measured here at $1500\,\angstrom$) and the Lyman $\alpha$ luminosities are shown in Fig.~\ref{fig:MW_MD_logLLya_M_UV}, with different colours for MUSE-Wide and -Deep. This plot demonstrates that the combination of a wide and a deep survey is essential to both cover the low-EW$_0$ and high-EW$_0$ tails of the LAE EW$_0$ distribution. 

While objects with strong Lyman $\alpha$ emission are rare (according to Lyman $\alpha$ luminosity functions, see e.g.  \citealp{Drake2017b, Herenz2019}), objects with faint UV magnitudes are harder to observe, which explains the lack of objects in the upper left part of the figure. However, this is exactly the range in which higher EW$_0$ objects can be found (lines of equal EW$_0$ are shown in the plot). Due to the larger survey area of MUSE-Wide, the high Lyman $\alpha$ luminosity part of the plot is mostly populated by MUSE-Wide LAEs, while the range in UV continuum magnitude covered is similar in both samples. MUSE-Deep is detecting more low Lyman $\alpha$ luminosity objects at similar UV continuum flux densities, resulting in many objects with smaller EW$_0$. 

It is interesting to note that some known LyC leaker galaxies at lower redshifts (Green Peas from \citealp{Izotov2016a, Izotov2016b, Izotov2018a, Izotov2018b}), marked in Fig.~\ref{fig:MW_MD_logLLya_M_UV} as black stars, occupy mostly the MUSE-Wide range of UV continuum magnitude and Lyman $\alpha$ luminosity values. MUSE-Wide LAEs might thus be ideal objects to search for LyC leakage at high redshift.

\begin{figure}
\centering
\includegraphics[width=0.49\textwidth]{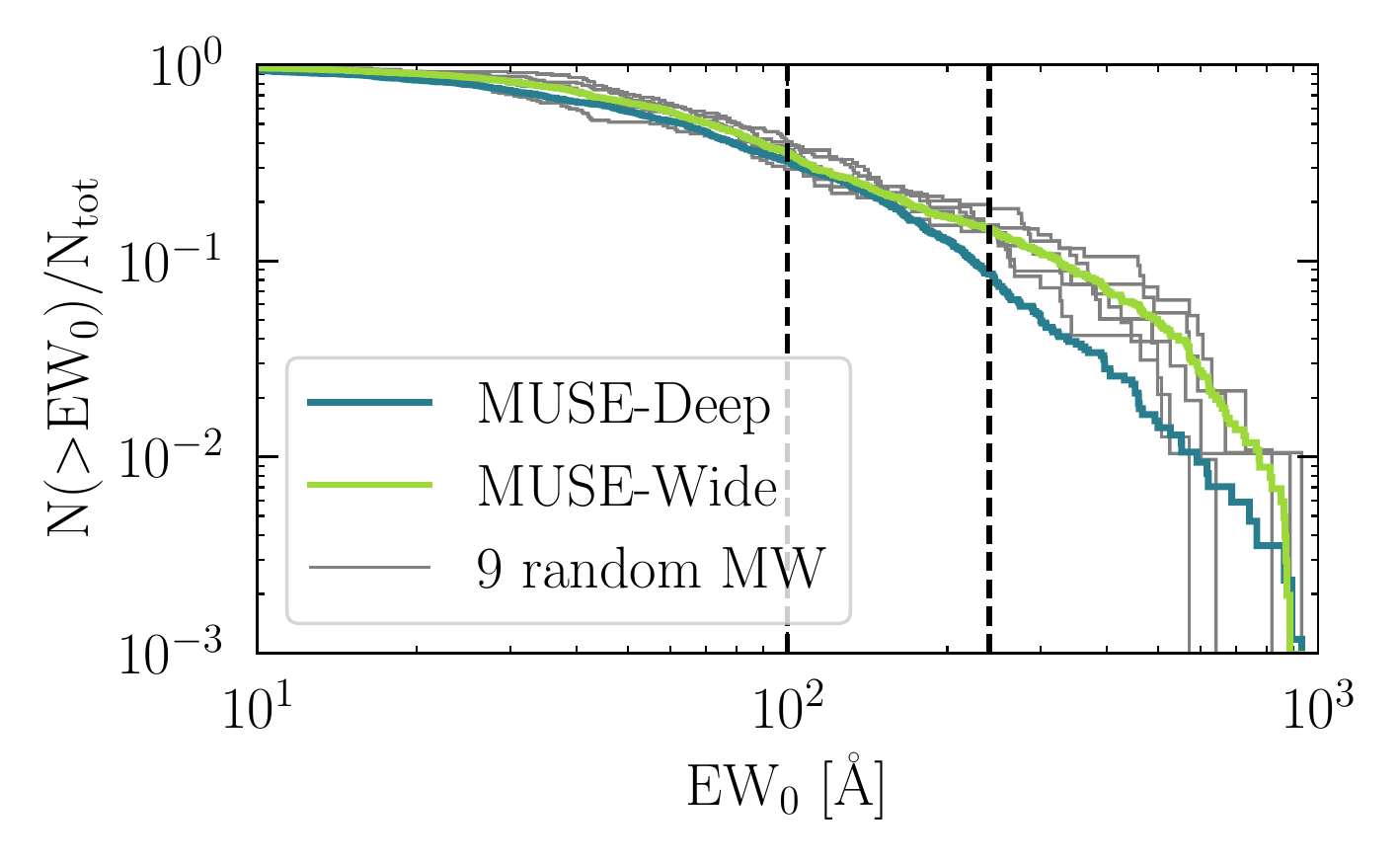}
\caption{Cumulative distribution of Lyman $\alpha$ EW$_0$ divided into MUSE-Wide (thick green line) and MUSE-Deep (thick blue line). The lines show the samples including objects with no detected UV continuum in the HST images, each step indicates an individual object. For the objects with UV continuum counterparts, the $1\,\sigma$ error was subtracted from the measured EW$_0$ in order to show the same lower limits as for objects without a continuum counterpart. The grey lines show the cumulative distributions of EW$_0$ for the MUSE-Wide data in an area of a similar size as MUSE-Deep, by picking nine MUSE-Wide fields at random. The black dashed lines indicate $\m{EW}_0 = 100\,\angstrom$ and $\m{EW}_0 = 240\,\angstrom$.}
\label{fig:EW_cumulative}
\end{figure}

To test the influence not only of the Lyman $\alpha$ detection limits but also of the UV continuum and survey area, we constructed a cumulative EW$_0$ distribution in Fig.~\ref{fig:EW_cumulative} similar to our statistical experiment in Fig.~\ref{fig:stat_frac_ews_cum}. The fraction of higher EW$_0$ is higher in MUSE-Wide than in MUSE-Deep, consistent with the discussion above. A Kolmogorov Smirnow (KS) test indicates that the two are likely not drawn from the same underlying distribution ($p=0.0014$). Since this plot shows the fraction of objects, the effect of the larger survey area of MUSE-Wide is accounted for. This is because a larger survey area will increase the total number of objects, but not the fraction of objects in certain EW$_0$ bins. To further test the possible influence of the survey area, we reduced the size of MUSE-Wide to nine fields, to simulate a similar size as MUSE-Deep. This can be seen as the grey lines in Fig.~\ref{fig:EW_cumulative}, each indicating nine randomly picked MUSE-Wide fields which follow the trend for the full MUSE-Wide survey (albeit with more scatter, since there are fewer objects). Since the distributions of EW$_0$ for the random MUSE-Wide fields do not match significantly better with the MUSE-Deep distribution (with a median p-value of $p=0.006$) than the real MUSE-Wide distribution, the influence of the survey area is likely to be small.

\begin{figure}
\centering
\includegraphics[width=0.49\textwidth]{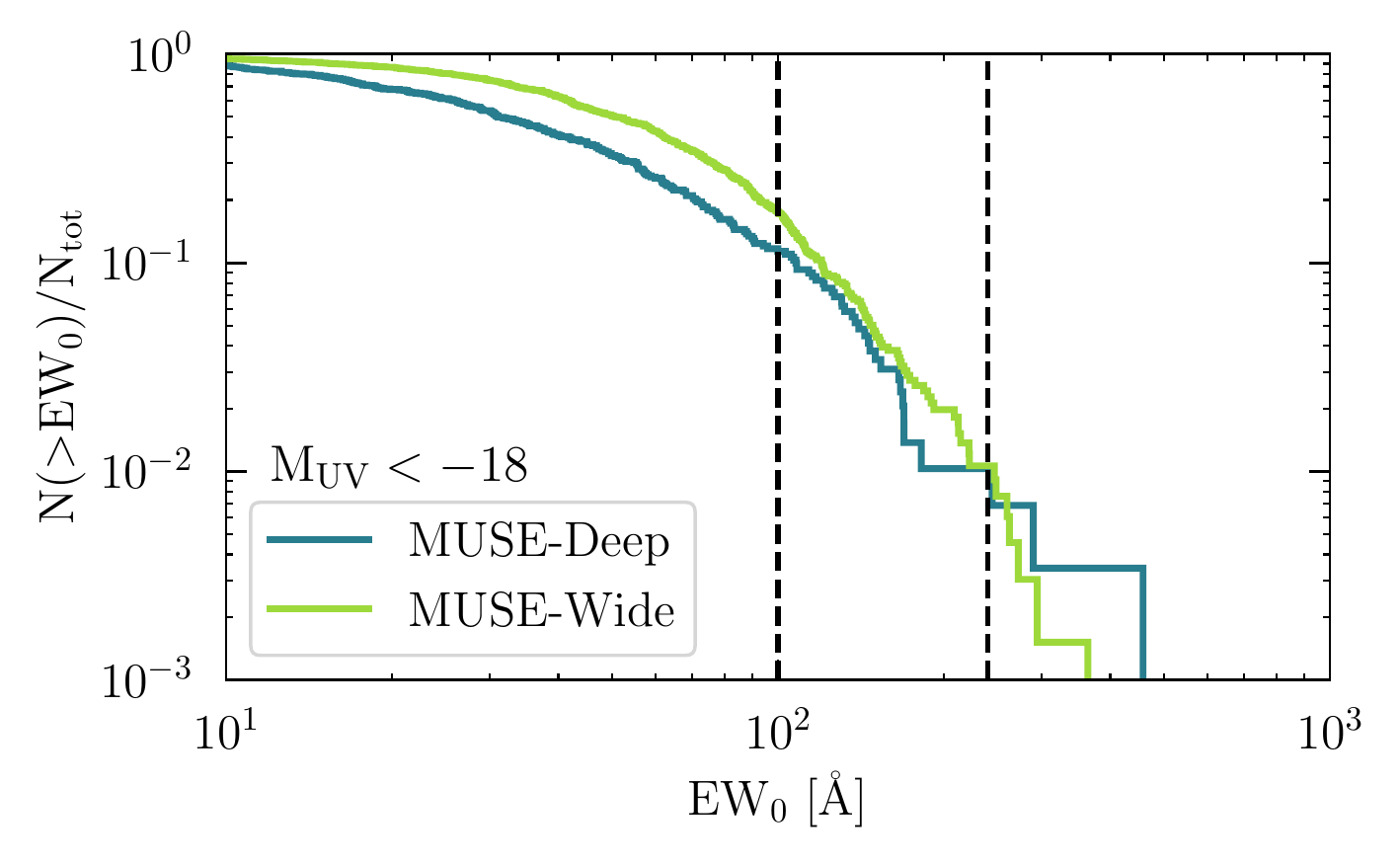}
\includegraphics[width=0.49\textwidth]{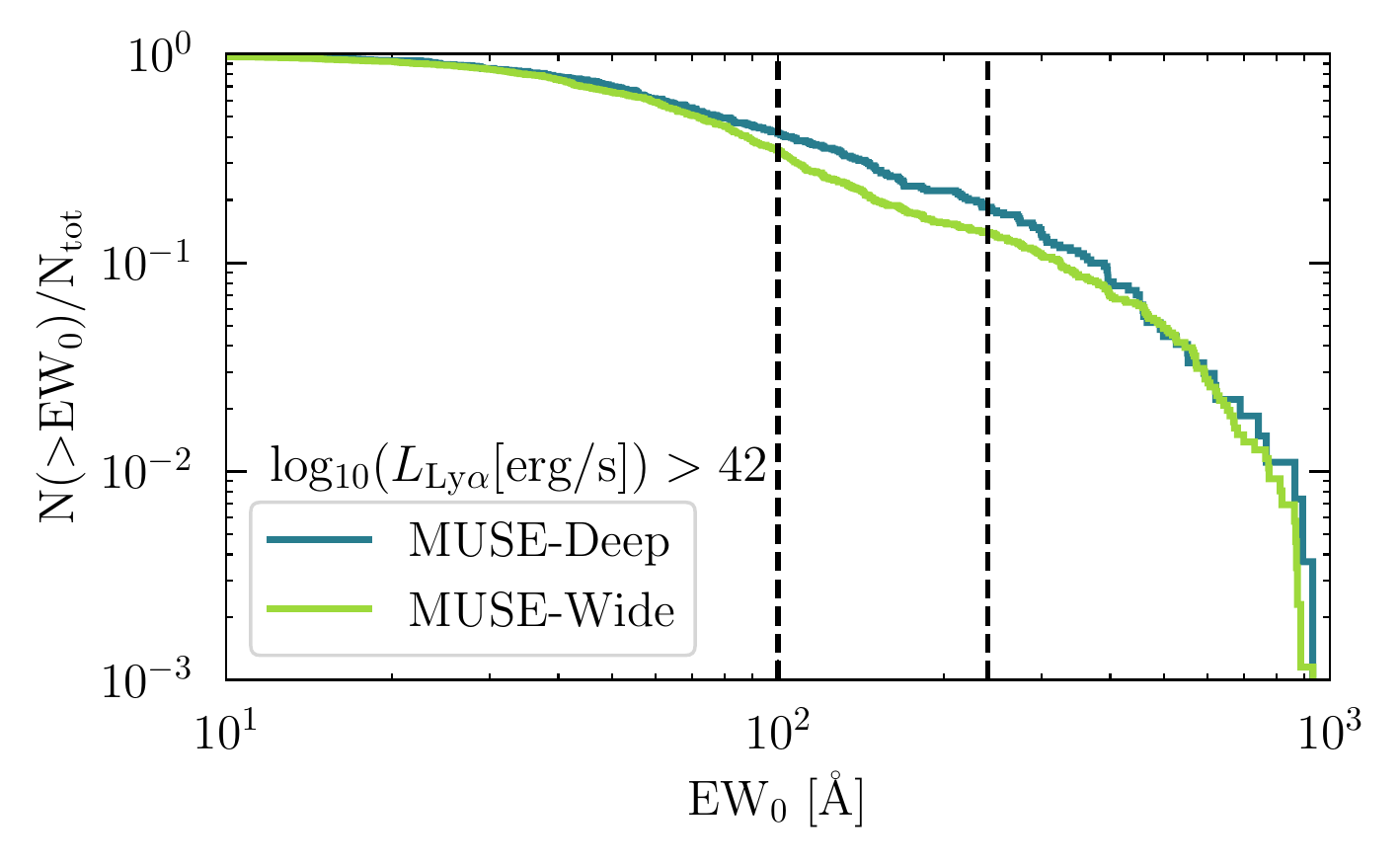}
\includegraphics[width=0.49\textwidth]{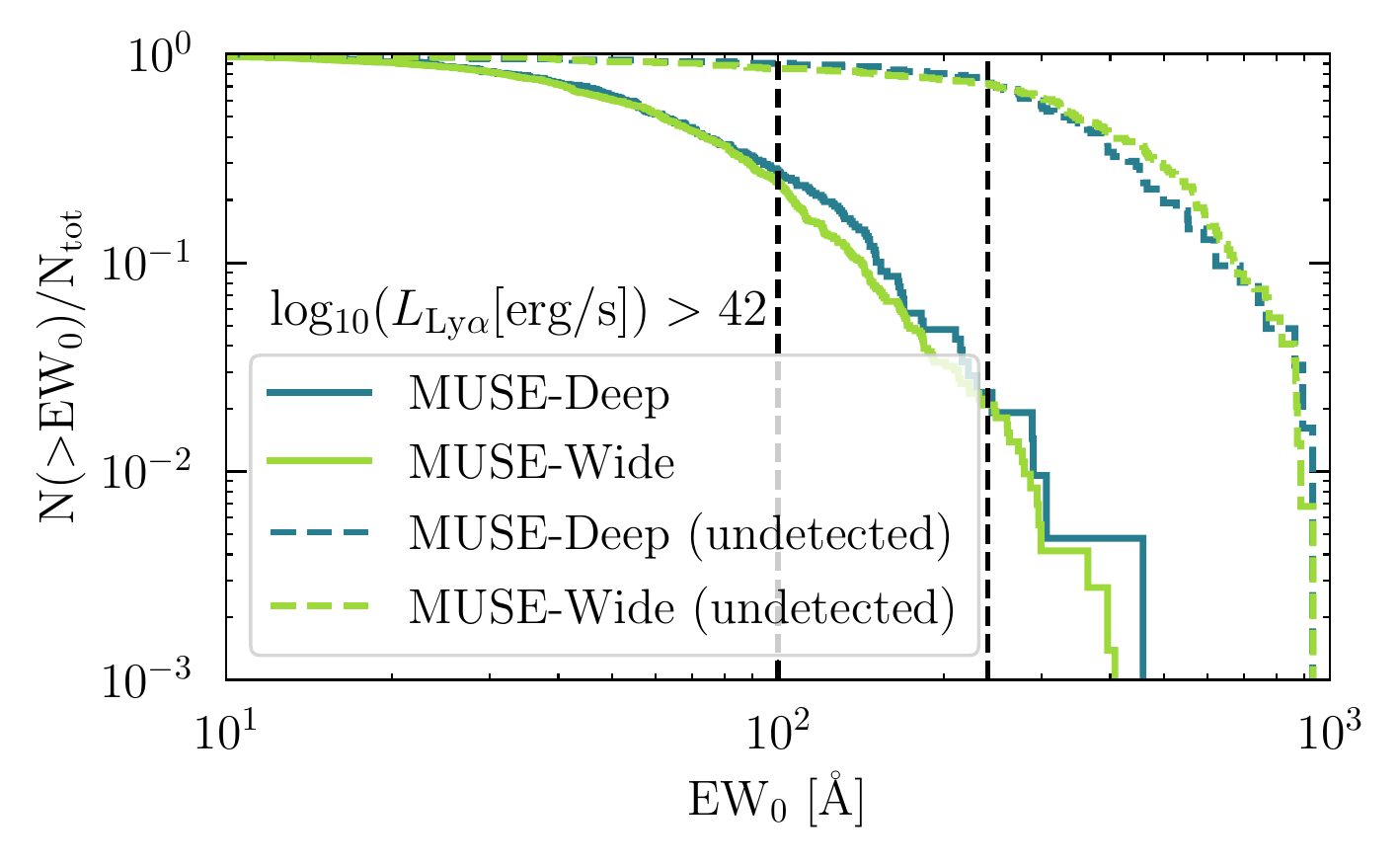}
\caption{Same plots as Fig.~\ref{fig:EW_cumulative} but with a cut in $\m{M}_{\m{UV}}<-18$ for the top panel and a cut in Lyman $\alpha$ luminosity of $\m{log_{10}}(L_{\m{Ly}\alpha}) >42$ for the middle and lower panel. Both cuts are marked as dashed lines in Fig.~\ref{fig:MW_MD_logLLya_M_UV}. In the lower panel, the sample is divided into objects with a UV continuum counterpart (solid lines) and without (dashed lines). As in Fig.~\ref{fig:EW_cumulative}, the black dashed lines indicate $\m{EW}_0 = 100\,\angstrom$ and $\m{EW}_0 = 240\,\angstrom$.}
\label{fig:EW_cumulative_cuts} 
\end{figure}

Next we tried to match the lines for MUSE-Wide and -Deep in the cumulative EW$_0$ distribution by introducing cuts in the absolute UV magnitude of $\m{M}_{\m{UV}} < -18$ (top panel of Fig.~\ref{fig:EW_cumulative_cuts}) and the Lyman $\alpha$ luminosity $\m{log}(L_{\m{Ly\alpha}}\,[\m{erg/s}]) >42$ (middle and bottom panels). Applying a cut in absolute UV magnitude does not remove but reduces the discrepancy between the two surveys, especially for the high EW$_0$ where more objects with fainter UV continua would be found. The cut in Lyman $\alpha$ luminosity, shown in the middle and bottom panels of Fig.~\ref{fig:EW_cumulative_cuts}, reduces the discrepancy especially in the low and very high range of EW$_0$. If we only look at objects with a UV continuum counterpart detected in HST, which effectively also introduces a cut in UV continuum depth, the curves still match well (with the additional Lyman $\alpha$ luminosity cut, see Fig.~\ref{fig:EW_cumulative_cuts}, bottom panel). 

This analysis leads to the conclusion that the difference in the EW$_0$ histogram is mostly due to the deeper Lyman $\alpha$ luminosity detection limit of MUSE-Deep, which leads to more objects with low EW$_0$. Thus, the construction of the survey has a significant influence on what kind of objects can be observed. Without correcting for these biases, simply looking at the distribution or fraction of high EW$_0$ (but also just the distribution of EW$_0$ as a whole) is not enough to understand the real occurrence of such objects. It would be useful to correct the distribution, taking the selection function into account and constructing an EW$_0$ distribution function, in the style of a luminosity function. We will exemplify such a procedure in a follow-up paper (Kerutt et al. in prep.). However, the wedding-cake approach taken here, using observations with different depth, is already an improvement over previous studies.

%%%%%%%%%%%%%%%%%%%%%%%%%%%%%%%%%%%%%%%%%%%%%%%%%%%%%%%%%%%%%%%%%%%%%%%%%%%%%%%%%%%%%%%%%%

\section{Connecting equivalent widths to line shape and morphological properties} \label{Connecting Equivalent Widths to other Properties}

In this section we study how the Lyman $\alpha$ EW$_0$ is connected to the shape of the line or of the UV continuum counterpart. In Sect.~\ref{Discussion} we go into a more detailed discussion of the properties of the highest EW$_0$ objects in particular.

%%%%%%%%%%%%%%%%%%%%%%%%%%%%%%%%%%%%%%%%%%%%%%%%%%%%%%%%%%%%%%%%%%%%%%%%%%%%%%

\subsection{Blue bump fraction and ratio} \label{Blue Blumps}

A common feature of Lyman $\alpha$ radiative transfer models is that they predict the existence of a weaker blue peak in the presence of outflowing gas (e.g. \citealp{Verhamme2015}). Due to the decreasing IGM transmission for photons to the blue of the rest-frame Lyman $\alpha$ wavelength, the observable fraction of blue bumps at higher redshifts will go down (\citealp{Laursen2011} but also \citealp{Hayes2020}), which is why a redshift cut is useful when discussing fractions of blue bumps. In total, around $33\%$ (324, see Table~\ref{tab:bump_stats}) of our objects below a redshift of $z<4$ have a blue bump. 

Objects with a blue bump tend to have a slightly larger EW$_0$ (see Table~\ref{tab:bump_stats}), although with a large spread in values. In fact, the EW$_0$ values of objects with a blue bump seem to have a larger spread, possibly due to the often smaller S/N of the blue bump, and thus could reach to higher values than the objects without a blue bump. If we assume that most emission lines (more than the approximately one-third we find) have a blue bump which is mostly eaten away by the intervening IGM, then this could explain why we see high $\m{EW}_0$ in cases where the blue bump was not absorbed by the IGM. However, because of the unknown IGM absorption it is not possible in this sample to draw a conclusion about the influence of the presence of a blue bump on a stronger Lyman $\alpha$ emission.

\begin{table*}
\begin{center}
\caption{Overview of the percentages of blue bumps in Lyman $\alpha$ lines and the median $\m{EW}_0$ for MUSE-Wide and MUSE-Deep, for objects with a redshift below 4. }
\begin{tabular}{ l | l l l}
\hline\hline
   & total & MUSE-Wide & MUSE-Deep \\ \hline
   percentage of objects with blue bump & 324 ($33.4 \%$) & 187 ($33.8 \%$) & 137 ($32.9 \%$) \\
   median $\m{EW}_0$ (no blue bump) & $83\, \pm 179\, \angstrom$ & $107\,\pm 214\, \angstrom$ & $70\,\pm105\, \angstrom$ \\
   median $\m{EW}_0$ (with blue bump) & $92\,\pm 409\, \angstrom$ & $104\,\pm 245\, \angstrom$ & $81\,\pm559\, \angstrom$ \\ \hline
\end{tabular}\label{tab:bump_stats}
\end{center}
\tablefoot{The percentages of the occurrence of blue bumps in the Lyman $\alpha$ line as well as the median EW$_0$ (with standard deviations) for the sample with and without a blue bump, separated into the full sample as well as the MUSE-Wide and MUSE-Deep survey, are given for a redshift cut of $z<4$.} 
\end{table*}

The high fraction of objects with blue bumps matches well with literature results. \citet{Kulas2012} find a similar result of $30\%$ for their LAEs at redshift $z \approx 2\--3$, \citet{Trainor2015} find $40\%$, while \citet{Yamada2012} even find a fraction as high as $50 \%$ in one of the largest overdense regions of LAEs at $z=3.1$. In a recent study of lensed LAEs at $z \approx 2\--3$, \citet{Cao2020} also find a blue bump fraction of $30-50\%$. 
However, most studies omit that the observed fraction of blue bumps in LAE spectra depends on the S/N of the line, with higher detection completeness and blue bump fractions for higher S/N. We can thus assume that most Lyman $\alpha$ spectra have a blue bump that is usually hidden in the noise or absorbed by the IGM. Indeed, \citet{Hayes2020} argue that the decline in the fraction of blue bump to main peak ratio with increasing redshift can be attributed entirely to the rise in the neutral fraction of the IGM. 
We can see this as well by looking at the fractions of blue bumps in different redshift bins. For the sample of LAEs below a redshift of $4$ we find $33\%$ of objects have blue bumps, while the fraction is only $16\%$ (107 LAEs) for objects between $4<z<5$ and $7\%$ (22 LAEs) for objects above $z>5$.

\begin{figure}
\centering
\includegraphics[width=0.49\textwidth]{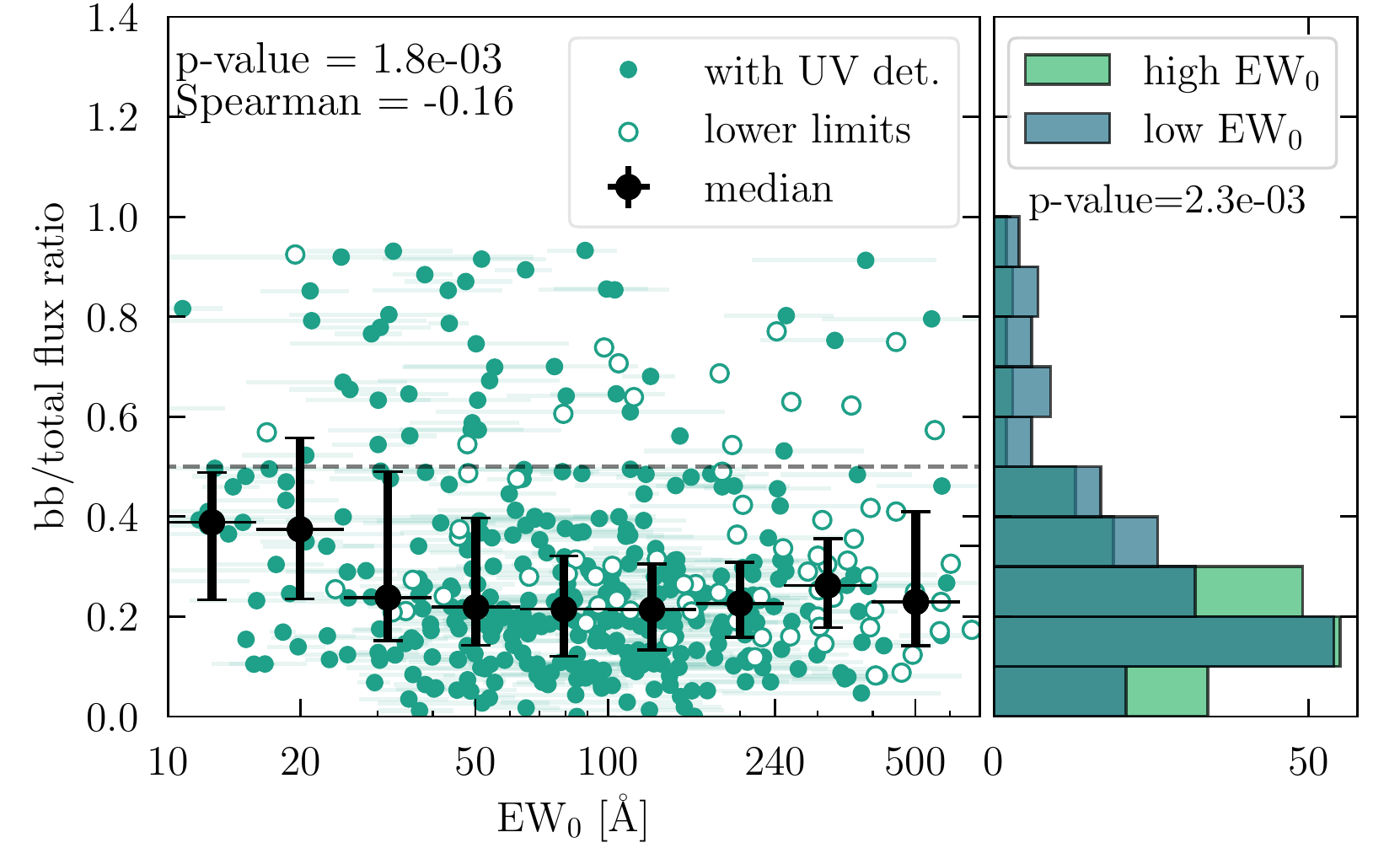}
\caption{Flux ratio of the peaks shown as a function of the $\m{EW}_0$ in \angstrom\, (with a logarithmic scale). Filled green dots show objects with UV continuum counterparts, empty green dots show objects with lower limits for $\m{EW}_0$. The flux ratio is defined as the fraction of the line flux of the blue bump compared to the total line flux (blue bump and red, main peak). The black dots show the median peak ratio (including lower limits) in bins of $\Delta(\m{log}_{10}(\m{EW}_0)) = 0.2$, the width of which is shown as the x-axis error bars. The y-axis error bars show the first and last quartiles. The Spearman rank correlation coefficient and p-value are given in the plot. The grey dashed line marks a ratio of 0.5, meaning above this value, the blue line is stronger than the red peak.
The panel on the right shows histograms for the flux ratios of the peaks divided into the half of the objects with higher EW$_0$ (green) and lower EW$_0$ (blue). The p-value based on the Kolmogorov Smirnow (KS) test is shown below the legend of the histogram, indicating the low likelihood that both distributions are taken from a common parent sample.}
\label{fig:EWs_0_ratio_peaks}
\end{figure}

Since the presence and strength of the blue bump are related to radiative transfer processes which can also have an influence on the EW$_0$, we wanted to test if we see a correlation between the two in our data. In Fig.~\ref{fig:EWs_0_ratio_peaks} we compare the ratio between blue bump and total line flux to the rest-frame EW and find no strong correlation between the two, with a p-value of only $p=0.0018$ (for a Spearman rank correlation test). It looks like the lowest EW$_0$ have preferentially large blue bumps with respect to the total line flux, but overall there is a wide spread in fractions. This can also be seen in the histogram, where we divide the sample into high and low EW$_0$. The p-value for the hypothesis that the two distributions are drawn from the same underlying sample is $p=0.0023$ (based on a KS-test), so unlikely but not impossible. Here lower EW$_0$ can more often be found with stronger blue bumps, which is not what for instance \citet{Erb2014} find for LAEs at $z \approx 2-3$. They use a slightly different definition, as they do not divide the lines into blue and red peaks, but blue- and redshifted emission along the systemic redshift and find a strong correlation between the fraction the EW$_0$ of the blue and red parts and EW$_0$, with a $6.3\,\sigma$ significance. A strong blue bump can indicate the presence of holes or low neutral hydrogen column densities in an outflowing gas cloud. Both could also facilitate the escape of Lyman continuum emission, which would otherwise get absorbed by the neutral hydrogen. Therefore, the strength of the blue bump or the fraction of Lyman $\alpha$ emission that is blueshifted can be used as a proxy for Lyman continuum leakage. It should be noted, though, that for low redshift Lyman continuum leakers, the peak flux ratio does not correlate well with the Lyman continuum escape (see \citealp{Verhamme17}), but that the strongest correlation is between Lyman continuum escape and the shift of the blue bump with respect to the systemic redshift (for which we would need additional emission lines to determine it precisely). 

One caveat of the analysis shown in Fig.~\ref{fig:EWs_0_ratio_peaks} is that it is difficult to account for objects without a detectable blue bump, which are not included in this figure. The detectability of a blue bump might itself depend on the ratio between the flux of the blue bump and the total flux, as well as on the S/N and possibly the EW$_0$, which is why this analysis should be seen as preliminary as these dependencies have to be studied further in a dedicated paper, see Vitte et al. in prep.

The median peak ratio (blue bump to total) is $f_{\m{bb}}/f_{\m{t}} = 0.237 \pm 0.205$ (consistent with the average profile found at $z\sim2$, see \citealp{Matthee2021}), which means the blue bump makes up around $20 \%$ of the total flux of the Lyman $\alpha$ line. This only holds for the objects where we could detect a blue bump, but if the ratio is very low and the blue bump thus weak, the line would not be included in this statistic. Therefore the real peak ratio might be lower than what we find here.
Notably, there are also 62 of 453 lines (for the full redshift range) where the blue bump is stronger than the red peak, which is a sign of infalling material or an outflow with a low neutral hydrogen column density, which enables the Lyman $\alpha$ photons to escape from the near side of the galaxy (\citealp{Erb2010, Alexandroff2015, Verhamme17}). 

\subsection{Connections between line properties} \label{Line Properties and their Connections}

In this subsection we focus on the possible connections between different line properties of the Lyman $\alpha$ line.

\subsubsection{Peak separation versus FWHM}

\begin{figure*}
\centering
\includegraphics[width=0.6\textwidth]{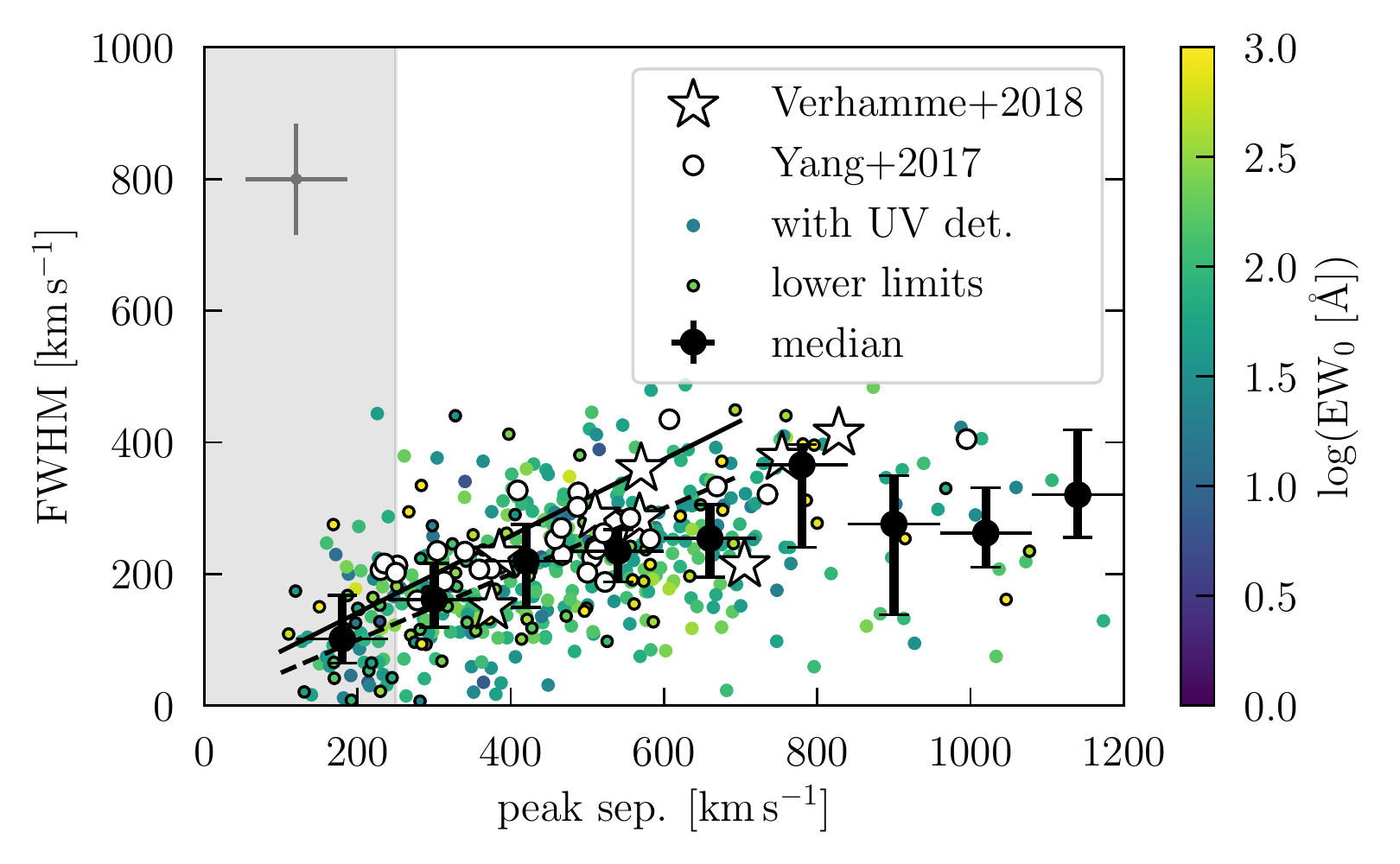}
\caption{FWHM of the main (red) peak of the Lyman $\alpha$ line plotted against the peak separation, both in $\m{km}\, \m{s}^{-1}$, colour-coded by the logarithmic EW$_0$. Coloured dots with black circles show objects without UV continuum counterparts. The grey bar in the top left corner shows the median errors. The black dots with error bars are binned median values. The x-axis error bars show the bin widths of $120\,\m{km/s}$, the y-axis error bars show the first and last quartiles.
The grey area is the peak separation range for which \citet{Izotov2018b} show a significant ($>10\%$) Lyman continuum escape fraction for their sample of low-redshift LyC leakers. The white stars indicate a sample of high-redshift LAEs compiled by \citet{Verhamme2018} (using those objects where the peak separation is given in their Table~1) with measured systemic redshifts. The black line shows their correlation, the dashed line the theoretical line where the peak separation is twice the FWHM. The white dots show a compilation of Green Peas from \citet{Yang2017} (values taken from their Table~3, peak separations are the sum of the individual peak shifts).}
\label{fig:peak_sep_fwhm}
\end{figure*}

As can be seen in Fig.~\ref{fig:peak_sep_fwhm}, the peak separation and the FWHM show a linear correlation (with a Spearman rank correlation coefficient of $0.5$ and a corresponding p-value of $5\times10^{-30}$), as lines with a larger FWHM have a higher peak separation. This is because both are tracers of the neutral hydrogen column density in the ISM of the galaxy. We expect this correlation from theoretical radiative transfer models (see \citealp{Verhamme2015}) and observed by \citet{Verhamme2018} for lower redshift galaxies, as indicated by the white stars. Since a low neutral hydrogen column density lets Lyman $\alpha$ photons escape with less scattering, leading to narrower lines and smaller peak separation, it also allows Lyman continuum emission to escape more easily (e.g. \citealp{Verhamme2015, Izotov2018b}). There seems to be a flattening of the relation up to higher peak separations, which might be influenced by small number statistics, though. The peak separation range for possible LyC leakers according to \citet{Izotov2018b} (up to $\sim250\,$km/s for a significant Lyman continuum escape fraction above $>10\%$), marked in grey in Fig.~\ref{fig:peak_sep_fwhm}, contains 65 objects of our sample (out of 455 objects with a double peak, so $\sim 12\%$, see also Vitte et al. in prep.). The reverse conclusion would be, that roughly $\sim88\%$ of our objects are not likely to be leaking Lyman continuum emission.

The mean peak separation of the full sample of our objects with a visible double peak is $481 \pm 244\, \m{km}\, \m{s}^{-1}$ (see Table~\ref{tab:line_stats} for an overview of the mean values for the different line properties divided into the high and low EW$_0$ sample), which can be translated approximately to a shift of the red peak with respect to the systemic redshift of $\Delta v \approx 241\, \m{km}\, \m{s}^{-1}$ (using the empirical correlation by \citealp{Verhamme2018}, based on the fact that most Lyman $\alpha$ lines with a double peak are symmetric around the systemic wavelength). In comparison, \citet{Shapley2003} find a shift of $\Delta v \approx 650\, \m{km}\, \m{s}^{-1}$ for a sample of LBGs, \citet{Kulas2012} find values with a mean of $\Delta v \approx 370\, \m{km}\, \m{s}^{-1}$ (for half the peak separation), which is closer to our value.
\citet{Turner2014} also look at a sample of LBGs and find a difference with respect to systemic redshift of $\Delta v = 220\, \m{km}\, \m{s}^{-1}$ for objects with only Lyman $\alpha$ emission and $\Delta v = 370\, \m{km}\, \m{s}^{-1}$ for those with Lyman $\alpha$ emission and interstellar absorption. 
\citet{Hashimoto2015} look at LAEs at $z \approx 2.2$ and find an even smaller offset between Lyman $\alpha$ and nebular lines of $\Delta v \approx 174\, \m{km}\, \m{s}^{-1}$, similar to $\Delta v \approx 180\, \m{km}\, \m{s}^{-1}$ from \citet{Song2014} and $\Delta v = 171\, \pm 8\, \m{km}\, \m{s}^{-1}$ (difference to systemic redshift) from \citet{Muzahid2020}. They both find an anti-correlation between the Lyman $\alpha$ $\m{EW}_0$ and the velocity offset. Since we do not know the systemic redshift of our LAEs, we have to approximate the velocity offset with the peak separation. 

\subsubsection{Peak separation versus \texorpdfstring{EW$_0$}{Lg}}

Since the peak separation and FWHM are connected to the neutral hydrogen column density, one would expect an anti-correlation between the peak separation and the EW$_0$ of the lines (see \citealp{Verhamme2015}), as Lyman $\alpha$ photons scatter in neutral hydrogen, increasing their path length and therefore their probability to be destroyed by dust. This has been shown for a sample of low-z Lyman continuum leaker candidates by \citet{Verhamme17}, but we do not see such a correlation here (see top panel of Fig.~\ref{fig:EWs_0_peak_sep}). In the top right panel in Fig.~\ref{fig:EWs_0_peak_sep} we show the histogram of peak separation values of the half of the sample having larger EW$_0$ and the half with smaller EW$_0$. The half with the higher EW$_0$ has a narrower distribution, with more objects at low to intermediate peak separations, but using a Kolmogorow-Smirnow-test we can not rule out the null-hypothesis of the same underlying distribution (with a p-value of $0.55$).

\subsubsection{FWHM versus \texorpdfstring{EW$_0$}{Lg}}

When it comes to the width of the line, the FWHM, the mean value we find is $\m{FWHM} = 217.89 \pm 102.26 \, \m{km}\,\m{s}^{-1}$, which in turn is smaller than what is found by other studies, for instance the average FWHM of Lyman $\alpha$ lines of LBGs at $z = 2\--3$ is found to be $\approx 650\, \m{km}\,\m{s}^{-1}$ by \citet{Steidel2010}, and \citet{Erb2010} find an LAE with a FWHM of $\approx 850\, \m{km}\, \m{s}^{-1}$. The latter study is analysing a single object though and among our sample of LAEs, the widest line has a FWHM that is even higher ($901.8 \pm 166\, \m{km}\,\m{s}^{-1}$). Another more recent study find FWHM values closer to ours, with FWHM$=212\pm32\,\angstrom$ for a sample of 3000 LAEs at $z\sim2$ (\citealp{Hashimoto2017a}).

In the bottom panel of Fig.~\ref{fig:EWs_0_peak_sep} we find no correlation between the FWHM and the EW$_0$. However, when we divide our sample again in two parts, one containing the higher EW$_0$ objects, one the lower, we can see in the histogram in the bottom right panel of Fig.~\ref{fig:EWs_0_peak_sep} that there is a small difference between the two samples (ruling out the null-hypothesis of the same underlying distribution with a p-value of $5\times10^{-6}$ using a KS-test).
For local LAE analogues, the velocity shift is as small as $\Delta v \approx 150\, \m{km}\, \m{s}^{-1}$ on average (\citealp{Verhamme17}).

\begin{table*} 
\begin{center} 
\caption{Overview of the mean, median, and standard deviation values of the measured emission line properties. } 
\begin{tabular}{ l | l | l l l | l l l | l l l } 
\hline\hline 
   & & \multicolumn{3}{c|}{all} & \multicolumn{3}{c|}{without lower limits} & \multicolumn{3}{c}{only lower limits} \\ \hline 
   & & mean & median & std & mean & median & std & mean & median & std \\ \hline 
$\m{EW}_0$ < 240\,\AA & peak ratio $f_{\m{bb}}/f_{\m{t}}$ & 0.29 & 0.23 & 0.21 & 0.28 & 0.22 & 0.21 & 0.37 & 0.28 & 0.19 \\  
& peak sep. [$\m{km}\, \m{s}^{-1}$] & 476.39 & 451.50 & 236.86 & 494.63 & 468.72 & 238.74 & 332.20 & 294.34 & 160.14 \\  
& FWHM [$\m{km}\, \m{s}^{-1}$] & 219.07 & 210.47 & 102.83 & 230.26 & 223.45 & 100.65 & 185.19 & 173.57 & 101.92 \\  
& a$_\m{asym.}$ & 0.13 & 0.16 & 0.14 & 0.14 & 0.16 & 0.14 & 0.11 & 0.13 & 0.15 \\  
& $\m{EW}_0$ [\AA] & 87.80 & 74.26 & 57.20 & 80.20 & 67.50 & 53.43 & 110.82 & 100.47 & 61.88 \\ \hline 
$\m{EW}_0$ > 240\,\AA & peak ratio $f_{\m{bb}}/f_{\m{t}}$ & 0.31 & 0.28 & 0.20 & 0.31 & 0.26 & 0.23 & 0.31 & 0.29 & 0.17 \\  
& peak sep. [$\m{km}\, \m{s}^{-1}$] & 504.33 & 484.77 & 275.77 & 524.48 & 495.33 & 338.91 & 492.00 & 423.92 & 227.79 \\  
& FWHM [$\m{km}\, \m{s}^{-1}$] & 212.27 & 193.03 & 99.02 & 222.64 & 213.05 & 80.96 & 209.14 & 191.00 & 103.66 \\  
& a$_\m{asym.}$ & 0.09 & 0.13 & 0.16 & 0.12 & 0.16 & 0.14 & 0.08 & 0.11 & 0.16 \\  
& $\m{EW}_0$ [\AA] & 550.79 & 393.74 & 493.81 & 363.23 & 321.57 & 148.84 & 607.46 & 427.21 & 544.97  \\ \hline 
\end{tabular}\label{tab:line_stats} 
\end{center} 
\tablefoot{The first columns give values for the full sample of 1920 objects, the second only for objects with secure $\m{EW}_0$ measurements (objects with a UV continuum counterpart) and the last columns for the objects that only have limiting $\m{EW}_0$ (without UV continuum counterparts). The first line gives the flux ratio between the blue bump ($f_{\m{bb}}$) and the total flux ($f_{\m{t}}$ including both the red main peak and the blue bump). The second line contains the peak separation in $\m{km}\, \m{s}^{-1}$, the third line the FWHM, also in $\m{km}\, \m{s}^{-1}$, the fourth line gives the asymmetry parameter a$_\m{asym.}$ of the red, main peak, and the last line the rest-frame EW$_0$ in \AA. The values above the middle division are given for the sub-sample of objects with an EW$_0$ below $240\, \angstrom$, while the lower values are for the sub-sample with higher EW$_0$.} 
\end{table*} 

\subsubsection{Flux ratio versus peak separation}
 
There is an anti-correlation (with a Spearman rank correlation coefficient of $-0.41$ and a corresponding p-value of $7.9\times10^{-20}$) between the contribution of the blue bump to the total line flux and the peak separation in Fig.~\ref{fig:peak_sep_ratio_peaks}; the peak separation decreases with increasing contribution by the blue bump to the line flux. There are several areas of the plot that have only few objects. One of which is the range of objects with narrow peak separations and small ratios. This could be caused by a bias, as small peak separations make it hard to detect the blue bump due to the line spread function of MUSE and also the intrinsic width of the lines. In addition, small peak ratios are hard to detect as well, which is why this combination might lead to missing objects in the bottom left corner of Fig.~\ref{fig:peak_sep_ratio_peaks}.
There are also no objects with a high peak ratio and a high peak separation.

One reason for this effect could be the IGM absorption of the blue bump, which would decrease its line flux, while the red main peak is shifted out of the resonance frequency. However, the green stars in Fig.~\ref{fig:peak_sep_ratio_peaks}, showing low redshift Green Peas (\citealp{Yang2016}), overlap with the measured peak ratios and separations of our sample of high redshift LAEs, highlighting the similarities between these objects. Since the low redshift Green Peas are not affected by absorption in the IGM, the trend can not be explained by this effect. On the one hand, ionised bubbles around LAEs and LBGs have been found even at high redshifts, which could allow a blue bump with a small peak separation to shift out of resonance before reaching the IGM (see \citealp{Matthee2018} for a blue bump at redshift $z=6.6$ and a peak separation of $220\pm20\,\m{km/s}$). The smaller the peak separation, the smaller the ionised bubble needs to be to allow the blue bump to shift enough to survive the IGM. On the other hand, it has been shown in simulations that the larger the peak separation, the more flux of the blue bump will be transmitted through the IGM (see Fig.~2 in \citealp{Laursen2011}). To learn more about the underlying processes of this anti-correlation, which might be linked to gas kinematics, we need more realistic radiative transfer models.

\subsubsection{Asymmetry versus peak separation}

The asymmetry of the red part of the line and the peak separation do not show a strong trend (see Fig.~\ref{fig:peak_sep_a1}, with a p-value of $0.04$). Only for small values, the peak separation seems to be smaller for higher asymmetries (with a red wing, which is the most common line shape for Lyman $\alpha$), which matches theoretical models (e.g. \citealp{Verhamme2015}). On average we can confirm the typically observed line shape of Lyman $\alpha$ in our sample, with an asymmetric line with a red tail and a median asymmetry parameter of $\m{a}_{\m{asym}}=0.156$.

\begin{figure}
\centering
\includegraphics[width=0.49\textwidth]{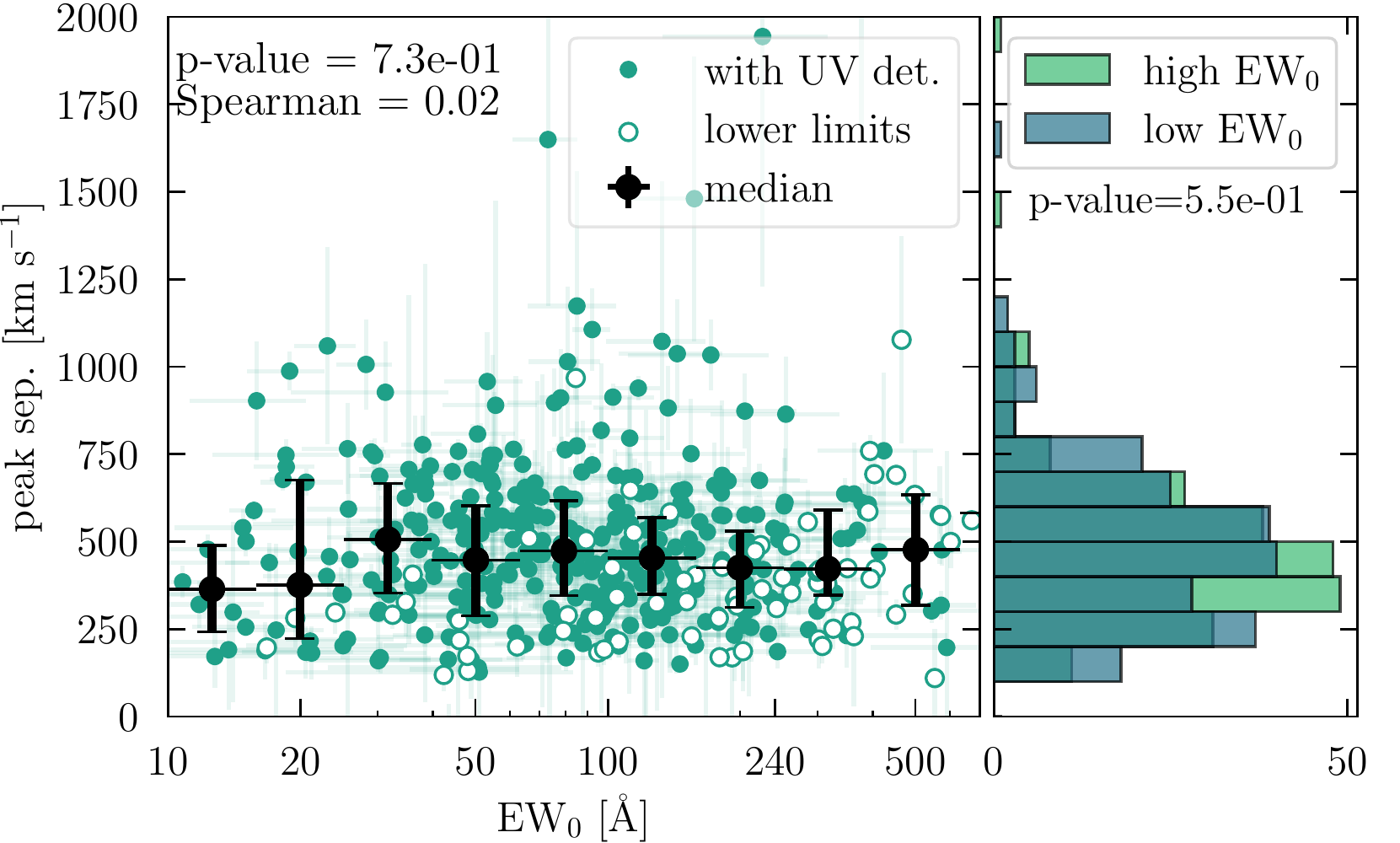}
\includegraphics[width=0.49\textwidth]{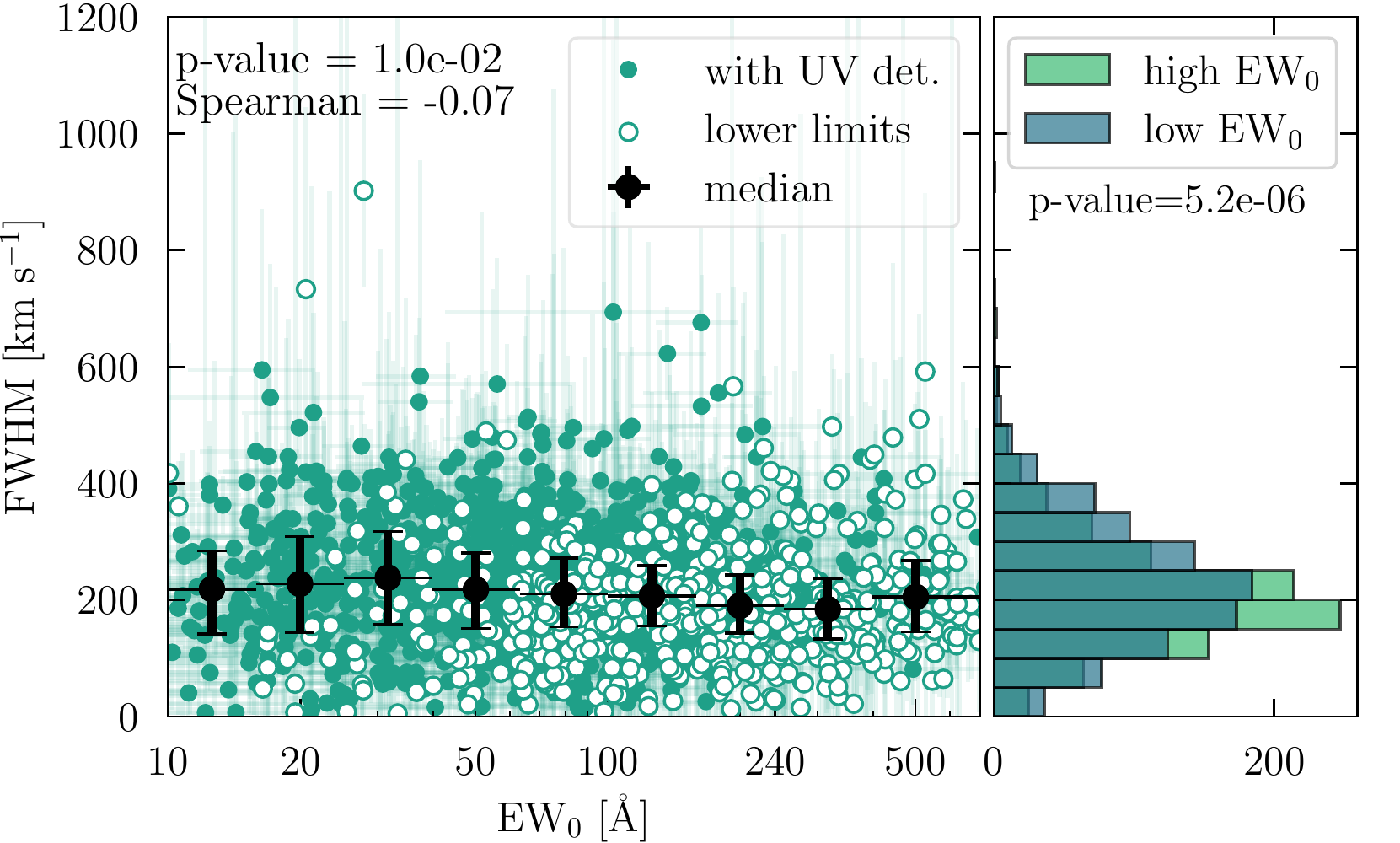}
\caption{Two properties of the Lyman $\alpha$ line plotted against EW$_0$ in \AA\, (in logarithmic scaling). The top panel shows the peak separation (for objects with a double peaked Lyman $\alpha$ line) in km/s, the bottom panel shows the FWHM of the main line (for all objects). The symbols are the same as in Fig.~\ref{fig:EWs_0_ratio_peaks}, the histograms to the right of the plots show again the distribution separated into the half with higher EW$_0$ (in green) and the half with lower EW$_0$ (in blue). The bin width for the histogram for the peak separation is $50\,\m{km/s}$, the bin width for the histogram for the FWHM is $100\,\m{km/s}$, the p-value is given based on a KS-test.}
\label{fig:EWs_0_peak_sep}
\end{figure}

\begin{figure}
\centering
\includegraphics[width=0.49\textwidth]{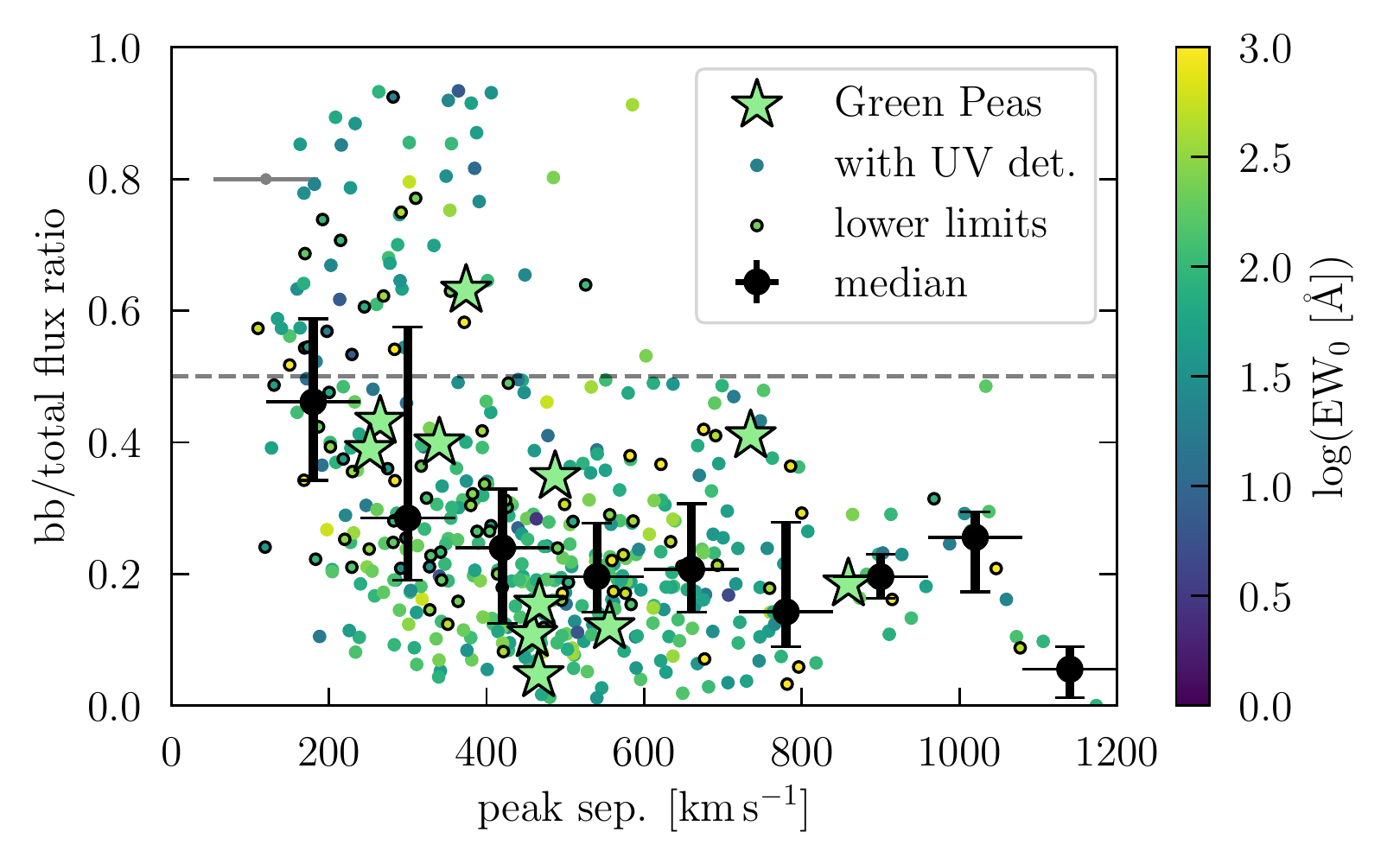}
\caption{Flux ratio of the two peaks plotted as a function of the peak separation, colour-coded by the logarithmic rest-frame EW. As in Fig.~\ref{fig:peak_sep_fwhm}, the coloured dots with black circles show objects without UV continuum counterparts and the grey bar in the top left corner shows the median errors. The black dots and error bars show the median flux ratio (including lower limits) in peak separation bins of $120\,\m{km/s}$. The x-axis error bars show the widths of the bins, the y-axis error bars show the first and last quartiles. The grey dashed line shows a ratio of 0.5, meaning above this value the blue bump dominates. The light green stars show Green Pea galaxies form \citet{Yang2016} in a redshift range between $z\approx 0.14 - 0.26$.}
\label{fig:peak_sep_ratio_peaks}
\end{figure}

\begin{figure}
\centering
\includegraphics[width=0.49\textwidth]{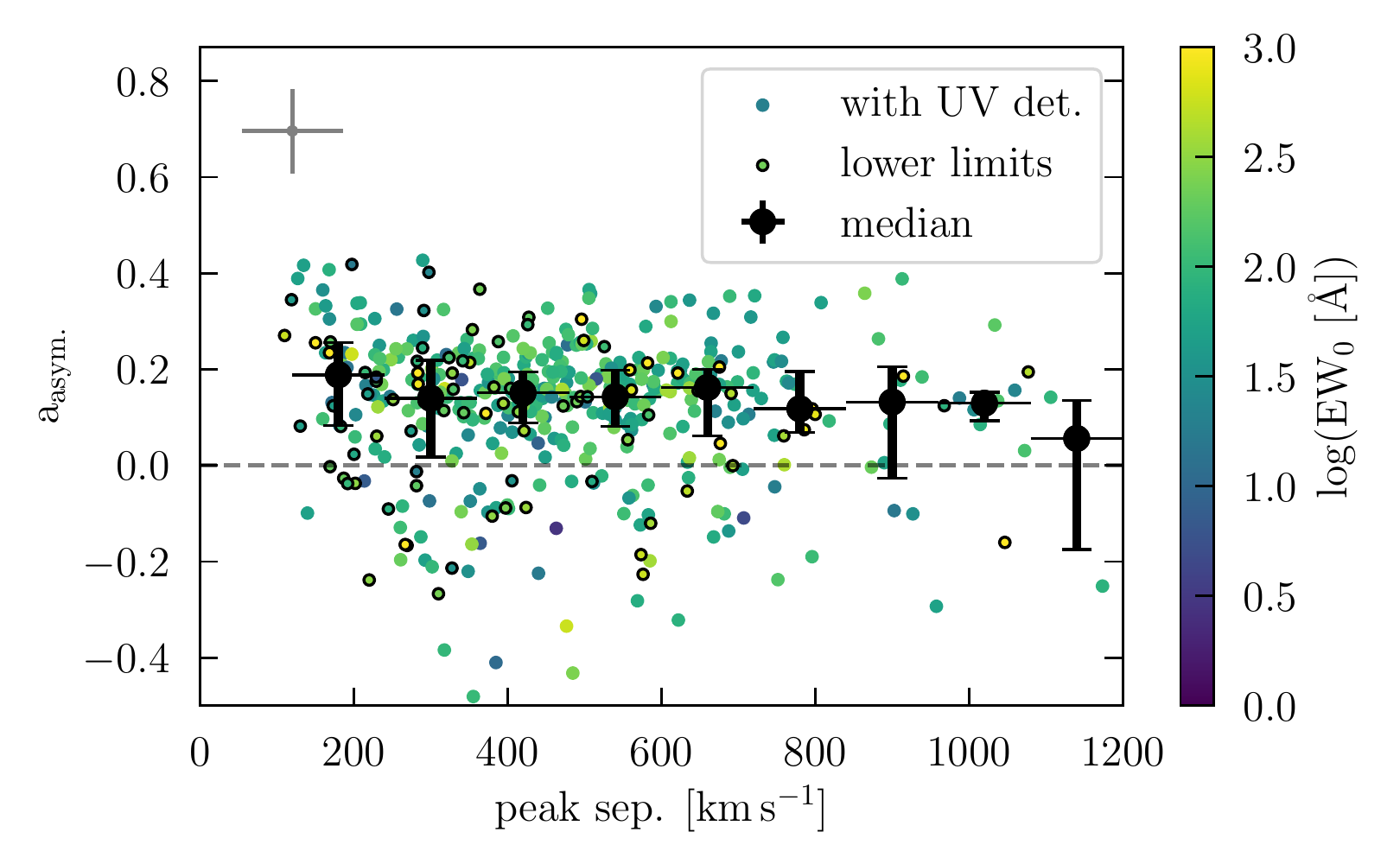}
\caption{Asymmetry of the red peak of the Lyman $\alpha$ line (asymmetry parameter a$_\m{asym.}$ in equation~\ref{asymmetry}) compared to the peak separation, colour-coded by the logarithmic EW$_0$. As in Fig.~\ref{fig:peak_sep_fwhm}, coloured dots with black circles show objects without UV continuum counterparts. The grey cross in the top left corner shows the median errors. The black dots and error bars show the median asymmetry (including lower limits) in peak separation bins. The x-axis error bars show the widths of the bins, the y-axis error bars show the first and last quartiles. The asymmetry is defined such that positive values mean the line has a red wing, while negative values mean the line has a blue wing (and zero means no asymmetry, shown as the dashed grey line).}
\label{fig:peak_sep_a1}
\end{figure}

In conclusion, we find that in our sample, LAEs have large peak separations on average, making them similar to values found for LBGs in the literature. Since the distinction between LBGs and LAEs is often an arbitrary cut in $\m{EW}_0$ of $\sim 20\, \angstrom$ which we do not make, this result could be caused by the ability to find faint emission lines in our MUSE data. The widths of our lines are smaller than found in other studies of LAEs, which hints at smaller neutral hydrogen column densities. However, the spread of values is large and we do find extreme values of FWHM $\sim 900 \, \m{km}\, \m{s}^{-1}$. 

\subsection{UV continuum morphology} \label{UV continuum morphology}

In this section we explore the connection between Lyman $\alpha$ EW$_0$ and the UV continuum morphological properties to see if objects with higher EW$_0$ can be distinguished from the rest of the sample. 
Since the main driver for the production of Lyman $\alpha$ photons are young, massive stars, an undergoing merger and thus experiencing a boost in star formation, can produce more Lyman $\alpha$ emission and thus higher Lyman $\alpha$ EW$_0$. In a similar vein, the orientation of the LAE (whether it is seen edge-on or face-on or the Lyman $\alpha$ emission is escaping through a hole in the ISM) could boost the Lyman $\alpha$ photons that are scattered into or out of the line of sight, which in turn influences EW$_0$ (\citealp{Verhamme2012, Behrens2014a, Behrens2014b, Shibuya2014b}). The orientation results in an elongation of the object, parameterised by the axis ratio which we can measure with \texttt{Galfit}.

Since we fitted the LAEs with a S\'ersic profile using \texttt{Galfit} (as explained in Sect.~\ref{Fitting the UV continuum}), we not only obtained the UV magnitudes of the LAEs but also morphological parameters such as the axis ratio, effective radius, position angle, and S\'ersic index. While it has been argued based on zoom simulations of high-redshift galaxies that properties such as axis ratio are of limited use when determining the state of the galaxy, such as whether it is merging (\citealp{Abruzzo2018}), it is interesting to see if the internal properties of the LAE that influence the strength of the Lyman $\alpha$ line, and thus the EW$_0$, are connected to its morphological properties.

The first morphological property we got from our fits is the number of components of the UV continuum counterpart. As shown in Fig.~\ref{fig:z_dist} and discussed in Sect.~\ref{Counterparts}, we modelled the counterparts consisting of one (1118 LAEs) or more components (166 LAEs) and found that objects with counterparts of only one component tend to have a larger EW$_0$ at $75_{-44}^{+128}\,\angstrom$ than objects with more components with EW$_0 = 49_{-28}^{+84}\,\angstrom$, however with a considerable spread (errors given as first and last quartiles).

The axis ratio can be used as a proxy for the viewing angle, with rounder objects (higher axis ratio) being viewed more face-on. It has been proposed that face-on galaxies should have a higher Lyman $\alpha$ escape fractions and thus a higher $\m{EW}_0$ (\citealp{Verhamme2012, Behrens2014b}), as the Lyman $\alpha$ photons do not have to travel through the dusty disc and can escape more easily. Observationally this has been shown for example by \citet{Shibuya2014a} and \citet{Paulino-Afonso2018}, who find a positive correlation between the axis ratio and the rest-frame EW of LAEs in a similar redshift range, however only going up to EW$_0$ of $300\,\angstrom$. 

\begin{figure}
\centering
\includegraphics[width=0.49\textwidth]{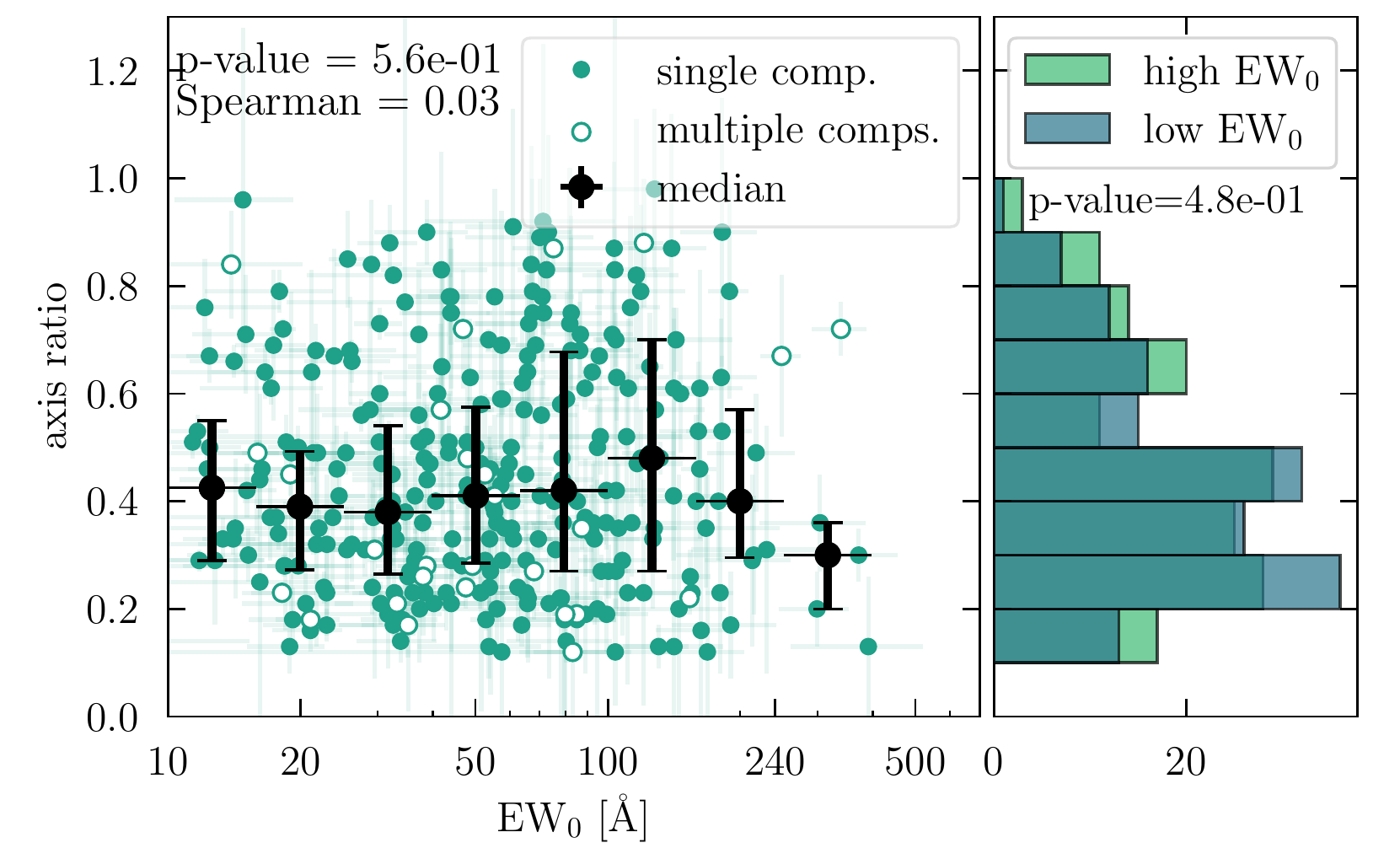}
\includegraphics[width=0.49\textwidth]{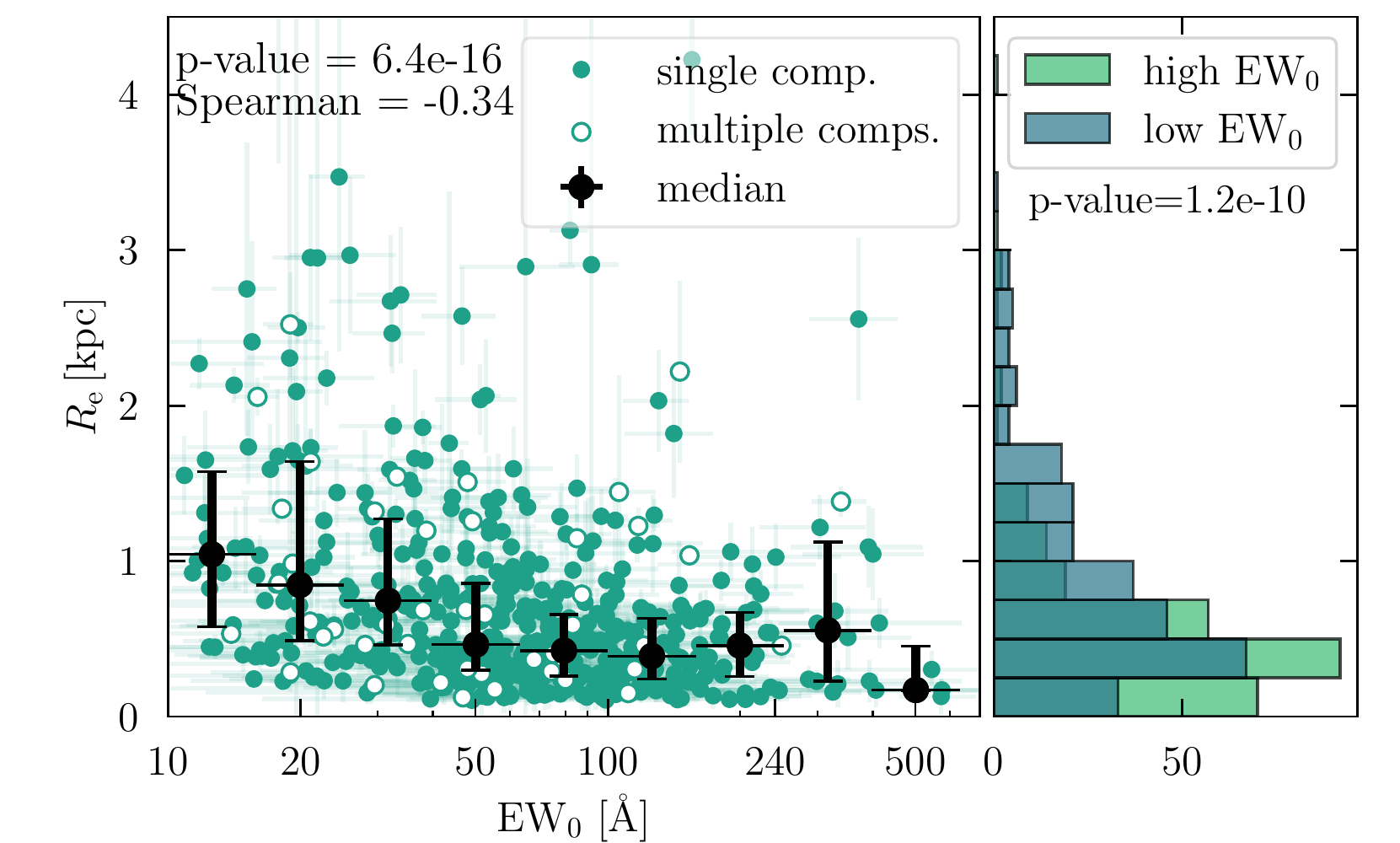}
\caption{Two different measures of properties of the UV continuum counterparts of the LAEs (measured in the HST photometry) compared to the rest-frame EW of the Lyman $\alpha$ line. The top panel shows the axis ratio over EW$_0$, the bottom panel the effective radius (in physical kpc) over EW$_0$. As in Fig.~\ref{fig:EWs_0_ratio_peaks}, the black dots show the median values in bins of $\Delta(\m{log}_{10}(\m{EW}_0)) = 0.2$, indicated by the widths of their error bars in $x$-direction with the first and last quartiles of the values in each bin indicated by the $y$-axis error bar. However, for these two plots, objects with a UV continuum counterpart consisting of only one component are shown as filled dots, objects with multiple components are shown in empty dots. For the latter, the brightest component was used to measure the size and axis ratio. The histograms show again the higher EW$_0$ (green) and lower EW$_0$ (blue) half of the sample, the axis ratio in bins of $0.1$, and the effective radius in bins of $0.2$.} 
\label{fig:axis_ratio_EW}
\end{figure}

Fig.~\ref{fig:axis_ratio_EW} shows a comparison between the rest-frame $\m{EW}_0$ and the axis ratio (top panel) and the effective radius $R_{\m{e}}$ (bottom panel). For those comparisons we use the main (brightest) component of the UV continuum counterpart in cases where there were several. 

While the axis ratio shows no significant correlation with the rest-frame EW, it is notable (seen from the histogram) that the highest EW$_0$ objects tend to have either elongated shapes or are very round, while objects with smaller EW$_0$ values are somewhere in between. In contrast, \citet{Shibuya2014a} find a trend that higher Lyman $\alpha$ rest-frame EW objects have lower UV ellipticity and argue that Lyman $\alpha$ photons can escape more easily from face-on galaxies.

Another noteworthy result is the distribution of the axis ratios (see the histogram in the top right panel of Fig.~\ref{fig:axis_ratio_EW}), with a median value of $\m{axis\,ratio} = 0.41$. The distribution has a maximum and is not flat as seen for local spiral galaxies (e.g.\ \citealp{Lambas1992, Padilla2008}).

\begin{figure}
\centering
\includegraphics[width=0.49\textwidth]{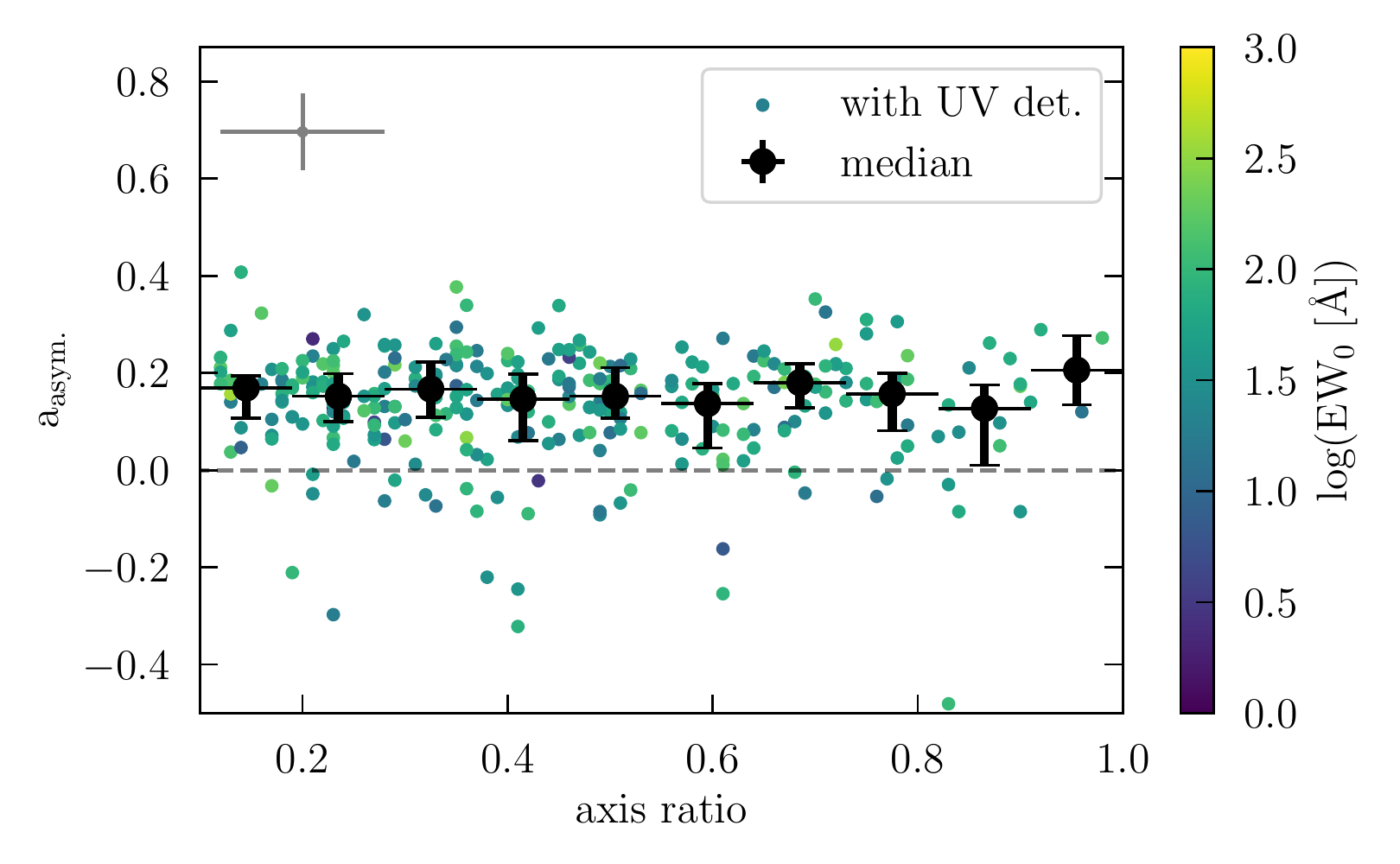}
\caption{Asymmetry $\m{a}_{\m{asym}}$ of the Lyman $\alpha$ line (see Sect.~\ref{Lyman alpha Line Properties}) as a function of the axis ratio of the UV continuum counterpart, similar to Figs.~\ref{fig:peak_sep_ratio_peaks} and \ref{fig:peak_sep_a1}. The black dots show the median values in bins, indicated by the widths of their error bars in $x$-direction with the first and last quartiles of the values in each bin indicated by the $y$-axis error bar. The top left cross shows the median errors in both directions, the grey dashed line indicates a symmetric line. Only objects with UV continuum counterparts are shown.} 
\label{fig:axis_ratio_a1}
\end{figure}

In Fig.~\ref{fig:axis_ratio_a1} the axis ratio of the LAEs is not correlated with the Lyman $\alpha$ line asymmetry (with a p-value of $0.6$). \citet{U2015} have found in their analysis that for more symmetric UV morphological properties, the Lyman $\alpha$ skewness (similar to the asymmetry shown in this work) has a larger scatter, which is not observed in our data. \citet{Childs2018} even caution that the measure of the asymmetry or skewness of the line is complicated by sky lines and resolution effects and should rather not be used for interpretations.

Another morphological property of LAEs is their UV continuum size. In the bottom panel of Fig.~\ref{fig:axis_ratio_EW} there is an anti-correlation (with a p-value of $6.4\times10^{-16}$ for a Spearman rank test) between the effective radius of the UV continuum counterpart and the EW$_0$. This could be caused by a bias, as faint UV counterparts are more difficult to fit, but will generally have higher $\m{EW}_0$. However, if we include a cut in magnitude of $\m{M}_{\m{UV}}<-18$ to minimise the influence of UV faint objects, the correlation persists. 
It is also notable that among the objects with the highest $\m{EW}_0$, most are compact with effective radii below $\sim 1\,\m{kpc\,(physical)}$, comparable for instance with the results in a study of LAE morphological properties by \citet{Paulino-Afonso2018}. Looking at the histograms of effective radii divided into the high and low EW$_0$ half of the sample confirms the anti-correlation, as low EW$_0$ tend to have larger effective radii.
One caveat of this study is that ideally we would need to compare to a sample of galaxies in the same redshift range that do not have Lyman $\alpha$ emission or even show it in absorption.

Looking at the values of the morphological parameters themselves, we find a large variation in the S\'ersic index values, indicating a variety of morphological properties in high redshift LAEs, but the median S\'ersic index is $n= 2\pm1.8$, meaning most LAEs are consistent with an exponential, disc-like profile (as found by e.g.\ \citealp{Gronwall2011} for LAEs at $z\approx3.1$). The median axis ratio is $b/a = 0.45 \pm 0.21$, matching with the value of $b/a = 0.45$ found by \citet{Gronwall2011}, and the distribution is skewed towards more elongated shapes. One should keep in mind, though, that for faint objects, which might be even unresolved in HST, the axis ratio derived from \texttt{Galfit} is influenced by the PSF estimation and the noise (see \citealp{Gronwall2011} for a discussion). 

%%%%%%%%%%%%%%%%%%%%%%%%%%%%%%%%%%%%%%%%%%%%%%%%%%%%%%%%%%%%%%%%%%%%%%%%%%%%%%%%%%%%%%%%%%

\section{Discussion of high EW\texorpdfstring{$_0$}{Lg} objects} \label{Discussion}
 
We find a significant number of LAEs with very high EW$_0$ among our sample, notably 306 LAEs with EW$_0$ of $\m{EW}_0>240\ \angstrom$ (including both objects with and without counterparts), one of the largest samples of high EW$_0$ LAEs. In this section we discuss the connections between Lyman $\alpha$ EW$_0$ and the spectral and morphological properties of the LAEs in order to determine what makes LAEs with high EW$_0$ (with $\m{EW}_0>240\ \angstrom$) special. 

Results on fractions of LAEs with high EW$_0$ from previous studies vary widely. Compared to most studies, our result of a fraction of $\approx 16\%$ of objects with EW$_0>240\,\angstrom$ is rather high. In a sample of $\sim3000$ narrow band selected LAEs with redshift $z \approx 2$, \citet{Hashimoto2017a} find a sub sample of six spectroscopically confirmed LAEs with an EW$_0$ of $\approx 200 - 400\, \angstrom$, which constitutes a lower limit to the fraction of high EW$_0$ LAEs of $\approx 0.2\%$. Similarly, \citet{Nilsson2009} show a fraction of $4\%$ of LAEs at redshift $z\sim3$ with $\m{EW}_0 > 240\,\angstrom$ and \citet{Santos2020} find only 45 in their SC4K (\citealp{Sobral2018}) sample of $\approx 4000$ LAEs ($\approx 1.1\%$) between $z\sim 2-6$. The SILVERRUSH (\citealp{Ouchi2018}) survey found a slightly larger fraction, $4\%$ at $z\simeq 5.7$ and $21\%$ at $z\simeq6.6$ (\citealp{Shibuya2018}). In contrast (at a redshift of $z = 4.5$), \citet{Malhotra2002} find a fraction of $60\%$ of LAEs exceeding $\m{EW}_0 > 240\,\angstrom$ in LAEs of the Large Area Lyman Alpha (LALA) survey (\citealp{Rhoads2000}), however effectively using an IGM correction for the Lyman $\alpha$ line.
Using emission line selected LAEs, \citet{Adams2011} find three LAEs ($\approx 3\%$) at $z\sim 2 - 4$ with $\m{EW}_0 > 240\,\angstrom$ among a sample of 105 galaxies and \citet{Shimasaku2006} find $30 \-- 40\%$ at $z=5.7$.
Studies that do not find high $\m{EW}_0$ are for example \citet{Gawiser2006}, \citet{Gronwall2007}, and \citet{Guaita2011} at redshifts $z\approx3.1$ and $z\approx2.1$.
In a recent paper, \citet{Maseda2020} use a sample of UV faint spectroscopically identified LAEs between $z = 3 - 4$ and stack their photometry to find a median $\m{EW}_0 = 249\,\angstrom$ with a high ionising photon production efficiency (in combination with H$\alpha$ measurements).
This also shows again that the survey properties are important to take into account when discussing fractions of high EW$_0$ LAEs, as we argued in Sect.~\ref{Differences between MUSE-Wide and MUSE-Deep}.

\subsection{Connection between high EW\texorpdfstring{$_0$}{Lg} and special UV continuum morphological properties}

We have seen in the previous Sect.~\ref{UV continuum morphology} that for our sample of LAEs, the half of our sample with larger EW$_0$ have indistinguishable UV morphological properties to the half with smaller EW$_0$ when it comes to S\'ersic index and axis ratio. As discussed above, the axis ratio could be a proxy for the viewing angle and we could expect that we see higher EW$_0$ from face-on galaxies. However, we do not find a correlation between axis ratio and EW$_0$, possibly because the Lyman $\alpha$ photons are indeed escaping primarily perpendicular to the disc's plane of the LAE, but are scattered back into the line of sight to make them visible from an edge-on view as well. Another explanation is that LAEs do not typically show the same morphological properties as low-redshift spiral galaxies. They are known to be clumpy, often merging systems (even at low redshifts, as can be seen in the LARS sample, e.g. \citealp{Ostlin2014, Hayes2014}, \citealp{Messa2019}) and the Lyman $\alpha$ morphological properties do not necessarily follow those of the UV continuum (\citealp{Ostlin2009, Hayes2014}). Therefore, the interpretation of the axis ratio as a viewing angle might not be accurate for LAEs at these high redshifts, where the galaxies are not well resolved and the axis ratio we measure could be due to irregularities more than the tilt of the LAEs' galaxy plane.

It has also been proposed that the Lyman $\alpha$ emission strength correlates with the physical size of the UV continuum. Small, compact galaxies could be young, less massive, and less dusty and allow an easier escape of Lyman $\alpha$ emission, such as through their stronger outflows. Our results show that the highest EW$_0$ objects (above $240\,\angstrom$) also have smaller effective radii, in agreement with this interpretation.
\citet{Bond2012} look at LAEs at redshifts $z = 2.1 \-- 3.1$ and find no correlation between the EW$_0$ and the half-light radius, while for example \citet{Marchi2019} show that UV compact sources have higher Lyman $\alpha$ EW$_0$ in their sample of star-forming galaxies at $z\approx3.5$. In a sample of LAEs stretching from redshift $z\sim2$ to $z\sim6$, \citet{Paulino-Afonso2018} find a similar trend between the effective radius and the EW$_0$ as shown in Fig.~\ref{fig:axis_ratio_EW} (bottom panel). Thus we can say that high EW$_0$ objects ($>240\ \angstrom$) tend to be compact and small, but small LAEs span a wide range of EW$_0$.

\subsection{Possible causes for high EW\texorpdfstring{$_0$}{Lg}} \label{What is the underlying cause for high EWs?}

From \citet{Verhamme2008} we learn that the FWHM and peak separation of the Lyman $\alpha$ line reflect the neutral hydrogen column density of the ISM in the LAEs. For expanding shell models, line widths of $\m{FWHM} > 500 \, \m{km}\, \m{s}^{-1}$ need neutral hydrogen column densities of $\m{N}>10^{20}\, \m{cm}^{-2}$ and vice versa. Following this model, this suggests that the majority of our objects has small neutral hydrogen column densities below $\m{N}<10^{19.5}\, \m{cm}^{-2}$, since their FWHM values are on average narrower than $\sim230\,$km/s (\citealp{Verhamme2008}).
It has been found in simulations that not only a low neutral hydrogen column density can lead to increased Lyman $\alpha$ EW$_0$. The production of ionising photons, which are necessary to produce Lyman $\alpha$ emission, is dictated by the internal properties of the galaxy, such as a high star formation rate, a more top-heavy IMF, and a low metallicity (e.g. \citealp{Schaerer2003, Raiter2010}). When comparing different IMF models, \citet{Raiter2010} show that for low metallicities (but without invoking PopIII stars yet), EW$_0$ of up to $500-700\,\angstrom$ can be reached, even up to $1500\,\angstrom$ when assuming top-heavy IMF models and a departure from Case B. The number of ionising photons produced in such simulations also depends on the stellar population synthesis models that were used. Models such as BPASS (\citealp{Eldridge2009}, which however assumes a binary fraction of $100\%$) can produce much higher numbers of ionising photons and thus also higher Lyman $\alpha$ EW$_0$, up to intrinsic EW$_0$ of over $1000\,\angstrom$ (see \citealp{Garel2021}). 
The hypothesis that indeed such unusual stellar populations are responsible for the high EW$_0$ could be investigated with further observations, for example probing young stellar ages with Balmer lines, massive stars with \rm{N\textsc{v}} P-Cygni profiles or measuring the metallicity (e.g. using R23~=~(\rm{O\textsc{ii}}~+~\rm{O\textsc{iii}})/H$\beta$), which is expected to be low for objects with high EW$_0$ ($\m{EW}_0>240\ \angstrom$). 

Directional boosting of Lyman $\alpha$ emission due to a clumpy, multi-phase medium as proposed by \citet{Neufeld1991} could also explain high EW$_0$ ($\m{EW}_0>240\ \angstrom$). It has been found to work only under special conditions, though, if the ISM is very clumpy, with clumps over five times more dense than typical, very dusty, and almost static with outflow velocities less than $10\,\m{km/s}$ (\citealp{Laursen2013,Gronke2014, Duval2014} and Blaizot et al. in prep.). 

To test this prediction, we can imagine that a static ISM would result in double peaked lines where both peaks contribute equally to the total flux, which means we could expect a trend that for a flux ratio of bb/total flux $\approx50\%$ the equivalent widths could be boosted, which we can not confirm (see Fig.~\ref{fig:EWs_0_ratio_peaks}). However, we do not take into account the possible absorption of the blue part of the line by the IGM, which could be a reason why we do not see a correlation here. However, we can see in Fig.~\ref{fig:peak_sep_ratio_peaks} that local galaxies which are not affected by IGM absorption seem to follow a similar trend as our high-redshift LAEs, when it comes to the flux ratio, which we infer from the comparison between the flux ratio and the peak separation which fall on the same correlation for our high-redshift LAEs and the local Green Peas from \citealp{Yang2016}. If the IGM absorption of the blue bump was significant, the high-redshift LAEs would fall below the correlation of the local Green Peas, as the flux ratio would be reduced. Therefore we can believe that indeed we do not see a boosted EW$_0$ from a static ISM in Fig.~\ref{fig:EWs_0_ratio_peaks}.

Instead of explaining high EW$_0$ with a multi-phase, possibly static ISM, \citet{Laursen2013} suggest alternative production mechanisms for large EW$_0$ values of Lyman $\alpha$ photons, such as collisional emission (e.g. \citealp{Dijkstra2017, Faucher2017}) and fluorescence (e.g. \citealp{Cantalupo2005, Kollmeier2010, Dijkstra2017}). However, directional boosting of Lyman $\alpha$ EW$_0$ can work (although with a large directional variability) when it is related to outflows caused by starbursts in combination with higher UV continuum absorption (\citealp{Smith2018}), resulting in narrower, single peaked lines. 

A small peak separation, possibly indicative of a lower neutral hydrogen column density (according to radiative transfer simulations), is related to the escape of Lyman continuum emission, which can produce an ionised region, facilitating the escape of blue bump photons. There could also be a selection bias in this plot, since weak blue bumps, resulting in a small peak ratio, that are also close to the main red peak (with a small peak separation), might not be easily detected due to instrumental limitations (see \citealp{Matthee2018} for a discussion). 

In that sense LAEs with high EW$_0$ ($\m{EW}_0>240\ \angstrom$) are also interesting for the study and search for Lyman continuum leakers, which are responsible for providing the ionising emission at the epoch of reionisation. LAEs with high EW$_0$ have been found to have elevated $\xi_{ion}$ (ionising photon production efficiency) values, lower metallicities, and younger ages (\citealp{Maseda2020}). We have seen for example in Fig.~\ref{fig:MW_MD_logLLya_M_UV} that the Green Pea galaxies from \citet{Izotov2016a,Izotov2016b,Izotov2018a,Izotov2018b}, all of which are leaking Lyman continuum, fall in a region of higher Lyman $\alpha$ EW$_0$.
Since a higher Lyman $\alpha$ EW$_0$ not only indicates a strong production of Lyman continuum photons, but also the possibility of a high escape fraction of Lyman $\alpha$ emission (especially considering that the half of our sample with higher and lower EW$_0$ have potentially different line widths, see Fig.~\ref{fig:EWs_0_peak_sep} and the discussion in Sect.~\ref{Line Properties and their Connections}), it has been proposed to use high Lyman $\alpha$ EW$_0$ as an indirect tracer of the Lyman continuum (see \citealp{Verhamme2015, Verhamme17, Marchi2018, Steidel2018, Fletcher2019}, but see also \citealp{Bian2020} who contest this). This makes our sample of 307 objects with $\m{EW}_0 > 240\,\angstrom$ all the more interesting for possible follow-up observations to confirm the presence of Lyman continuum leakage.

\subsection{The highest EW\texorpdfstring{$_0$}{Lg} LAE} \label{The highest EW LAE}

Among our sample, the highest robust EW$_0$ is $\m{EW}_0 = 588.9 \pm 193.4\, \angstrom$ (which is the highest EW$_0$ that is measured with a $3\,\sigma$ certainty), with the ID 123007091. The highest lower limit EW$_0$, where no HST counterpart could be identified, is $\m{EW}_0=4464\, \angstrom$.
Again, the choice of $\beta$ influences the EW$_0$. For $\beta=-1.57$ (which is the upper quartile of the distribution of measured $\beta$ values, see Sect.~\ref{Distribution of Equivalent Widths}) the object with the highest robust measurement would have $\m{EW}_0 = 689.26 \pm 260.5\ \angstrom$, while for $\beta=-2.29$ (the lowest quartile) it would have $\m{EW}_0 = 519.3 \pm 153.0\ \angstrom$. The Lyman $\alpha$ luminosity is $\m{log}_{10}(L_{\m{Ly}\alpha}[\m{erg/s}]) = 42.56\pm 0.26$ and with a UV magnitude of $\m{M}_{\m{UV}}=-17.2\pm0.3$ it is rather faint.

\citet{Kashikawa2012} find an even higher EW$_0$ of $\m{EW}_0 = 872^{+844}_{-298}\ \angstrom$ at a redshift of $z=6.5$, which could hint at a candidate population III galaxy. Their measured a Lyman $\alpha$ luminosity of $(1.07\pm0.30)\times10^{43}\,\m{erg/s}$. Assuming a fixed $\beta=2$, they get a UV continuum flux density at the Lyman $\alpha$ wavelength of $f_{\m{cont}} = (7.12\pm3.41)\times10^{-21}\,\m{erg/s/cm}^2\m{/}\angstrom$, compared to $f_{\m{cont}} = (15.22\pm4.68)\times10^{-21}\,\m{erg/s/cm}^2\m{/}\angstrom$ for our object. 

However, they apply an IGM correction and assume that around half of the Lyman $\alpha$ emission line flux was absorbed by the IGM. Their uncorrected value is at $\m{EW}_0 = 436^{+422}_{-149}\, \angstrom$, lower than our highest value and with a large error, indicating that it could potentially be much higher. 

\begin{figure*}
\centering
\includegraphics[width=\textwidth]{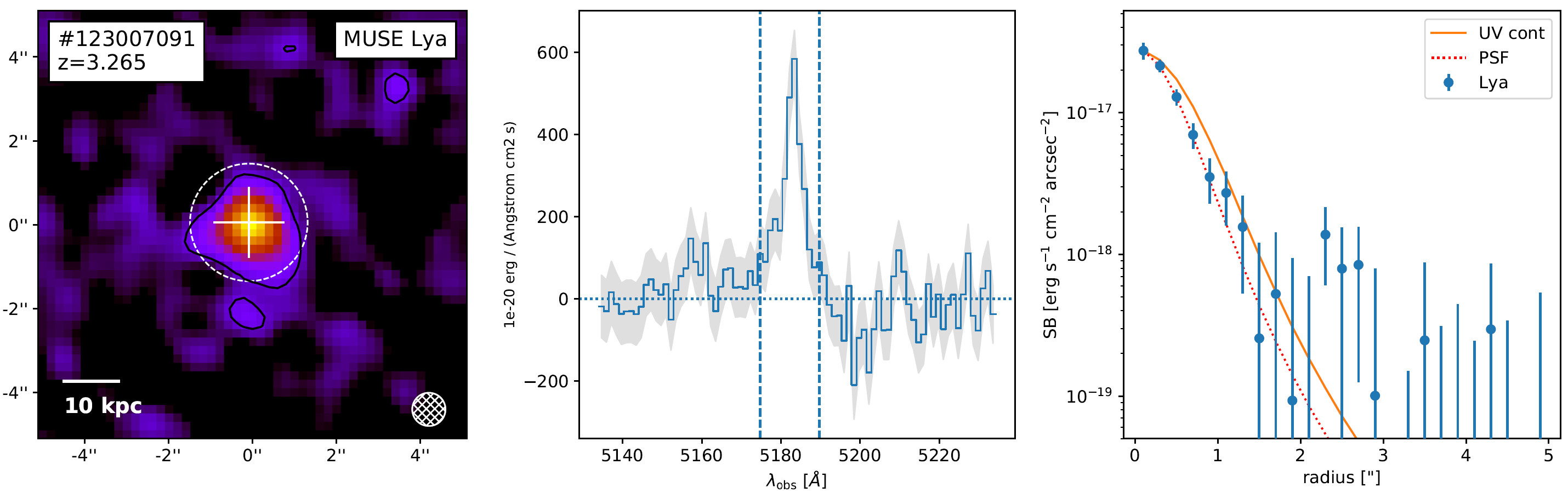}
\caption{Measurements of the Lyman $\alpha$ halo of object 123007091 as in \citet{Leclercq2017}. The left panel shows the MUSE Lyman $\alpha$ narrow-band image, smoothed slightly with the PSF FWHM Gaussian kernel, indicated by the white hatched circle in the lower right corner. The black lines show the $2\,\sigma$ surface brightness limit, the white dashed circle shows the extraction aperture for the spectrum of the Lyman $\alpha$ emission line. The latter is shown in the middle panel, with vertical lines indicating the width of the narrow-band image. The right panel shows the extent of the Lyman $\alpha$ halo in blue and the continuum component in orange, as well as the PSF (red dotted line).}
\label{fig:2020-04-10}
\end{figure*}

\begin{figure*}
\centering
\includegraphics[width=\textwidth]{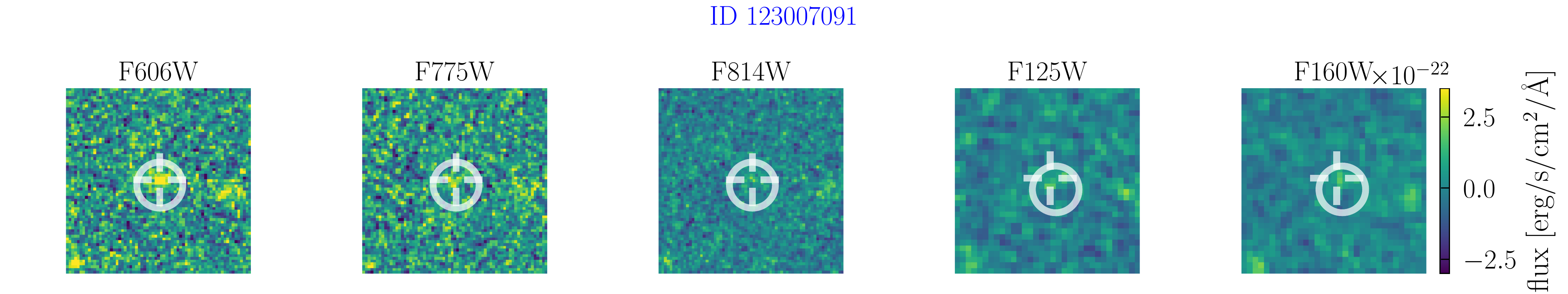}
\caption{HST cutouts of the bands ACS F606W, F775W, F814W and WFC3 F125W and F160W. For the object shown in Fig. \ref{fig:2020-04-10}. As in Fig. \ref{fig:example_cutouts}, the cross is centred on the UV continuum position, the circle is centred on the Lyman $\alpha$ position from \texttt{LSDCat}, and has have a diameter of $0\farcs5$.}
\label{fig:cutout_acs_814w_123007091}
\end{figure*}

It is interesting to look more closely at the properties of our object 123007091, which has a redshift of z$=3.26$ (see Fig.~\ref{fig:2020-04-10} for the Lyman $\alpha$ halo measurements and Fig.~\ref{fig:cutout_acs_814w_123007091} for the UV continuum). The spectrum shows no significant UV emission lines apart from Lyman $\alpha$. The Lyman $\alpha$ emission line has a tentative close blue bump at a peak separation of $197$~km/s and a FWHM of the main, red peak of FWHM=$178\pm110$~km/s (corrected for the MUSE LSF). Both the line separation and the line width are smaller than the average of the full sample (shown in Table~\ref{tab:line_stats}), although the FWHM is within the one sigma range of the average. Still, based on the peak separation, this indicates that the Lyman $\alpha$ emission did not scatter much to escape the galaxy, possibly due to a low neutral hydrogen column density or clear paths in the ISM in our direction of observation, making it relatively easy for the Lyman $\alpha$ emission and possibly also the Lyman continuum emission to escape. This object has a spatially compact Lyman $\alpha$ emission with no measurable extended halo beyond the UV continuum (see Fig.~\ref{fig:2020-04-10}), which supports the assumption that Lyman $\alpha$ (and also Lyman continuum) escapes more easily from objects with compact Lyman $\alpha$ emission, producing a higher EW$_0$. 

%%%%%%%%%%%%%%%%%%%%%%%%%%%%%%%%%%%%%%%%%%%%%%%%%%%%%%%%%%%%%%%%%%%%%%%%%%%%%%%%%%%%%%%%%%

\section{Summary and conclusion} \label{Summary and Conclusion}

In this paper we presented the derived EW$_0$ for a sample of $\sim2000$ LAEs with homogeneous measurements of their properties. The LAE sample was constructed using a consistent procedure for the detection and fitting of emission lines (without applying an IGM correction) in data from the MUSE-Wide and MUSE-Deep GTO surveys. We showed EW$_0$ histograms (Figs.~\ref{fig:EW_dist} and \ref{fig:EW_dist_shifted_MW_MD_one}) and an analysis of the Lyman $\alpha$ and morphological properties of the LAEs. Furthermore we connected the measured EW$_0$ to other spectral and UV continuum morphological properties and discussed the discovery of numerous high EW$_0$ objects (with $\m{EW}_0>240\ \angstrom$). The main results of our work are the following:

\begin{itemize}
 \item A third of our LAEs do not have UV continuum counterparts visible in the HST broad-band data. Of the remaining objects, $13\%$ have multi-component counterparts, hinting at merging, clumpy morphology or star formation (see Figs.~\ref{fig:example_cutouts} and \ref{fig:z_dist}).
 \item A fraction of $\approx16\%$ (306 objects) of the full sample of our LAEs have $\m{EW}_0>240\, \angstrom$, which is comparable to other studies in the literature, especially when considering Lyman $\alpha$ EW$_0$ studies that, like us, do not apply an IGM correction (see Sect.~\ref{Discussion}). Due to our careful measurements of the EW$_0$ values, we establish firmly the existence of a population of galaxies with very high EW$_0$ (with $\m{EW}_0>240\ \angstrom$ and even much above, see e.g. Figs.~\ref{fig:EW_cumulative}, \ref{fig:EWs_0_ratio_peaks} and \ref{fig:EWs_0_peak_sep}).
 \item In addition we show that such fractions of high $\m{EW}_0>240\ \angstrom$ are influenced by the Lyman $\alpha$ line flux depth of the survey, since we find $\approx 20\%$ for the MUSE-Wide LAEs and $\approx 11\%$ for the ones from MUSE-Deep.
 \item There is a discrepancy between the EW$_0$ distributions of the MUSE-Wide and MUSE-Deep survey LAEs (Fig.~\ref{fig:EW_dist_shifted_MW_MD_one}), with the deeper data having a smaller fraction of high $\m{EW}_0>240\ \angstrom$ objects than the shallower survey with a larger survey area. We propose as an explanation the ability to detect fainter Lyman $\alpha$ emission in the deeper data. Thus a deeper Lyman $\alpha$ survey at fixed UV continuum depth unveils a larger fraction of the galaxy population which typically has a lower EW$_0$ (e.g. \citealp{Shapley2003} find only an EW$_0$ of $14.3\,\angstrom$ in their stack of LBGs at $z\sim3$). 
 \item Smaller Lyman $\alpha$ luminosities allow to detect the numerous UV-bright objects with small EW$_0$ rather than the rare UV faint extreme EW$_0$ objects.
 \item We find an object with an extremely high EW$_0$ of $\m{EW}_0 = 588.93 \pm 193.4\, \angstrom$, hinting at unusual stellar populations. Among the objects with high EW$_0$, there is a large spread in emission line shapes (as exemplified in Fig.~\ref{fig:overview}), which implies a large diversity of the influence of radiative transfer processes and the interplay between gas dynamics and dust content. Therefore, the underlying cause of high EW$_0$ ($\m{EW}_0>240\ \angstrom$) is probably rooted less in extreme transfer processes, but more in the properties of the stellar populations with respect to age, IMF, and metallicity.
 \item We confirm the correlation reported in the literature between the effective radius of the UV continuum counterpart and the Lyman $\alpha$ EW$_0$ (bottom panel of Fig.~\ref{fig:axis_ratio_EW}), but do not find a correlation between Lyman $\alpha$ EW$_0$ and axis ratio (top panel of Fig.~\ref{fig:axis_ratio_EW}), which should be a proxy for the viewing angle. We conclude that compact galaxies make it easier for Lyman $\alpha$ to escape, while the axis ratio of the UV continuum counterpart might not be correlated to the viewing angle, but influenced by irregular shapes or star formation regions.
 \item Objects belonging to the half of the sample with higher EW$_0$ have potentially different line width properties than the half with lower EW$_0$ (Fig.~\ref{fig:EWs_0_peak_sep}).
 \item In general we find that objects with large EW$_0$ (if we consider the half of our sample with higher EW$_0$) can have a wide range of spectral line shapes and UV morphological properties (but preferentially small sizes), which means underlying properties of the LAEs can also be varied and there is no only one driving force for producing high EW$_0$ values.
\end{itemize}

As we have discussed in this paper, the EW$_0$ distribution and the fraction of high EW$_0$ ($>240\,\angstrom$) objects depend heavily on the survey design that the sample of LAEs was taken from. The observation depth and the area of the survey determine the number of Lyman $\alpha$ faint and bright sources that can be detected. A wedding-cake strategy as used here with the MUSE-Wide and -Deep surveys allows us to find a wider range of LAE properties. A deep survey will unveil more faint LAEs with lower EW$_0$, which is useful for constraining the faint-end of the LF (e.g. \citealp{Drake2017b}), while a larger survey area at lower exposure times will uncover more rare objects with large EW$_0$ (given that the UV continuum depth stays the same). Therefore, to understand the true nature of the LAE population with respect to EW$_0$, one has to take into account the survey properties. For this it would be ideal to construct an EW$_0$ distribution function giving the number of LAEs in bins of EW$_0$ per volume, similar to a luminosity function, thus correcting for the selection function in the two surveys. We will explore this possibility in an upcoming paper. 

It is also interesting to study the reasons for the high EW$_0$ of $\m{EW}_0 > 240\, \angstrom$ which we find in 307 LAEs. They are ideal candidates to observe with the upcoming JWST instrument to better understand their properties, for example using H$\alpha$ and H$\beta$ to infer the produced Lyman $\alpha$ luminosity and thus to get the Lyman $\alpha$ escape fraction and learn about the nature of the stellar population. To explore the line shape properties in full it will also be useful to obtain systemic redshifts from JWST.
We would also be able to search for rest-frame nebular lines and obtain metallicities, stellar ages, and dust content to see if we can explain their astonishingly high Lyman $\alpha$ EW$_0$ by extreme ages and metallicities or if we need even more exotic scenarios.

\vspace{20pt}

\begin{acknowledgements}
The authors wish to thank the anonymous referee for the thorough comments and insight which have helped to significantly improve the quality of the manuscript.
We would further like to thank Peter Weilbacher for carefully reading this manuscript and providing suggestions to greatly improve the legibility and conciseness. We also thank Pascal Oesch, Daniel Schaerer, and Moupiya Maji for helpful discussions. Special thanks go to T. Dumont for a fruitful discussion about cinnamon stars. No (LSD)cats were harmed in the making of this paper. The plots in this paper were created using \texttt{Matplotlib} (\citealp{Hunter2007}) and we also used the Python packages \texttt{Numpy} (\citealp{vanderWalt2011}) and \texttt{Astropy} (\citealp{Astropy2013}). JK, AV, HK, and FL are supported by the SNF Professorship PP00P2\_176808. AV, VM, and TG are supported by the ERC Starting Grant 757258 'TRIPLE'. JK, ECH, TU, KS, and LW acknowledge funding from the Competitive Fund of the  Leibniz Association through grants SAW-2013-AIP-4 and SAW-2015-AIP-2. HK acknowledges support from the Swiss Government Excellence Scholarships and Japan Society for the Promotion of Science (JSPS) Overseas Research Fellowship. TH was supported by Leading Initiative for Excellent Young Researchers, MEXT, Japan (HJH02007). We also want to thank the staff at the VLT during the observations of the GTO data.
\end{acknowledgements}

\bibliographystyle{aa}
\bibliography{references}

\appendix

\section{Assessment of the influence of halo sizes} \label{Halo Sizes}

\begin{figure}
\centering
\includegraphics[width=0.49\textwidth]{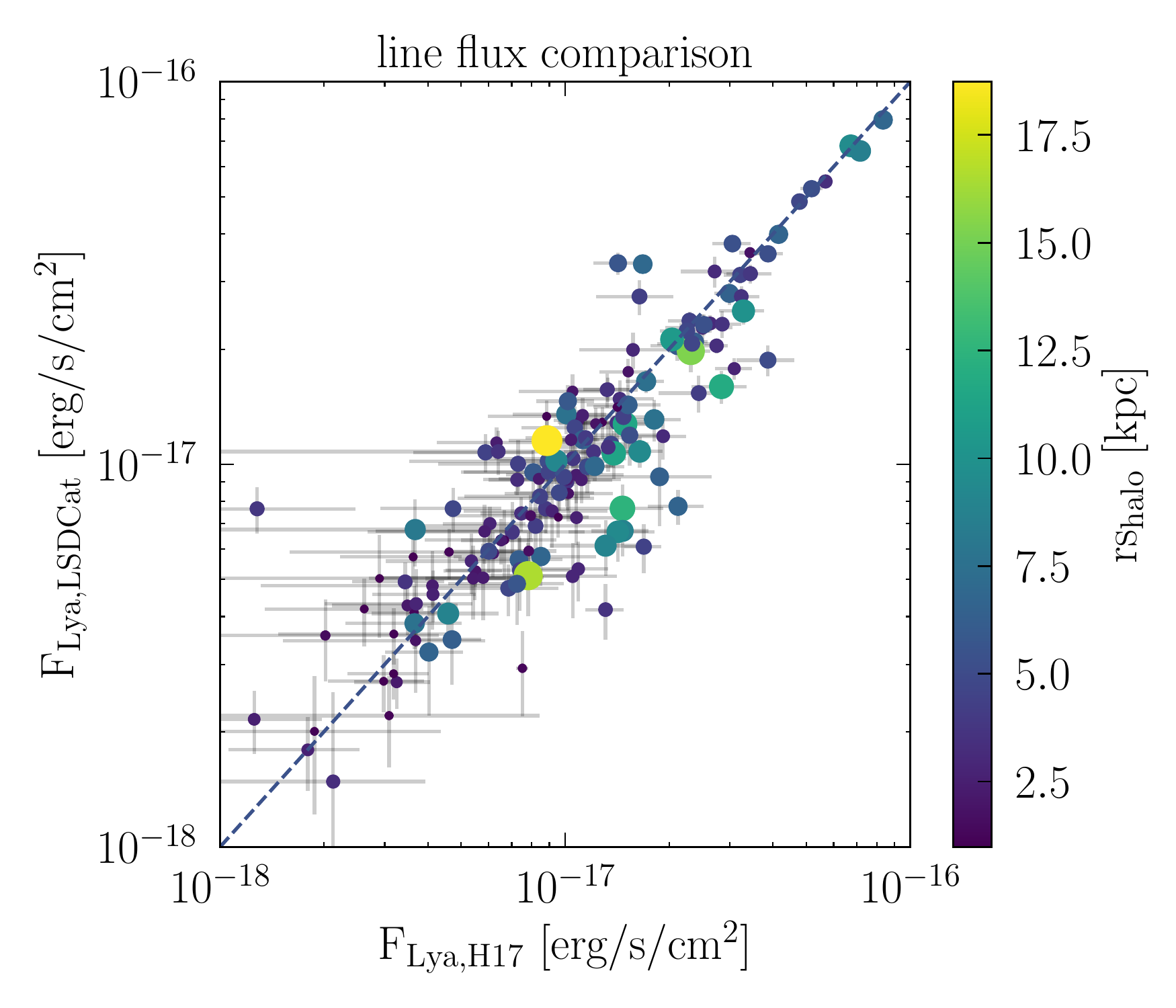}
\includegraphics[width=0.49\textwidth]{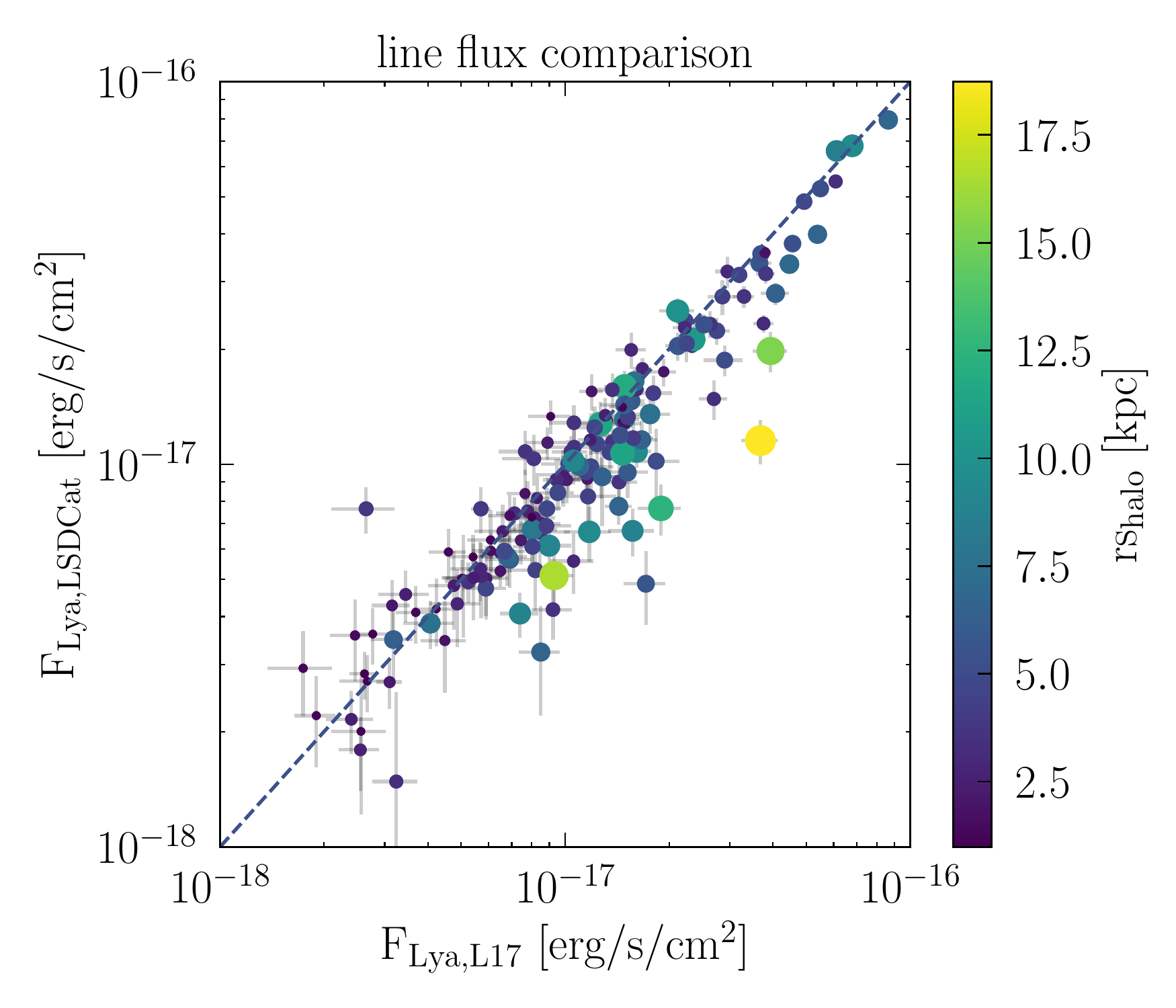}
\caption{Comparison of line flux measurements from different catalogues and methods. The Lyman $\alpha$ line flux from \citet{Hashimoto2017} (top panel, on the x-axis) and the Lyman $\alpha$ line flux from \citet{Leclercq2017} (bottom panel) are compared to the measurements from \texttt{LSDCat} (y-axis, in the same fields using the same observational depths). The one-to-one relation is shown as the dashed line and the colour-coding and sizes of the dots correspond to the halo scale lengths $\m{rs}_{\m{halo}}$ from \citet{Leclercq2017}. The line flux measurement from \citet{Hashimoto2017} is based on a curve-of-growth measurement from narrow-band images as described by \citet{Drake2017b,Drake2017a}, similar to \citet{Leclercq2017} who use an additional two-dimensional decomposition for the halo and continuum components of the LAEs.}
\label{fig:HLya_comp} 
\end{figure}

It has been shown in many studies that high redshift Lyman $\alpha$ emitters usually have extended Lyman $\alpha$ halos (\citealp{Steidel2011, Momose2014, Momose2016, Wisotzki2016, Leclercq2017, Wisotzki2018, Leclercq2020}) that can carry the majority of the emitted Lyman $\alpha$ flux ($\approx 65\%$ as shown by \citealp{Leclercq2017}). Since the EW$_0$ depends on correct measurements of the full Lyman $\alpha$ flux from a galaxy, we need to understand if we are capturing the contribution of the halo emission reasonably well. Although there exist measurements of the line flux emission that include the Lyman $\alpha$ halo for some of the objects in this paper, not all are bright enough to perform a curve-of-growth measurement as was done for example by \citet{Leclercq2017, Drake2017a, Drake2017b}, and \citet{Hashimoto2017}. In order to use a consistent method to obtain the emission line fluxes for all objects in the sample of this paper (both the objects in the MUSE-Wide fields as well as the MUSE-Deep fields), we compared the line flux measurements of \texttt{LSDCat} (which are available for all objects in our sample) to the selected sample of objects from \citet{Leclercq2017}, where the halo and UV continuum contributions of the Lyman $\alpha$ halo were analysed separately to obtain the full line flux as well as the halo sizes and the contribution of the halo to the line flux. In Fig.~\ref{fig:HLya_comp} we show a comparison between the line flux values from \citet{Hashimoto2017} and \citet{Leclercq2017} that specifically include the halo contribution and line flux measurements within three Kron radii from \texttt{LSDCat} (the same method that was used for the MUSE-Wide fields by \citealp{Urrutia2019}). It can be seen that the measurements from \texttt{LSDCat} match well for most objects, for both the \citet{Hashimoto2017} and \citet{Leclercq2017} line flux values. However, there is a slight (and expected) trend that for objects with a large halo scale length (in green and yellow in Fig.~\ref{fig:HLya_comp}), the line flux measurements from \texttt{LSDCat} are smaller and thus do not capture the full halo contribution. Nevertheless, except for those few extreme objects, the line flux measurements match well, which is why we use the \texttt{LSDCat} measurements throughout this paper in order to keep the method consistent. One should keep in mind, though, that the EW$_0$ estimates are on average slightly underestimated for objects with large halos.

%%%%%%%%%%%%%%%%%%%%%%%%%%%%%%%%%%%%%%%%%%%%%%%%%%%%%%%%%%%%%%%%%%%%%%%%%%%%%%%%%%%%%%%%%%

\section{Table of measurements}

A full table containing the measurements shown in this study can be found in the online version of this paper. An overview of the different columns and an example of 34 objects is shown in Table \ref{tab:all_measurements}. 

\begin{landscape}
\begin{table} 
\begin{center} 
\caption{Measurements from our analysis for the 34 objects with EW$_0>240\angstrom$ and a S/N>3.} 
\begin{tabular}{ l l l l l l l l l l l l l l} \hline\hline 
ID & I17 ID & RA & DEC & z & Conf. & EW$_0$ & $M_{\mathrm{UV}}$ & log($L_{\mathrm{Ly}\alpha}$) & Peak Sep. & FWHM & Asym. & R$_\mathrm{e}$ & Axis Ratio\\ 
 &  & [deg] & [deg] &  &  & [\AA] & [AB] & [erg/s] & [km/s] & [km/s] &  & [kpc] & \\ \hline 
461040 & 7159 & 53.16792 & -27.81231 & 2.99 & 3 & $298.3\pm88.47$ &-17.42 & 42.38 & 0.0 & 190.05 & 0.0 & 0.98 & 1.0\\ 
2241793 & 7199 & 53.15928 & -27.7609 & 3.33 & 3 & $254.36\pm71.03$ &-17.01 & 42.15 & 484.77 & 150.91 & -0.43 & 1.0 & 1.0\\ 
3112085 & 2515 & 53.13791 & -27.78773 & 3.47 & 3 & $303.47\pm82.19$ &-17.89 & 42.58 & 0.0 & 233.65 & 0.07 & 5.52 & 0.36\\ 
5872881 & 1969 & 53.16738 & -27.76681 & 4.08 & 3 & $298.8\pm54.6$ &-18.75 & 42.92 & 301.16 & 220.27 & 0.2 & 2.9 & 0.2\\ 
7543361 & 3699 & 53.13719 & -27.78732 & 4.4 & 3 & $285.85\pm70.27$ &-17.53 & 42.41 & 248.23 & 122.37 & 0.22 & 1.2 & 1.0\\ 
7953461 & 0 & 53.18046 & -27.77458 & 4.47 & 2 & $242.97\pm78.02$ &-17.76 & 42.43 & 185.59 & 210.92 & 0.21 & 1.0 & 1.0\\ 
9993755 & 3946 & 53.17794 & -27.77705 & 4.93 & 2 & $572.11\pm113.47$ &-18.39 & 43.06 & 318.11 & 181.02 & 0.16 & 0.68 & 1.0\\ 
22831048 & 0 & 53.1657 & -27.78485 & 5.47 & 3 & $338.66\pm48.41$ &-19.06 & 43.1 & 508.63 & 222.49 & 0.26 & 7.68 & 0.72\\ 
108009107 & 0 & 53.09144 & -27.83961 & 3.38 & 2 & $423.85\pm124.29$ &-16.94 & 42.34 & 759.82 & 407.82 & 0.0 & 1.0 & 0.5\\ 
110003005 & 0 & 53.11131 & -27.86294 & 3.41 & 2 & $369.09\pm122.58$ &-18.25 & 42.81 & 532.72 & 244.11 & 0.12 & 2.61 & 0.28\\ 
123007091 & 0 & 53.14519 & -27.83468 & 3.26 & 2 & $588.93\pm193.38$ &-17.12 & 42.56 & 197.36 & 177.38 & 0.23 & 1.56 & 0.2\\ 
132051105 & 0 & 53.09537 & -27.75587 & 5.12 & 1 & $271.93\pm85.04$ &-18.31 & 42.7 & 0.0 & 335.82 & 0.24 & 3.78 & 0.26\\ 
134006016 & 0 & 53.1284 & -27.798 & 3.15 & 3 & $437.89\pm141.26$ &-17.66 & 42.65 & 0.0 & 214.55 & 0.08 & 1.0 & 1.0\\ 
137026047 & 0 & 53.12434 & -27.76322 & 3.6 & 2 & $269.86\pm85.98$ &-17.17 & 42.24 & 0.0 & 228.07 & -0.1 & 1.0 & 1.0\\ 
143007016 & 0 & 53.0691 & -27.79423 & 3.57 & 3 & $240.98\pm75.12$ &-17.93 & 42.49 & 0.0 & 213.72 & 0.06 & 1.0 & 1.0\\ 
145004011 & 0 & 53.05244 & -27.80156 & 3.09 & 1 & $253.79\pm82.72$ &-17.44 & 42.32 & 864.34 & 120.84 & 0.36 & 1.0 & 1.0\\ 
146025262 & 0 & 53.0538 & -27.77564 & 3.77 & 2 & $247.25\pm75.78$ &-17.4 & 42.29 & 530.07 & 222.77 & 0.18 & 1.87 & 0.53\\ 
148007034 & 0 & 53.15991 & -27.75133 & 3.71 & 1 & $245.04\pm77.78$ &-17.84 & 42.47 & 0.0 & 213.05 & -0.08 & 0.78 & 1.0\\ 
152010066 & 0 & 53.20174 & -27.78535 & 3.24 & 2 & $574.33\pm166.44$ &-17.73 & 42.79 & 476.6 & 348.21 & -0.33 & 0.76 & 1.0\\ 
153046123 & 0 & 53.2103 & -27.78908 & 4.92 & 3 & $324.6\pm52.13$ &-18.81 & 42.98 & 0.0 & 178.43 & 0.17 & 0.83 & 1.0\\ 
154014100 & 0 & 53.17977 & -27.7603 & 3.77 & 3 & $280.31\pm71.58$ &-18.04 & 42.61 & 0.0 & 191.92 & 0.09 & 1.0 & 1.0\\ 
155030086 & 0 & 53.18333 & -27.79594 & 4.95 & 2 & $283.32\pm92.61$ &-18.43 & 42.77 & 0.0 & 149.79 & 0.1 & 2.01 & 0.53\\ 
157028277 & 0 & 53.04116 & -27.80352 & 3.95 & 2 & $522.33\pm157.92$ &-18.1 & 42.9 & 0.0 & 243.51 & -0.17 & 1.0 & 1.0\\ 
160005048 & 0 & 53.05919 & -27.84718 & 3.22 & 3 & $248.15\pm25.93$ &-19.46 & 43.12 & 395.45 & 289.74 & 0.18 & 2.02 & 0.67\\ 
209023164 & 0 & 150.10097 & 2.21398 & 3.82 & 3 & $252.7\pm38.45$ &-19.28 & 43.06 & 612.51 & 189.55 & 0.3 & 5.35 & 0.55\\ 
214002011 & 0 & 150.19263 & 2.21984 & 3.08 & 3 & $421.36\pm126.34$ &-18.77 & 43.08 & 782.51 & 192.1 & 0.4 & 0.83 & 0.38\\ 
218004058 & 0 & 150.13919 & 2.23506 & 3.12 & 3 & $345.4\pm82.89$ &-18.64 & 42.94 & 636.52 & 267.67 & 0.16 & 1.0 & 1.0\\ 
218025107 & 0 & 150.13344 & 2.22748 & 4.17 & 2 & $328.02\pm106.13$ &-18.27 & 42.77 & 353.47 & 133.78 & -0.16 & 3.31 & 1.0\\ 
259025087 & 0 & 150.11522 & 2.32879 & 3.94 & 2 & $244.43\pm79.11$ &-18.41 & 42.69 & 505.89 & 214.85 & 0.17 & 0.81 & 1.0\\ 
301008548 & 0 & 53.2639 & -27.69422 & 3.1 & 2 & $400.14\pm123.2$ &-17.19 & 42.42 & 0.0 & 266.55 & 0.13 & 4.57 & 0.2\\ 
301010551 & 0 & 53.25956 & -27.68777 & 3.43 & 2 & $390.8\pm130.12$ &-17.61 & 42.58 & 606.51 & 177.51 & 0.16 & 4.93 & 0.13\\ 
302037137 & 0 & 53.24913 & -27.67301 & 4.9 & 2 & $371.56\pm84.8$ &-17.83 & 42.64 & 284.42 & 170.82 & 0.1 & 13.43 & 0.3\\ 
303024064 & 0 & 53.23744 & -27.68124 & 3.63 & 2 & $270.86\pm87.64$ &-17.03 & 42.19 & 0.0 & 119.86 & 0.05 & 1.0 & 1.0\\ 
403021094 & 0 & 53.26579 & -27.85941 & 4.39 & 3 & $261.46\pm49.32$ &-18.51 & 42.76 & 0.0 & 216.2 & 0.23 & 0.91 & 0.59\\ 
\hline 
\end{tabular}\label{tab:all_measurements} 
\end{center} 
\tablefoot{Explanations of columns: ID: Identifier from the MUSE-Wide survey. I17 ID: Identifier from \citet{Inami2017} (0 if there is no match). RA and DEC: coordinates of UV continuum counterpart fitted with \texttt{Galfit}. z: Redshift based on the Lyman $\alpha$ line, corrected using the FWHM or peak separation where available using the correlations by \citet{Verhamme2018}. Conf.: Confidence of the classification as an LAE. EW$_0$: Rest-frame EW in \AA\, based on $\beta = -1.97$. $M_{\m{UV}}$: Absolute UV magnitude at $1500\,\angstrom$. log$_{10}$($L_{\m{Ly}\alpha}$): Lyman $\alpha$ luminosity given in erg/s. Peak Ratio: Ratio of the blue peak over the total line emission, zero if there is no blue bump. Peak Sep.: Separation between the two peaks in km/s. FWHM: Full width at half maximum of the main, red peak, given in km/s and corrected for the MUSE LSF. Asym.: Asymmetry parameter of the asymmetric Gaussian fit to the main red peak. R$_\m{e}$: Effective radius in (physical) kpc measured with the \texttt{Galfit} software. Axis Ratio: Morphological parameter of the UV continuum counterpart, obtained using \texttt{Galfit}.} 
\end{table}  
\end{landscape}

\end{document}